\documentclass[aps, prxq, reprint]{revtex4-2}

\usepackage{amsmath}
\usepackage{amssymb}
\usepackage{bm}
\usepackage{braket}
\usepackage[notrig]{physics}

\usepackage{float}
\usepackage{xcolor}
\usepackage{graphicx}
\usepackage[mode=buildnew]{standalone}
\usepackage[caption=false]{subfig}

\usepackage[colorlinks=true, allcolors=blue]{hyperref}
\usepackage[capitalise]{cleveref}
\crefname{section}{Sec.}{Secs.}
\AddToHook{cmd/appendix/before}{%
  \crefalias{section}{appendix}%
  \crefalias{subsection}{appendix}%
  \crefalias{subsubsection}{appendix}%
}
\usepackage{orcidlink}

\newcommand{\Z}{\mathbb{Z}}
\newcommand{\eps}{\varepsilon}

\newcommand{\ext}{_\mathrm{ext}}
\newcommand{\maxx}{_\mathrm{max}}
\newcommand{\minn}{_\mathrm{min}}
\newcommand{\intt}{_\mathrm{int}}
\newcommand{\coup}{_\mathrm{c}}

\newcommand{\vect}[1]{\mathbf{\bm{#1}}}
\newcommand{\T}{^\mathrm{T}}
\newcommand{\inv}{^{-1}}

\newcommand{\ketbasisbra}[3]{\ket*{#1}_{#2}\!\!\bra*{#3}}

\newcommand{\rhoh}{\hat{\rho}}

\renewcommand{\H}{\hat{H}}
\newcommand{\U}{\hat{U}}
\newcommand{\V}{\hat{V}}

\newcommand{\x}{\hat{x}}
\newcommand{\p}{\hat{p}}
\newcommand{\n}{\hat{n}}
\newcommand{\thetah}{\hat{\theta}}
\newcommand{\phih}{\hat{\phi}}
\newcommand{\varphih}{\hat{\varphi}}
\newcommand{\zetah}{\hat{\zeta}}
\newcommand{\alphah}{\hat{\alpha}}

\begin{document}

\title{Protected phase gate for the 0-\texorpdfstring{$\vect\pi$}{π} qubit using its internal modes}

\author{Xanda~C.~Kolesnikow}
\email{xkol5336@uni.sydney.edu.au}
\affiliation{Centre for Engineered Quantum Systems, School of Physics, University of Sydney, Sydney, NSW 2006, Australia.}

\author{Thomas~B.~Smith}
\affiliation{Centre for Engineered Quantum Systems, School of Physics, University of Sydney, Sydney, NSW 2006, Australia.}

\author{Felix~Thomsen}
\affiliation{Centre for Engineered Quantum Systems, School of Physics, University of Sydney, Sydney, NSW 2006, Australia.}

\author{Abhijeet~Alase}
\affiliation{Centre for Engineered Quantum Systems, School of Physics, University of Sydney, Sydney, NSW 2006, Australia.}

\author{Andrew~C.~Doherty}
\affiliation{Centre for Engineered Quantum Systems, School of Physics, University of Sydney, Sydney, NSW 2006, Australia.}

\date{\today}

\begin{abstract}
    Protected superconducting qubits such as the $0$-$\pi$ qubit promise to substantially reduce physical error rates.
    However, a key challenge in the field is designing gates for these qubits that do not compromise their protection, or become infeasibly slow as the protection of the qubit is improved. 
    In this work we propose a protected phase gate that is compatible with the protected regime of the $0$-$\pi$ qubit, and does not suffer from spurious coupling to additional circuit modes.
    Our gate utilises an internal mode of the circuit as an ancilla, and is achieved by varying the qubit-ancilla coupling via a tunable Josephson element.
    Through numerical simulations, we study how the gate error scales with the circuit parameters of the $0$-$\pi$ qubit and the tunable Josephson element that enacts the gate.
    Ultimately, we find that a protected gate with the $0$-$\pi$ qubit is possible with near-term circuit parameters.
    Our work opens up the possibility of performing protected gates on protected superconducting qubits, which may significantly reduce hardware overheads for quantum computation.
\end{abstract}

\maketitle

\section{Introduction}

Superconducting quantum circuits are a leading platform to achieve the low error rates demanded by quantum error-correcting codes~\cite{Acharya2023,Acharya2024}.
Qubits are typically encoded in single-mode circuits, such as the transmon~\cite{Koch2007} and fluxonium~\cite{Manucharyan2009}.
These relatively simple circuits can be engineered to highly suppress either bit-flips or phase-flips, but not both simultaneously.
Substantial reduction of error rates beyond the state-of-the-art may require qubits encoded in more complex multimode circuits that are capable of suppressing both types of errors~\cite{Gyenis2021a}. 

\begin{figure}[t]
    \centering%
    \includegraphics[scale=1]{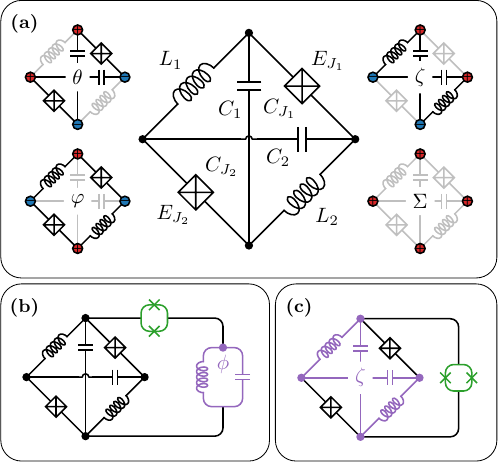}%
    \subfloat{\label{fig:zero-pi-circuit}}%
    \subfloat{\label{fig:zero-pi-ancilla-circuit-intro}}%
    \subfloat{\label{fig:zero-pi-zeta-circuit-intro}}%
    \caption{
        Protected gates for the $0$-$\pi$ qubit.
        (a) Circuit diagram for the $0$-$\pi$ qubit; two capacitors, two inductors, and two Josephson junctions are connected in the geometry shown.
        The circuit has four quadrupole circuit modes; the $\theta$ and $\varphi$ modes encode the qubit, the $\zeta$ mode is harmonic, and $\Sigma$ is nondynamical.
        Red/blue nodes represent positive/negative contributions to each mode.
        (b) Implementation of a protected phase gate using an external oscillator~\cite{Kitaev2006,Brooks2013}.
        A tunable Josephson element (green) couples the $0$-$\pi$ qubit to an oscillator mode $\phi$ (purple). 
        (c) Our scheme for a protected phase gate, which utilises the internal harmonic mode $\zeta$ (purple) instead of an external oscillator.
        }
    \label{fig:intro-fig}
\end{figure}

The $0$-$\pi$ qubit shown in \cref{fig:zero-pi-circuit} is a promising example of one such protected qubit, theoretically proposed in Refs.~\cite{Kitaev2006, Brooks2013}.
In an appropriate regime of the circuit parameters, this circuit encodes a nearly degenerate qubit that is resilient to both bit-flips and phase-flips~\cite{Dempster2014}.
This protection arises from the multimode nature of the encoding; the $0$-$\pi$ qubit is encoded in two nonlinearly coupled modes of its circuit, labelled $\theta$ and $\varphi$ in \cref{fig:zero-pi-circuit}.
The transmon-like mode $\theta$ is concatenated with the fluxonium-like mode $\varphi$, affording the qubit more protection than would be possible with a single mode.
Despite its apparent complexity, the $0$-$\pi$ circuit has been experimentally realised, albeit in a `soft' parameter regime where the encoded qubit is partially protected~\cite{Gyenis2021}.
Current experimental efforts are dedicated to improving qubit protection by pushing into the `hard' parameter regime~\cite{Kim2024,Hassani2024}.

A consequence of increased qubit protection is an increased difficulty to perform quantum gates by conventional means.
Rabi-style gates are infeasible for protected qubits since the matrix elements between computational states are small by design, leading to exponentially long gate times. 
This can be overcome in the near-term by utilising noncomputational states for Raman-style gates~\cite{Gyenis2021,Paolo2019}.
However, this strategy is only feasible for $0$-$\pi$ qubits in the partially protected parameter regime, and not in the protected regime because higher excited states decouple from the qubit manifold in the protected regime.
With imminent experimental improvements in protected qubits, new methods for performing gates in this protected regime will be crucial.

In this paper, we propose a protected phase gate for the $0$-$\pi$ qubit that is compatible with its protected regime.
The gate utilises the internal harmonic mode of the $0$-$\pi$ circuit, labelled $\zeta$ in \cref{fig:zero-pi-circuit}, as an ancilla.
Our implementation is inspired by the protected gate proposed in Refs.~\cite{Kitaev2006,Brooks2013}, shown in~\cref{fig:zero-pi-ancilla-circuit-intro}, but crucially does not utilise an additional ancillary oscillator.

As identified in Ref.~\cite{Paolo2019}, the gate proposed in Refs.~\cite{Kitaev2006,Brooks2013} is complicated by the multimode nature of the $0$-$\pi$ qubit.
In particular, the $\zeta$ mode interferes with the coupling between the qubit and the oscillator.
Due to the harmonic nature of the $\zeta$ mode, and the nonlinear coupling that arises from the tunable Josephson element [see \cref{eqn:zero-pi-interaction}], the desired interaction between the qubit and oscillator becomes exponentially suppressed in the impedance of the $\zeta$ mode [see \cref{eqn:zero-pi-interaction-exp-suppression}].
We show that this scaling is in direct competition with the protected regime of the $0$-$\pi$ qubit, which benefits from a large $\zeta$-mode impedance. 
This renders the gate infeasible for a protected $0$-$\pi$ qubit.

We circumvent this problem by utilising the $\zeta$ mode \textit{as} the requisite ancilla, since it does not play a role in the qubit encoding.
Previous studies have either neglected this mode~\cite{Brooks2013} or have considered it problematic for the lifetime of the qubit~\cite{Groszkowski2018,Paolo2019}.
Our implementation is shown in \cref{fig:zero-pi-zeta-circuit-intro}.
In this case, we show that the coupling is no longer exponentially suppressed in the $\zeta$-mode impedance and is therefore compatible with the protected regime of the $0$-$\pi$ qubit.
We also reduce the hardware requirements for the gate by eliminating the need for an external high-impedance oscillator.

We analyse the efficacy of this scheme using numerical simulations, and compare to the case of using an external oscillator.
Our simulations show that a protected gate using an external oscillator would require excessively large Josephson coupling energies, whereas using the internal $\zeta$ mode does not.
We also find that a large $\zeta$-mode impedance in combination with the large charging energy ratio of the $0$-$\pi$ qubit leads to an even larger $\varphi$-mode impedance (which is the largest impedance of the $0$-$\pi$ circuit) required for a protected gate.
In particular, we find that both schemes require a $\varphi$-mode impedance of at least fifty times the resistance quantum, but that unprotected gates with error rates below $10^{-3}$ may be obtained at smaller impedances.
Furthermore, we analyse the effect of circuit disorder and show that our gate is robust to circuit parameter asymmetries of up to 25\%.

The structure of this paper is as follows.
In \cref{sec:BKP} we analyse a toy model for a phase gate with a protected qubit, and determine the parameter regime in which the gate is protected.
In \cref{sec:zero-pi} we turn our attention to the implementation of a protected phase gate with the $0$-$\pi$ qubit.
We describe our proposal that utilises the $\zeta$ mode as an ancilla, and 
quantify the performance of the gate through numerical simulations, comparing to the case of using an external oscillator.
We show that our proposal is compatible with the protected regime of the $0$-$\pi$ qubit, whereas using an external oscillator is not.
We also introduce a one-dimensional model for the $0$-$\pi$ qubit, and discuss the possibility of a gate that uses the $\varphi$ mode as an ancilla.
In \cref{sec:hardware-constraints} we discuss the hardware requirements of our proposal.
This includes simulations of the gate with $0$-$\pi$ circuit disorder and flux noise, and a compilation of results that quantify the constraints on circuit parameters for both of the $0$-$\pi$ circuit and the tunable Josephson element. 
We also discuss the role of cooling and photon loss.
In \cref{sec:other-gates} we propose possible extensions of our scheme to protected non-Clifford gates and protected two-qubit gates.
We conclude in \cref{sec:conclusions}.

\section{Protected gate for an ideal qubit}\label{sec:BKP}

In this section, we analyse a toy model of a protected phase gate for a protected qubit, inspired by the concept introduced in Refs.~\cite{Kitaev2006,Brooks2013}.
This allows us to focus on the gate itself, agnostic to the particular physical implementation of the protected qubit.
The setup is shown in \cref{fig:BKP-circuit};
a protected qubit mode $\theta$ is coupled to a harmonic oscillator mode $\phi$ via a tunable Josephson element (depicted here as a SQUID).
The basic principle of the gate is as follows.

When the qubit-oscillator coupling is turned on, the oscillator evolves into an approximate Gottesman-Kitaev-Preskill (GKP) qubit codeword~\cite{Gottesman2001} that is dependent on the state of the protected qubit.
The oscillator is then allowed to evolve for a time $\tau$ in order to perform a logical phase gate on the GKP state encoded in the oscillator.
When the qubit-oscillator coupling is turned off, the oscillator returns to its original state, but with an accrued phase.
At the end of this sequence, the system has acquired a qubit-state-dependent phase due to the evolution of the oscillator.
We provide a detailed description of this gate sequence in the following sections.

In \cref{sec:BKP-hamiltonian} we describe the Hamiltonian for this circuit, which we find is most conveniently understood in terms of GKP stabiliser generators and logical operators.
Following this, in \cref{sec:BKP-gate} we detail the evolution of the system throughout the gate sequence.
Finally, in \cref{sec:BKP-simulations} we determine through numerical simulations the circuit parameters required of the oscillator and tunable coupler to obtain a protected gate.

\subsection{Hamiltonian}\label{sec:BKP-hamiltonian}

\begin{figure*}
    \centering%
    \includegraphics[scale=1]{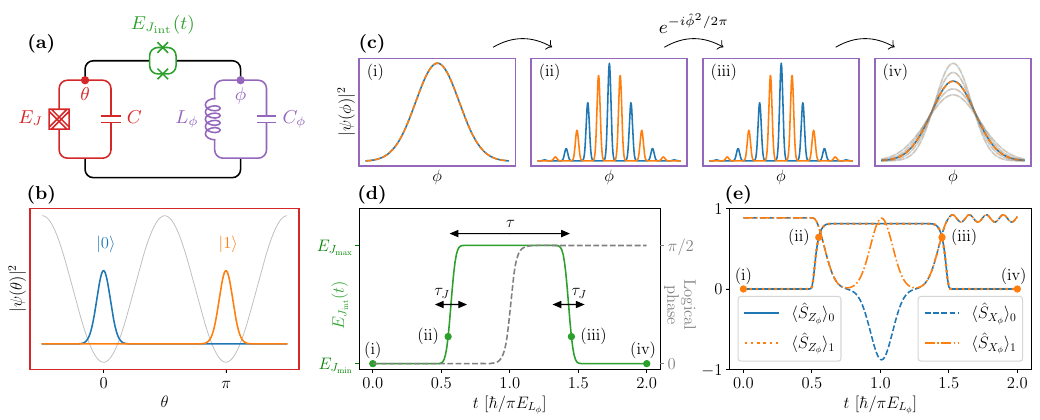}
    \subfloat{\label{fig:BKP-circuit}}%
    \subfloat{\label{fig:BKP-theta-wavefunctions}}%
    \subfloat{\label{fig:BKP-phi-wavefunctions}}%
    \subfloat{\label{fig:BKP-pulse}}%
    \subfloat{\label{fig:BKP-stabilisers}}%
    \caption{
        Protected phase gate for a protected qubit. 
        (a) Circuit diagram for a protected qubit mode $\theta$ (red) coupled to a harmonic oscillator mode $\phi$ (purple) via a tunable Josephson element (green).
        (b) Approximate eigenstates of the protected qubit GKP logical operator $\bar Z_\theta$.
        Blue and orange wavefunctions correspond to $\langle \bar Z_\theta \rangle \approx \pm 1$, respectively, and the potential energy for the qubit is shown in grey.
        (c) Qubit-state-dependent oscillator wavefunctions at various points in the gate sequence.
        (d) Pulse schedule for the tunable Josephson coupling (green); characterised by a wait-time $\tau$ and a ramp-time $\tau_J$.
        Also shown, is the relative logical phase (grey) acquired by the states $\ket{0}$ and $\ket{1}$ throughout the gate sequence.
        (e) GKP stabiliser generator expectation values of the oscillator throughout the gate sequence. 
        }
    \label{fig:BKP-gate}
\end{figure*}

For this discussion we assume an ideal protected qubit, labelled $\theta$ in \cref{fig:BKP-circuit}.
The circuit consists of a $\pi$-periodic Josephson element with energy $E_J$ shunted by a capacitor with capacitance $C$.
The $\pi$-periodic Josephson element only allows for the tunnelling of pairs of Cooper pairs.
As we will show in \cref{sec:effective-model}, this circuit serves as a good approximation to the $0$-$\pi$ circuit in its maximally protected regime.

The Hamiltonian for this circuit is
\begin{equation}\label{eqn:BKP-qubit}
    \hat H_\theta 
    = 
    4 E_C \hat n_\theta^2 
    - 
    E_J \cos {2\thetah}
    ,
\end{equation}
where $[\thetah, \hat n_\theta] = i$, $E_C = e^2/2C$, and the spectrum of $\thetah$ resides in the compact domain $[0, 2\pi)$.
This Hamiltonian possesses a double-well potential with minima at $\theta = 0$ and $\theta = \pi$, as illustrated in \cref{fig:BKP-theta-wavefunctions}.
The ideal protected qubit is distinct from the transmon, which has a single nondegenerate ground state localised near $\theta=0$. 
In contrast, for $E_J \gg E_C$, \cref{eqn:BKP-qubit} has a pair of nearly degenerate ground states localised near $\theta=0$ and $\theta=\pi$ that correspond to an encoded qubit.

These ground states are approximate GKP qubit codewords~\cite{Gottesman2001,Smith2020}. 
GKP qubit states are defined by the stabiliser generators,
\begin{equation}\label{eqn:GKP-stabilisers}
    \hat S_{X_\theta} = e^{-2i\pi \hat n_\theta},
    \quad 
    \hat S_{Z_\theta} = e^{2i \hat \theta},
\end{equation}
and the logical operators 
\begin{equation}\label{eqn:GKP-logicals}
    \bar X_\theta = e^{-i\pi \hat n_\theta},
    \quad 
    \bar Z_\theta = e^{i \hat \theta}.
\end{equation}
Whilst ideal GKP states are simultaneous $+1$ eigenstates of both stabiliser generators, physically realisable GKP states are only approximate $+1$ eigenstates of at least one of the stabiliser generators.
The $X$-type stabiliser $\hat S_{X_\theta}$ enforces $\pi$-periodicity of the wavefunction. 
The most commonly used version of the GKP code is one whose domain is unbounded in both conjugate variables.
However, GKP states may instead be encoded into a rotor, whose periodic boundary conditions grant that any state in the Hilbert is automatically a $+1$ eigenstate of $\hat S_{X_\theta} $~\cite{Gottesman2001,Smith2020,Vuillot2024}. 
We will refer to these codes as \emph{rotor-GKP codes}. 

Our model assumes periodic boundary conditions on $\theta$, meaning that every state in the Hilbert space of \cref{eqn:BKP-qubit} satisfies $\langle \hat S_{X_\theta} \rangle = 1$. 
Furthermore, for $E_J \gg E_C$ the ground states also satisfy $\langle \hat S_{Z_\theta} \rangle \approx \exp{(-\sqrt{2E_C/E_J})}$.
This can be shown by Taylor-expanding the potential at $\theta = 0$ or $\theta = \pi$ and calculating the expectation value of the approximately harmonic wavefunctions.
Therefore, when $E_J \gg E_C$ the circuit passively stabilises a rotor-GKP qubit.
This results in a highly protected superconducting qubit~\cite{Kitaev2006}.

The protected qubit is coupled to a high-impedance harmonic oscillator $\phi$ by a tunable Josephson element.
The circuit Hamiltonian is $\hat H (t) = \hat H_\theta + \hat H_\phi + \hat H \intt (t).$
The oscillator is described by the Hamiltonian
\begin{equation}\label{eqn:BKP-oscillator}
    \hat H_\phi 
    =
    4 E_{C_\phi} \hat n_\phi^2
    +
    \frac{E_{L_\phi}}{2} \hat \phi^2 
    ,
\end{equation}
where $[\phih, \hat n_\phi] = i$, $E_{C_\phi} = e^2/2C_\phi$, $E_{L_\phi} = \phi_0^2/L_\phi$, and $\phi_0 = \hbar/2e$.
The tunable Josephson element connecting $\theta$ and $\phi$ leads to a time-dependent interaction of the form
\begin{equation}\label{eqn:BKP-interaction}
    \hat H \intt (t) 
    =
    -E_{J \intt} (t)
    \cos{\big( 
        \hat \phi - \hat \theta
    \big)}
    ,
\end{equation}
where $E_{J \intt}(t)$ is the tunable Josephson energy. 
For simplicity, we assume the tunable Josephson element has no capacitance, and postpone an analysis of nonzero capacitance to \cref{sec:tunable-josephson-element}. 
We also note that both inductive loops of the circuit must be threaded with a time-dependent external flux in order to recover the interaction in \cref{eqn:BKP-interaction} (see the circuit quantisation in \cref{app:circuit-quantisation}). 

If the coupling is turned on at the correct rate, an approximate GKP qubit codeword is prepared in the oscillator.
With this in mind, we rewrite \cref{eqn:BKP-interaction} as
\begin{equation}\label{eqn:BKP-interaction-GKP}
    \hat H \intt (t) 
    =
    -\frac{E_{J \intt} (t)}{2}
    \big( 
        \bar Z_\phi \bar Z_\theta \inv + \bar Z_\theta \bar Z_\phi \inv  
    \big)
    ,
\end{equation}
where $\bar Z_\phi = e^{i\hat\phi}$ is the $Z$-type GKP logical operator for the oscillator mode.
This form of the interaction Hamiltonian elucidates the nature of the tunable Josephson element; it acts as a GKP logical $ZZ$ interaction between the protected qubit and the oscillator.
When the qubit-oscillator coupling is turned on, the oscillator evolves to an eigenstate of $\bar Z_\phi$ that is dependent on the expectation value of the qubit logical operator $\bar Z_\theta$.
This interaction is the foundation of the protected phase gate, which we describe in the following section.

\subsection{Gate sequence}\label{sec:BKP-gate}

The protected phase gate is most easily understood by tracking the evolution of the oscillator.
Let us assume that the protected qubit is in an eigenstate of $\bar Z_\theta$, such that $\langle \bar Z_\theta \rangle = \pm 1$.
The Hamiltonian for the oscillator has the simplified form
\begin{equation}\label{eqn:BKP-ideal}
    \hat H (t)
    =
    4 E_{C_\phi} \hat n_\phi^2
    +
    \frac{E_{L_\phi}}{2} \hat \phi^2
    \mp \frac{E_{J \intt} (t)}{2}
    \big( 
        \bar Z_\phi + \bar Z_\phi \inv  
    \big)
    .
\end{equation}
We neglect the qubit mode $\theta$ for this discussion since it is expected to be minimally affected during the gate.
This is because the qubit is approximately stabilised by the operators $\hat S_{X_\theta}$ and $\hat S_{Z_\theta}$.
Since the interaction Hamiltonian in \cref{eqn:BKP-interaction-GKP} commutes with both of these stabilisers, the qubit remains protected throughout the gate.

\Cref{fig:BKP-phi-wavefunctions} shows the qubit-state-dependent oscillator wavefunctions at four key moments throughout the gate, labelled (i) through to (iv).
These are also indicated on \cref{fig:BKP-pulse}, which shows the time-dependent Josephson coupling pulse that enacts the gate, and \cref{fig:BKP-stabilisers} which shows the GKP stabiliser expectation values of the oscillator.
We colour-code the plots based on the initial state of the qubit.
Blue corresponds to a protected qubit state localised at $\theta = 0$ with $\langle \bar Z_\theta \rangle = +1$, and orange corresponds to a protected qubit state localised at $\theta = \pi$ with $\langle \bar Z_\theta \rangle = -1$.
We stress that the expectation values plotted in \cref{fig:BKP-stabilisers} pertain to the dynamics of the oscillator mode $\phi$, and not the qubit mode $\theta$.

The start of the gate sequence is denoted (i).
At this point, the Josephson coupling is at a minimum value $E_{J \minn} \ll E_{C_\phi}, E_{L_\phi}$ such that the oscillator is approximately in the ground state of \cref{eqn:BKP-oscillator}.
This ground state has a large variance in $\phi$ owing to the high impedance of the oscillator. 
The GKP stabiliser expectation values at this point in the sequence are
\begin{equation}\label{eqn:phi-stabilisers-impedance}
    \langle \hat S_{X_\phi} \rangle = e^{-\pi R_Q/2Z_\phi},
    \quad
    \langle \hat S_{Z_\phi} \rangle = e^{-2\pi Z_\phi/R_Q},
\end{equation}
where $\hat S_{X_\phi} = e^{-2i\pi \hat n_\phi}$ and $\hat S_{Z_\phi} = e^{2i \hat \phi}$ are the GKP stabiliser generators for the oscillator, and $Z_\phi = \sqrt{L_\phi/C_\phi}$ is the impedance of the oscillator.
Consequently, we have that $\langle \hat S_{X_\phi} \rangle \approx 1$ and $\langle \hat S_{Z_\phi} \rangle \approx 0$.

The Josephson coupling is then turned on at a rate characterised by the ramp-time $\tau_J \approx \hbar/\sqrt{8 E_{C_\phi} E_{L_\phi}}$.
The ramp-time must be slow enough to prevent plasmonic excitations within each well of the cosine potential, but fast enough to preserve the envelope of the initial state. 
Once the Josephson coupling reaches a value $E_{J\intt}(t) \gg  E_{C_\phi}, E_{L_\phi}$, the logical $ZZ$ interaction between the protected qubit and oscillator dominates.
This causes the oscillator to evolve to an approximate eigenstate of $\bar Z_\phi$, and $\langle \hat S_{Z_\phi} \rangle$ increases.
At point (ii) in \cref{fig:BKP-stabilisers}, the expectation values of both the $X$- and $Z$-type stabiliser generators are close to one.
This signifies that the state of the oscillator is in the approximate GKP codespace.

From here, the GKP state prepared in the oscillator starts to evolve under the inductive term in \cref{eqn:BKP-oscillator}.
The unitary evolution after a time $t$ is given by the operator
\begin{equation}\label{eqn:BKP-unitary}
    \hat U (t)
    =
    e^{ -i E_{L_\phi} \hat \phi^2 t/2\hbar}
    .
\end{equation}
This operator commutes with $\hat S_{Z_\phi}$, but not with $\hat S_{X_\phi}$.
The effect of this can be observed in \cref{fig:BKP-stabilisers}: the mid-pulse expectation values of the $X$-type stabiliser differ depending on the initial qubit logical state.
This leads to a fault-tolerant logical phase gate for the GKP state encoded in the oscillator after an evolution time $t = \hbar/\pi E_{L_\phi}$~\cite{Gottesman2001}.
While the gate time increases linearly with the inductance of the resonator $L_\phi$, we show in \cref{sec:BKP-impedance} that the gate error decreases exponentially with the impedance of the oscillator $\sqrt{L_\phi/C_\phi}$. 
At point (iii), the oscillator is back in the approximate GKP codespace.
However, the system acquires a relative phase dependent on the initial state of the qubit.
The evolution of this relative phase is plotted in grey in \cref{fig:BKP-pulse}.

The end of the gate sequence is denoted (iv).
When the coupling is turned off at the rate characterised by $\tau_J$, the oscillator returns to an approximate ground state of \cref{eqn:BKP-oscillator}, but may be left with some excess entropy depending on the exact timing of the pulse.
This mistiming error leaves the oscillator `ringing', which is observed in the oscillations of $\langle \hat S_{X_\phi} \rangle$ in \cref{fig:BKP-stabilisers}. 
Importantly, this extra entropy may be extracted by cooling the oscillator mode, leaving the qubit unaffected.

This gate is protected for three reasons.
First, the qubit-oscillator coupling commutes with the stabiliser generators of the protected qubit. 
Second, the high impedance of the oscillator along with the strong Josephson coupling allows for the preparation of an approximate GKP logical state in the oscillator.
Third, the GKP state encoded in the oscillator possesses a fault-tolerant logical phase gate.

The second condition places constraints on the oscillator impedance $Z_\phi$ and the maximum Josephson coupling $E_{J\maxx}$.
If they are not sufficiently large, then the state in the oscillator will be a poor approximation of a GKP state.
We require the following energy hierarchy:
\begin{equation}\label{eqn:BKP-energy-constraint}
    E_{L_\phi} 
    \ll E_{C_\phi} 
    \ll E_{J_\text{max}}
    .
\end{equation} 
The third condition provides flexibility in the details of the pulse shape $E_{J\intt}(t)$.
Due to the fault-tolerant nature of the GKP logical phase gate, the gate is robust to errors in the pulse wait-time $\tau$ and ramp-time $\tau_J$.
The degree to which the gate is robust is determined by the quality of the GKP state.
Thus, lower tolerance to errors can be traded for reduced hardware requirements on $Z_\phi$ and $E_{J\maxx}$ and vice versa.
In the following section, we numerically quantify constraints on these parameters to obtain a protected gate. 

\subsection{Numerical simulations}\label{sec:BKP-simulations}

Here we simulate the dynamics of the protected gate under the Hamiltonian $\hat H (t)= \hat H_\theta + \hat H_\phi + \hat H \intt (t)$ where the three terms are given by \cref{eqn:BKP-qubit,eqn:BKP-oscillator,eqn:BKP-interaction}, respectively.
When $E_{J \intt} (t) = 0$, the two ground states of this Hamiltonian have a tensor product structure:
\begin{equation}\label{eqn:BKP-0-1}
    \ket*{+} 
    =
    \ket*{+}_\theta \otimes \ket{0}_\phi
    ,
    \quad
    \ket*{-} 
    =
    \ket*{-}_\theta \otimes \ket{0}_\phi
    ,
\end{equation}
where $\ket*{\pm}_\theta$ are the $\pm 1$ eigenstates of the rotor-GKP logical operator $\bar X_\theta$ and the ground states of \cref{eqn:BKP-qubit}, and $\ket{0}_\phi$ is the harmonic oscillator ground state of \cref{eqn:BKP-oscillator}.
We emphasise that the following simulations treat both the qubit and oscillator as dynamical variables, and do not assume a perfect qubit like the previous section.
We provide the details of our numerical methods in \cref{app:numerical-simulations}.

For our simulations, we initialise the system in the ground state $\ket{\psi_\mathrm{i}} = \ket*{+}_\theta \otimes \ket{0}_\phi$.
We vary the Josephson coupling following the error-function-shaped pulse shown in \cref{fig:BKP-pulse}. 
As described in the previous section, this should result in a logical phase and the final state $\ket{\psi_\mathrm{f}} = \ket*{-i}_\theta \otimes \ket{0}_\phi$, where $\ket*{-i}_\theta$ is the $-1$ eigenstate of the rotor-GKP logical operator $\bar Y_\theta$. 

We use a metric for the gate performance that takes into account the passive error-correcting properties of the $0$-$\pi$ qubit.
Whilst the logical operator $\bar{Z}_\theta$ may be used, this is overly pessimistic since superpositions of the ground states of \cref{eqn:BKP-qubit} are only $\pm 1$ eigenstates of this operator in the $E_J/E_C \to \infty$ limit. 
Due to the properties of the rotor-GKP code, the modular value of $\theta$ can be used to determine the qubit state, whereby states whose support lies in the range $\theta = [-\pi/2,\pi/2)$ are deemed to be correctable to $\ket{0}$ and states whose support lies in $\theta = [\pi/2, 3\pi/2)$ are deemed to be correctable to $\ket{1}$. 
To this end, in \cref{app:subsystem-decomp} we use a subsystem decomposition~\cite{Pantaleoni2020,Shaw2024} to define effective qubit operators $\bar X_\text{eff}$, $\bar Y_\text{eff}$, and $\bar Z_\text{eff}$, whereby the $\pm 1$ eigenstates of $\bar Z_\text{eff}$ are precisely the states that can be corrected to produce logical eigenstates of $\bar{Z}_\theta$. 
We stress that this simply amounts to a redefinition of the qubit and does not imply any additional error-correction protocol is required.

In \cref{app:dnorm} we show that the diamond norm deviation of the gate from its ideal action may be expressed in terms of these operators, and is approximately 
\begin{equation} \label{eqn:dnorm-0}
    \eps_\diamond 
    \approx 
    \frac{1}{2}
    \sqrt{
        (1 + \langle \bar Y_\text{eff} \rangle)^2 + 
        \langle \bar X_\text{eff} \rangle^2
        }
    ,
\end{equation}
where the expectation values are taken with respect to the final state after the gate. 
Note that because the effective qubit operators act as the identity on the oscillator Hilbert space, this metric is agnostic to the state of the oscillator. 
We use the diamond norm deviation rather than the average gate fidelity because it aligns with the gate error in threshold theorems~\cite{Kitaev1997,Aharonov1997,Knill1998}. 
For coherent errors, which we focus on here, the average gate fidelity potentially underestimates the effect of an error in the context of fault-tolerant quantum computing~\cite{Sanders2015,Kueng2016}.

We describe the qubit-oscillator system with the following parameters for our simulations.
The qubit is characterised by the tunnelling energy ratio $E_J/E_{C_\phi}$, and the oscillator by its impedance $Z_\phi/R_Q$.
The relative energy scale of the two systems is determined by the charging energy ratio $E_{C_\phi}/E_C$, and the Josephson coupling between them is captured by the minimum and maximum coupling ratios $E_{J\minn}/E_{C_\phi}$ and $E_{J\maxx}/E_{C_\phi}$, respectively.

\begin{figure*}
    \centering%
    \includegraphics[scale=1]{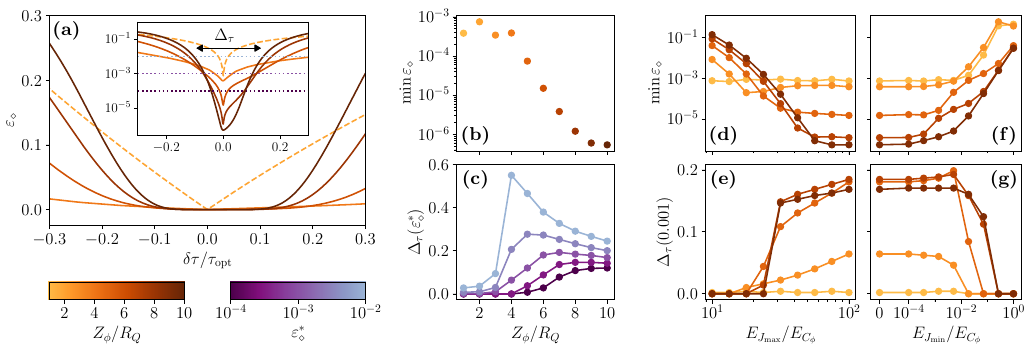}%
    \subfloat{\label{fig:gate-errors}}%
    \subfloat{\label{fig:imprecision-Z}}%
    \subfloat{\label{fig:robustness-Z}}%
    \subfloat{\label{fig:imprecision-EJ-max}}%
    \subfloat{\label{fig:robustness-EJ-max}}%
    \subfloat{\label{fig:imprecission-EJ-min}}%
    \subfloat{\label{fig:robustness-EJ-min}}%
    \caption{
        Performance of a protected phase gate for an ideal qubit.
        (a) Gate error $\varepsilon_\diamond$ (measured by the diamond norm deviation) as a function of deviations in the pulse wait-time $\delta \tau$, where $\delta \tau$ is given by \cref{eqn:delta-tau}, calculated for different values of the oscillator impedance $Z_\phi$.
        Protected gates are marked in solid lines whereas the dashed line denotes an unprotected gate. 
        Inset: identical data on a log-linear plot. 
        Dotted lines correspond to threshold gate errors $\eps_\diamond^*$ for (c).
        (b) Gate imprecision (measured by the minimum gate error $\min \varepsilon_\diamond$) as a function of $Z_\phi$. 
        (c) Gate robustness (measured by the maximum range of $\delta \tau$ to remain below a threshold error rate $\eps_\diamond^*$) as a function of $Z_\phi$.
        (d) Imprecision, and (e) robustness as a function of the maximum Josephson coupling $E_{J\maxx}$.
        (f) Imprecision, and (g) robustness as a function of the minimum Josephson coupling $E_{J\minn}$.
        For these simulations, $E_{C_\phi}/E_C = 100$, $E_J/E_{C_\phi} = 1$, and in panels (a) -- (c), $E_{J \minn} = 0$ and $E_{J \maxx} / E_{C_\phi} = 100$.
        }
    \label{fig:BKP-numerics}
\end{figure*}

\subsubsection{Impedance of the oscillator}\label{sec:BKP-impedance}

Our first set of simulations interrogate the protected nature of the gate for different values of the oscillator impedance.
For these simulations we set $E_{J \minn}/E_{C_\phi} = 0$ and $E_{J \maxx}/E_{C_\phi} = 100$.
In the following section we show that these constraints can be relaxed while maintaining a protected gate.

To analyse the protection of the phase gate, we consider the effect of a mistimed pulse on the gate error $\varepsilon_\diamond$. 
An unprotected gate (e.g. a Rabi pulse) is one in which a mistimed gate leads to a linear increase in the gate error when measured by the diamond norm. 
In contrast, a protected gate is one whose sensitivity to gate mistiming is exponentially suppressed in some system parameter.

In \cref{fig:gate-errors} we plot the gate error $\varepsilon_\diamond$ as a function of the mistiming parameter
\begin{equation}\label{eqn:delta-tau}
    \delta \tau = \tau - \tau_\mathrm{opt}, 
\end{equation}
where $\tau$ is the pulse wait-time as indicated in \cref{fig:BKP-pulse}, and $\tau_\mathrm{opt}$ is the optimal wait-time.
We plot the gate error for multiple values of $Z_\phi/R_Q$ and
observe two qualitatively different types of gates.
For $Z_\phi/R_Q = 2$ (dashed line), the gate error increases linearly around $\delta \tau = 0$, signifying an unprotected gate.
For $Z_\phi/R_Q = 4$--$10$ (solid lines), the gate error is extremely flat near $\delta \tau = 0$, indicating that the gate is protected against mistiming.

We consider two characteristic parameters to quantify the imprecision and robustness of the protected gate.
Imprecision is captured by the minimum gate error $\min(\eps_\diamond)$ at $\delta \tau = 0$.
Robustness is captured by the maximum range of mistiming $\Delta_\tau (\varepsilon_\diamond^*)$ to remain below a particular gate error threshold $\varepsilon_\diamond^*$.
A protected gate should have an imprecision that is exponentially suppressed in some circuit parameter, along with some nonzero robustness at a reasonable threshold error rate.
We plot both of these quantities as a function $Z_\phi/R_Q$ in \cref{fig:imprecision-Z,fig:robustness-Z}, respectively.
In \cref{fig:robustness-Z} we plot the robustness for different threshold error rates $\varepsilon_\diamond^*$.

\Cref{fig:imprecision-Z} shows that the imprecision is exponentially suppressed with the impedance of the oscillator for $Z_\phi/R_Q \ge 4$, and plateaus around $Z_\phi/R_Q = 10$. 
This exponential suppression is indicative of a protected gate.
The plateau arises due to the finite charging energy ratio between the oscillator and the protected qubit, which we have set to $E_{C_\phi} / E_C = 100$. This leads to a nonzero tunnelling rate between the two computational states during the gate.
In \cref{app:charging-energy-ratio} we show that the imprecision continues to decrease for larger $E_{C_\phi} / E_C$.
In the limit that $E_C \rightarrow 0$ then $E_J/E_C \rightarrow \infty$ and the superposition of the ground states of \cref{eqn:BKP-qubit} are the $\pm 1$ eigenstates of the logical operator $\bar Z_\theta$.
In this case, the system reduces to the ideal model \cref{eqn:BKP-ideal}, and the gate error continues to be exponentially suppressed at arbitrarily large $Z_\phi/R_Q$.

\Cref{fig:robustness-Z} shows that the robustness increases from zero at approximately $Z_\phi/R_Q = 4$ for multiple values of $\varepsilon_\diamond^*$.
This indicates that the gate has appreciable tolerance to mistiming errors, providing further evidence that the gate is protected for $Z_\phi/R_Q \ge 4$.

The dependence of the robustness on $Z_\phi/R_Q$ seen in \cref{fig:robustness-Z} is nonmonotonic and for each $\eps_\diamond^*$, there exists an optimal value of $Z_\phi/R_Q$ that maximises the robustness of the gate.
This arises from the fact that there is an optimal value of $\expval*{\hat{S}_{X_\phi}}$ to minimise the gate error for a given mistiming error (see \cref{app:phase-latching}). 
Since this expectation value is controlled by the impedance of the oscillator [\cref{eqn:phi-stabilisers-impedance}], this leads to an optimal value of the impedance.  

From these results, we conclude that a minimum oscillator impedance of $Z_\phi/R_Q \approx 4$ is required to obtain a protected gate.
Note that while here we have considered the pulse wait-time $\tau$ as the basic metric to study the protection of the gate, in \cref{app:ramp-time} we show that the protected gate is similarly robust to deviations in the pulse ramp-time $\tau_J$.

\subsubsection{Dynamic range of the Josephson coupling}\label{sec:BKP-dynamic-range}

In the previous section we assumed that the tunable Josephson coupling has minimum and maximum values of $E_{J\minn}/E_{C_\phi} = 0$ and $E_{J \maxx} / E_{C_\phi} = 100$.
The most obvious candidate to achieve a tunable Josephson coupling is a SQUID, which consists of two Josephson junctions in a loop.
The effective Josephson energy is tunable via the external magnetic flux threading the loop, ideally going to zero at half-flux. 
However, inevitable asymmetry in the Josephson energies of the junctions leads to a nonzero effective Josephson energy at half-flux~\cite{Koch2007,Blais2021}.
Motivated by this limitation, here we quantify the dynamic range of the Josephson coupling $E_{J\maxx}/E_{J\minn}$ required for a protected gate.

In \cref{fig:imprecision-EJ-max,fig:robustness-EJ-max,fig:imprecission-EJ-min,fig:robustness-EJ-min} we plot the imprecision and robustness of the gate (as defined previously) as a function of both $E_{J\minn}/E_{C_\phi}$ and $E_{J \maxx}/E_{C_\phi}$. 
Each curve corresponds to a different value of $Z_\phi/R_Q$.
For the robustness parameter, we choose a fixed threshold error rate of $\varepsilon_\diamond^* = 10^{-3}$. 
For the remainder of the paper, we choose this value when quantifying robustness.

\Cref{fig:imprecision-EJ-max} shows that, for $Z_\phi/R_Q \ge 4$, the imprecision is exponentially suppressed with increasing $E_{J_\text{max}}/E_{C_\phi}$ and then plateaus.
The point at which this plateau occurs increases with increasing $Z_\phi/R_Q$.
This means that increasing the maximum Josephson coupling beyond a particular value is only beneficial for a sufficiently large oscillator impedance.
\Cref{fig:robustness-EJ-max} shows that, for $Z_\phi/R_Q \ge 4$, the gate possesses nonzero robustness for $E_{J \maxx}/E_{C_\phi} \ge 30$. 
We use this as a rough estimate of the minimum value of $E_{J \maxx}/E_{C_\phi}$ for a protected gate.

\Cref{fig:imprecission-EJ-min} shows that, for $Z_\phi/R_Q \ge 4$, the imprecision is exponentially suppressed with decreasing $E_{J \minn}/E_{C_\phi}$ and then plateaus.
This means that decreasing the minimum Josephson coupling below a particular value is only beneficial for a sufficiently large oscillator impedance.
\Cref{fig:robustness-EJ-min} shows that, for $Z_\phi/R_Q \ge 4$, the gate possesses nonzero robustness for values below $E_{J \minn}/E_{C_\phi} \approx 0.005$, and for $Z_\phi/R_Q \ge 8$ this value increases to $E_{J \minn}/E_{C_\phi} \approx 0.08$.

From these results, we find that a protected gate requires a tunable tunable Josephson element with a minimum dynamic range of $E_{J_\text{max}}/E_{J_\text{min}} \approx 6 \times 10^3$ for $Z_\phi/R_Q \ge 4$.
This requirement is slightly alleviated for a larger oscillator impedance; we require $E_{J \maxx} / E_{J \minn} \approx 4 \times 10^2$ for $Z_\phi/R_Q \ge 8$. 
In \cref{sec:tunable-josephson-element} we discuss some ideas for how to achieve these large dynamic ranges.

These hardware requirements are obtained assuming an ideal protected qubit.
In the following section, we describe our implementation of this protected gate for the $0$-$\pi$ qubit, and discuss how the hardware requirements of the circuit are changed.

\section{Protected gate for the \texorpdfstring{$\vect0$}{0}-\texorpdfstring{$\vect\pi$}{π} qubit}\label{sec:zero-pi}

In this section, we describe our proposal for a protected gate with the $0$-$\pi$ qubit.
In \cref{sec:zero-pi-hamiltonian} we introduce the $0$-$\pi$ qubit, and explain the difficulties encountered when utilising an external oscillator as an ancilla for a protected phase gate.
We propose a modified gate that instead utilises an internal harmonic mode.
In \cref{sec:effective-model} we describe an effective model for the $0$-$\pi$ qubit that facilitates a numerical investigation of the hardware requirements of the gate, which we present in \cref{sec:numerical-simulations-zero-pi}.
Finally, in \cref{sec:phi} we briefly discuss the possibility of utilising another internal mode of the $0$-$\pi$ qubit for a protected gate.

\subsection{Hamiltonian}\label{sec:zero-pi-hamiltonian}

The circuit for the $0$-$\pi$ qubit is shown in \cref{fig:zero-pi-circuit}.
It is a four-node circuit that consists of a pair of capacitors with capacitances $C_{1,2}$, a pair of inductors with inductances $L_{1,2}$, and a pair of Josephson junctions with Josephson energies $E_{J_{1,2}}$ and  capacitances $C_{J_{1,2}}$.
The Hamiltonian for this circuit is most conveniently expressed in terms of the modes $\theta$, $\varphi$ and $\zeta$.
These modes are quadrupole combinations of the original circuit nodes that are illustrated in \cref{fig:zero-pi-circuit}.
A fourth mode $\Sigma$ does not appear in the Hamiltonian and is discarded.
We derive the full Hamiltonian in \cref{app:zero-pi-circuit}.

For symmetric circuit parameters between each pair of elements (i.e. $C_1 = C_2 = C$, $L_1 = L_2 = L$, etc.) the Hamiltonian for the $0$-$\pi$ qubit is $\hat H = \hat H_{\theta\varphi} + \hat H_\zeta$.
The qubit is encoded in the two-mode Hamiltonian 
\begin{equation}\label{eqn:zero-pi-theta-phi}
    \hat H_{\theta\varphi}
    =
    4 E_{C_\theta} \n_\theta^2 
    + 4 E_{C_\varphi} \n_\varphi^2
    + E_L 
    \varphih 
    ^2
    - 2 E_J\cos\thetah\cos\varphih
    ,
\end{equation}
where $[\thetah, \hat n_\theta] = i$, $[\varphih, \hat n_\varphi] = i$, $E_{C_\theta} = e^2/4(C+C_J)$, $E_{C_\varphi} = e^2/4C_J$, and $E_L = \phi_0^2/L$.
The $\theta$ and $\varphi$ modes are decoupled from the harmonic mode $\zeta$:
\begin{equation}\label{eqn:zero-pi-zeta}
    \hat H_\zeta
    =
    4 E_{C_\zeta} \n_\zeta^2 
    + E_L \zetah^2
    ,
\end{equation}
where $[\zetah, \hat n_\zeta] = i$ and $E_{C_\zeta} = e^2/4C$.
The two-mode Hamiltonian in \cref{eqn:zero-pi-theta-phi} has an egg-carton-shaped potential with local minima at values $\theta + \varphi = 0 \pmod{2\pi}$ and local maxima at $\theta + \varphi = \pi \pmod{2\pi}$, as shown in \cref{fig:zero-pi-potential-energy}.
In the parameter regime where
\begin{equation}\label{eqn:zero-pi-energy-constraint}
    E_L, E_{C_\theta} 
    \ll 
    E_J, E_{C_\varphi}
    ,
\end{equation}
the circuit has a pair of nearly degenerate ground states that are simultaneously localised in $\theta$ ($E_J \gg E_{C_\theta}$) and delocalised in $\varphi$ ($E_L \ll E_{C_\varphi}$) as shown in \cref{fig:zero-pi-wavefunctions}.
In this parameter regime, the $0$-$\pi$ qubit is protected from both dephasing and 
relaxation~\cite{Gyenis2021a}, and its Hamiltonian is well-described by an effective one-dimensional model that approaches the Hamiltonian in \cref{eqn:BKP-qubit}.
We explain this in further detail in \cref{sec:effective-model}.

The $0$-$\pi$ qubit is characterised by a charging energy ratio $E_{C_\theta}/E_{C_\varphi}$, a Josephson energy $E_J/E_{C_\varphi}$ and an inductive energy $E_L/E_{C_\varphi}$. 
The protected regime corresponds to the limit of $E_{C_\theta}/E_{C_\varphi} \rightarrow 0$ and $E_L/E_{C_\varphi} \rightarrow 0$. 
However, at nonzero values of $E_{C_\theta}/E_{C_\varphi}$ and $E_L/E_{C_\varphi}$, there exists an optimal value of $E_J/E_{C_\varphi}$ that maximises the protection of the qubit~\cite{Dempster2014}.
We analyse this optimal value in \cref{app:optimal-EJ}.

We now consider a protected gate for the $0$-$\pi$ qubit. 
The most straightforward extension of the scheme discussed in \cref{sec:BKP} is to replace the ideal qubit with the $0$-$\pi$ circuit, resulting in the circuit shown in \cref{fig:zero-pi-ancilla-circuit-intro}. 
The Hamiltonian is $\hat H (t)= \hat H_{\phi} + \hat H_{\theta\varphi} + \hat H_{\zeta} + \hat H^\phi \intt (t)$, where the first three terms are given by \cref{eqn:zero-pi-theta-phi,eqn:zero-pi-zeta,eqn:BKP-oscillator}, respectively, and the interaction term is~\cite{Paolo2019}
\begin{equation}\label{eqn:zero-pi-interaction}
    \hat H^\phi \intt (t) 
    =
    -E_{J \intt} (t)
    \cos{\big( 
        \hat \phi - \hat \theta + \hat \zeta
    \big)}
    ,
\end{equation}
as we show in \cref{app:zero-pi-oscillator-circuit}.
The harmonic $\zeta$ mode is involved in the qubit-oscillator interaction.
This is problematic for the protected gate.
To see this, we assume the $\zeta$ mode is in the ground state of \cref{eqn:zero-pi-zeta}. This is a generous assumption considering that the mode is naturally low-frequency in the protected regime described by \cref{eqn:zero-pi-energy-constraint}, and will have increasingly high thermal population.
Nevertheless, the interaction term becomes 
\begin{equation}\label{eqn:zero-pi-interaction-exp-suppression}
    \hat H^\phi \intt (t) 
    =
    -E_{J \intt} (t)
    e^{-{\pi Z_\zeta}/{2 R_Q}}
    \cos{\big( 
        \hat \phi - \hat \theta 
    \big)}
    ,
\end{equation}
where $Z_\zeta = \sqrt{L/4C}$ is the $\zeta$-mode impedance.
The effective Josephson coupling between the $0$-$\pi$ qubit and the oscillator is exponentially suppressed in the impedance of the $\zeta$ mode, which makes it very difficult to achieve the $E_{J \maxx}$ required for a protected gate.

To solve this issue, we propose a gate that utilises the $\zeta$ mode in place of an external oscillator.
This is achieved by shunting the $0$-$\pi$ circuit with a tunable Josephson element as shown in \cref{fig:zero-pi-zeta-circuit-intro}.
In \cref{app:zero-pi-zeta-circuit} we show that this leads to an interaction Hamiltonian of the form 
\begin{equation}\label{eqn:zero-pi-zeta-interaction}
    \hat H^\zeta \intt (t) 
    =
    -E_{J \intt} (t)
    \cos{\big( 
        \hat \zeta + \hat \theta
    \big)}
    .    
\end{equation}
Comparing this expression to \cref{eqn:zero-pi-interaction-exp-suppression} we can see that $\zeta$ takes the role of the ancilla, replacing the oscillator mode $\phi$.
Importantly, the qubit-ancilla interaction is no longer suppressed in the impedance of the $\zeta$ mode.

In the following sections, we investigate the circuit parameters required of the $0$-$\pi$ qubit to obtain a protected gate.
We show that the interaction described by \cref{eqn:zero-pi-zeta-interaction} is compatible with the protected regime of the $0$-$\pi$ qubit, while the interaction described by \cref{eqn:zero-pi-interaction-exp-suppression} is not.
First, however, we describe a single-mode effective model for the two-mode Hamiltonian in \cref{eqn:zero-pi-theta-phi}.
This allows us to reduce the overhead of our numerical simulations.

\subsection{Effective model}\label{sec:effective-model}

We seek to reduce the time-dependent Hamiltonian $\H(t)$ from three modes to two modes.
To achieve this, we approximate the Hamiltonian in \cref{eqn:zero-pi-theta-phi} with a single-mode effective model. 
Single-mode models for the $0$-$\pi$ qubit have been explored previously~\cite{Shen2015,Paolo2019,Hassani2024}. 
Here we use a tight-binding approximation in the discrete local minimum basis~\cite{Koch2009,Smith2022}.
This is valid in the parameter regime $2E_J > E_{C_\varphi},E_{C_\theta}, 2E_L$.
A derivation of the effective model is provided in \cref{app:effective-model}, along with a discussion of an effective model when $2E_J$ is not the dominant circuit parameter.

\begin{figure}
    \centering%
    \includegraphics[scale=1.0]{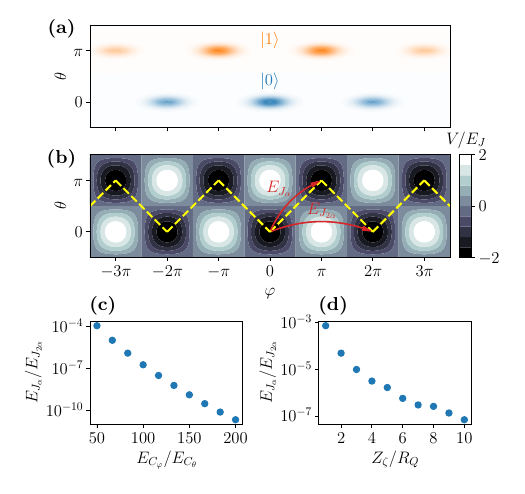}%
    \subfloat{\label{fig:zero-pi-wavefunctions}}%
    \subfloat{\label{fig:zero-pi-potential-energy}}%
    \subfloat{\label{fig:tunnelling-rates-mass-ratio}}%
    \subfloat{\label{fig:tunnelling-rates-Z-zeta}}%
    \caption{
        Effective one-dimensional model for the $0$-$\pi$ qubit. 
        (a) Wavefunctions for the $0$-$\pi$ qubit logical states.
        (b) Two-dimensional potential of the $0$-$\pi$ qubit, with the effective one-dimensional model (yellow) constructed from the local minima.
        The nearest-neighbour and next-nearest-neighbour tunnelling rates (red) are $E_{J_\alpha}$ and $E_{J_{2\alpha}}$, respectively.
        (c) Ratio of the tunnelling rates $E_{J_\alpha}/E_{J_{2\alpha}}$ as a function of the $0$-$\pi$ qubit charging energy ratio $E_{C_\varphi}/E_{C_\theta}$.
        (d) Ratio of the tunnelling rates as a function of the $\zeta$-mode impedance $Z_\zeta$, with $E_J/E_{C_\varphi}$ given by the value that maximises the protection of the $0$-$\pi$ qubit at each impedance.
        Throughout all results, $E_{C_\varphi}/E_{C_\theta} = 100$, and in panels (a) -- (c), $E_J/E_{C_\varphi} = 5$ and $Z_\zeta / R_Q = 10$.
        }
\end{figure}

We construct a one-dimensional system based on the local minima of the two-dimensional potential, as illustrated in \cref{fig:zero-pi-potential-energy}.
The effective Hamiltonian for this one-dimensional system is
\begin{equation}\label{eqn:H-alpha}
    \hat H_\alpha 
    =
    4 E_{C_\alpha} \n_\alpha^2 
    - E_{J_\alpha} \cos\alphah
    - E_{J_{2\alpha}} \cos2\alphah 
    ,
\end{equation}
where $[\hat \alpha, \hat n_\alpha] = i$ and the effective energies are given by
\begin{align}
    E_{C_\alpha} 
    &= 
    \frac{\pi^2 E_L}{4}, \label{eqn:EC-alpha}\\
    E_{J_\alpha} 
    &\sim 
    e^{-2(\sqrt{2} - 1) \sqrt{2E_J/E_{C_\theta}} - \pi(\sqrt{2} - 1)\sqrt{E_J/2 E_{C_\varphi}}}, \label{eqn:EJ1}\\
    E_{J_{2\alpha}} 
    &\sim
    e^{-4\sqrt{E_J / E_{C_\varphi}}} \label{eqn:EJ2}
    .
\end{align}
This effective Hamiltonian is similar to \cref{eqn:BKP-qubit}, with the addition of a term $\cos\alphah$ that splits the near-degeneracy of the two qubit states.
In the language of the logical operators for the rotor-GKP code, this additional term acts like the $Z$-type logical operator $\bar Z_\alpha = e^{i\hat \alpha}$.

The expressions for the effective charging energy $E_{C_\alpha}$ and tunnelling energies $E_{J_\alpha}$ and $E_{J_{2\alpha}}$ are derived from the tight-binding model and depend on the circuit parameters of the $0$-$\pi$ qubit. 
The charging energy is proportional to the inductive energy $E_L$ of the $0$-$\pi$ qubit because flux-like variables get mapped to charge-like variables in the effective model.
We stress that $\alphah$ may therefore not be regarded as either of the phase variables $\thetah$ or $\varphih$ of the $0$-$\pi$ qubit.
The exponential scalings for the tunnelling rates are calculated from the classical action over the tunnelling paths.
Their prefactors, which are polynomials in the circuit parameters, may be found using WKB or instanton methods~\cite{Smith2020,Zwiebach2022}.

First we consider $E_{J_{2\alpha}}$, which describes next-nearest-neighbour tunnelling in the effective model.  
Since this tunnelling occurs solely in the $\varphi$ direction and there is a large separation in the charging energies ($E_{C_\varphi} \gg E_{C_\theta}$), the qubit wavefunctions are confined in the $\theta$ direction.
Thus, $E_{J_{2\alpha}}$ may be approximated using the one-dimensional tunnelling rate for a transmon~\cite{Koch2007}; see \cref{eqn:t2} for its full expression.
We find this to be a good approximation for $E_{C_\varphi} / E_{C_\theta} \ge 50$.

Now we consider $E_{J_{\alpha}}$, the nearest-neighbour tunnelling rate.
Unlike next-nearest-neighbour tunnelling, nearest-neighbour tunnelling occurs through a nontrivial path in $\theta$ and $\varphi$, which is computed in \cref{app:tunnelling-rates}.
Due to the component of the tunnelling path in the $\theta$ direction, its rate is exponentially suppressed in $\sqrt{E_{C_\varphi}/E_{C_\theta}}$ relative to $E_{J_{2\alpha}}$.
Therefore, $E_{J_\alpha}$ is suppressed relative to $E_{J_{2\alpha}}$ for the protected regime of the $0$-$\pi$ qubit [\cref{eqn:zero-pi-energy-constraint}].
For simplicity, we calculate the prefactor for $E_{J_\alpha}$ numerically by fitting \cref{eqn:H-alpha} to \cref{eqn:zero-pi-theta-phi}, with $E_{C_\alpha}$ and $E_{J_{2\alpha}}$ fixed. 

In \cref{fig:tunnelling-rates-mass-ratio} we plot the ratio of the tunnelling energies as a function of $E_{C_\varphi}/E_{C_\theta}$.
We see that $E_{J_\alpha}/E_{J_{2\alpha}}$ is suppressed, as expected.
Furthermore, in \cref{fig:tunnelling-rates-Z-zeta} we show that the nearest-neighbour tunnelling rate is also suppressed in the impedance of the internal $\zeta$ mode, for $E_{C_\varphi}/E_{C_\theta} = 100$.
Thus, a higher $\zeta$-mode impedance leads to a more protected $0$-$\pi$ qubit.
This directly conflicts with the suppression of the effective qubit-oscillator coupling in \cref{eqn:zero-pi-interaction-exp-suppression}, but not the interaction term when using the $\zeta$ mode in \cref{eqn:zero-pi-zeta-interaction}.

Finally, we can rewrite the interaction Hamiltonians in \cref{eqn:zero-pi-interaction-exp-suppression,eqn:zero-pi-zeta-interaction} in terms of the effective mode $\alpha$ by letting $\thetah = \pi \n_\alpha$,
\begin{align}
    \label{eqn:zero-pi-interaction-exp-suppression-alpha}
    \hat H^\phi \intt (t) 
    &=
    -E_{J \intt} (t)
    e^{-{\pi Z_\zeta}/{2 R_Q}}
    \cos(\phih - \pi\n_\alpha),
    \\
    \label{eqn:zero-pi-zeta-interaction-alpha}
    \hat H^\zeta \intt (t)
    &=
    - E_{J \intt}(t) 
    \cos(\zetah - \pi\n_\alpha)
    .
\end{align}
In the following section, we use this effective model to simulate a protected gate for the $0$-$\pi$ qubit.

\subsection{Numerical simulations}\label{sec:numerical-simulations-zero-pi}

\begin{figure}
    \centering
    \includegraphics[scale=1]{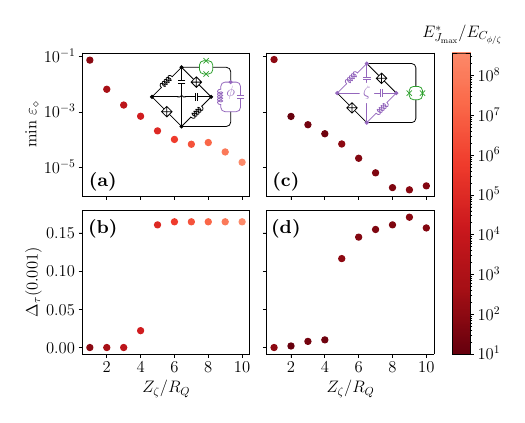}%
    \subfloat{\label{fig:imprecision-zero-pi-ancilla}}%
    \subfloat{\label{fig:robustness-zero-pi-ancilla}}%
    \subfloat{\label{fig:imprecision-zero-pi-zeta}}%
    \subfloat{\label{fig:robustness-zero-pi-zeta}}%
    \caption{
        Comparison of two different implementations of a protected phase gate for the $0$-$\pi$ qubit.
        (a) Imprecision and (b) robustness of a gate that uses an external oscillator $\phi$ as a function of the $\zeta$-mode impedance $Z_\zeta$ of the $0$-$\pi$ qubit.
        We have set $Z_\phi / R_Q = 10$.
        (c) Imprecision and (d) robustness of a gate that uses the $\zeta$ mode of the $0$-$\pi$ qubit, instead of an external oscillator.
        For each data point, a different value of $E_{J \maxx}$ is chosen, denoted $E^*_{J \maxx}$.
        This is defined to be the minimal value of $E_{J \maxx}$ for which $\Delta_\tau (0.001)>0$, if it exists, or the value that maximises the precision, otherwise.
        Its units are in $E_{C_\phi}$ for the left column and $E_{C_\zeta}$ for the right column.
        Throughout all simulations, $E_{C_\varphi}/E_{C_\theta} = 100$ and $E_{J \minn} = 0$.
        The circuit diagrams corresponding to each implementation of the gate are shown as insets in the top panels of each column.
        }
    \label{fig:zero-pi-gate-errors}
\end{figure}

Here we simulate the dynamics of a protected phase gate for the $0$-$\pi$ qubit.
We contrast two models: one that uses an external oscillator mode $\phi$, and our approach that uses the $\zeta$ mode of the $0$-$\pi$ circuit. 
We simulate the Hamiltonian $\hat H (t)= \hat H_\phi + \hat H_\alpha + \hat H \intt (t)$, where the first two terms are given by \cref{eqn:BKP-oscillator,eqn:H-alpha}, and the final term is given by either \cref{eqn:zero-pi-interaction-exp-suppression-alpha} or \cref{eqn:zero-pi-zeta-interaction-alpha}, depending on the model of interest.
Our numerical approach is similar to the simulations in \cref{sec:BKP-simulations}, though there are some slight differences that we discuss in \cref{app:numerical-simulations}.

In \cref{fig:imprecision-zero-pi-ancilla,fig:robustness-zero-pi-ancilla} we plot the imprecision and robustness of a gate that uses an external oscillator with $Z_\phi/R_Q = 10$.
In \cref{fig:imprecision-zero-pi-zeta,fig:robustness-zero-pi-zeta} we plot the same metrics for a gate that uses the internal $\zeta$ mode.
In both cases, the metrics are plotted as a function of the $\zeta$-mode impedance $Z_\zeta/R_Q$.
We fix the charging energy ratios to $E_{C_\phi}/E_{C_\theta} = E_{C_\varphi}/E_{C_\theta} = 100$ and for each value of $Z_\zeta/R_Q$, we set $E_J/E_{C_\varphi}$ such that the $0$-$\pi$ qubit protection is maximised.
We have also set $E_{J \minn} = 0$, while the maximum Josephson coupling is set to a critical value denoted $E^*_{J \maxx}$.
This is the minimum value of $E_{J \maxx}$ required to obtain a protected gate with robustness $\Delta_\tau(0.001) > 0$.
If that is not possible, then $E^*_{J \maxx}$ is simply the value that minimises the imprecision of the gate.
This is intended to reflect the hardware requirements of the tunable Josephson element for a protected gate, as we vary $Z_\zeta/R_Q$.

\Cref{fig:zero-pi-gate-errors} shows that both versions of the gate require $Z_\zeta/R_Q \ge 5$ to acquire appreciable robustness.
This is due to the suppression of the symmetry-breaking $\cos \alpha$ term in \cref{eqn:H-alpha} that occurs for large $Z_\zeta$, which may be seen in \cref{fig:tunnelling-rates-Z-zeta}.
A larger $\zeta$-mode impedance is therefore preferable, as it recovers the ideal qubit Hamiltonian in \cref{eqn:BKP-qubit}.
However, in this protected regime the value of $E^*_{J \maxx}$ required to achieve a robust gate becomes very large for a gate that uses the external oscillator, reaching values of $E_{J \maxx}^*/E_{C_\phi} > 10^8$ for $Z_\zeta/R_Q = 10$.
This aligns with the exponential suppression of the Josephson coupling in \cref{eqn:zero-pi-interaction-exp-suppression-alpha}.
By contrast, a gate that uses the $\zeta$ mode has $E^*_{J \maxx}/E_{C_\zeta} \le 100$ for all $Z_\zeta$.
For comparison, in \cref{sec:BKP} we found that $E^*_{J \maxx}/E_{C_\phi} \approx 30$ for the case of an ideal qubit. 

Performing a protected gate with the $0$-$\pi$ qubit using an external oscillator is very challenging due to the two competing effects of increasing the $\zeta$-mode impedance.
On the one hand, a large $\zeta$-mode impedance coincides with the protected regime of the $0$-$\pi$ qubit.
On the other hand, it exponentially suppresses the qubit-oscillator interaction. 
Using the $\zeta$ mode in place of the external oscillator solves this issue because the interaction strength is no longer suppressed with the $\zeta$-mode impedance. 

\subsection{\texorpdfstring{$\vect \varphi$}{φ} mode} \label{sec:phi}

An alternative to using the internal $\zeta$ mode for a protected gate with the $0$-$\pi$ qubit is to use the internal $\varphi$ mode.
This could be achieved by replacing the static Josephson junctions in the $0$-$\pi$ qubit with tunable Josephson elements, facilitating a tunable coupling between the $\theta$ and $\varphi$ modes. 

In \cref{app:phi-mode-gate}, we simulate this version of the gate but find that a protected gate is not feasible for near-term experimental charging energy ratios. 
The underlying reason is that tuning the coupling between the $\theta$ and $\varphi$ modes of the qubit to perform the gate simultaneously tunes the barrier height in both the $\varphi$ and $\theta$ directions of the qubit. 
In \cref{sec:BKP-simulations} we showed a requirement of $E_{J \minn} < E_{C_\phi}$ to obtain a protected gate. 
When using the internal $\varphi$ mode, this translates to $E_{J \minn} < E_{C_\varphi}$.
However, we find that this small value leads to unwanted tunnelling between the two logical states throughout the gate. 
At larger values of $E_{J \minn}$, this unwanted tunnelling is reduced, but the protection of the gate is compromised due to the smaller dynamic range. 

The tunnelling between the two qubit states could be reduced whilst maintaining a sufficiently small $E_{J_\text{min}}$ for a protected gate by increasing the charging energy ratio $E_{C_\varphi}/E_{C_\theta}$.
However, given that increasing $E_{C_\varphi} / E_{C_\theta}$ is experimentally challenging  
(which we discuss in \cref{sec:charging-energy-ratio}),
the $\zeta$ mode is likely to be an easier path to a protected gate.

\section{Hardware constraints} \label{sec:hardware-constraints}

In this section we delve further into the hardware constraints for implementing a protected gate.
In \cref{sec:circuit-disorder,sec:flux-noise} we analyse the impacts of circuit disorder and flux noise on the performance of the gate, respectively.
In \cref{sec:tunable-josephson-element} we consider the hardware requirements for the tunable Josephson element to achieve a protected gate.
\Cref{sec:charging-energy-ratio,sec:ancilla-impedance} examine the constraints on the charging energy ratio and ancilla impedance for a protected gate, respectively.
Finally, in \cref{sec:cooling,sec:photon-loss}, we discuss the role of cooling and photon loss, respectively.

\subsection{Circuit disorder} \label{sec:circuit-disorder}

So far we have ignored the circuit disorder of the $0$-$\pi$ qubit; in \cref{eqn:zero-pi-theta-phi,eqn:zero-pi-zeta} we assumed each pair of circuit components in the $0$-$\pi$ qubit are identical. 
In practice, this symmetry will be broken due to limitations of device fabrication.
In \cref{app:zero-pi-circuit}, we derive the Hamiltonian for the $0$-$\pi$ circuit in the presence of disorder.
Here we analyse the impact of disorder on our proposed gate.

\begin{figure}
    \centering
    \includegraphics[width=\linewidth]{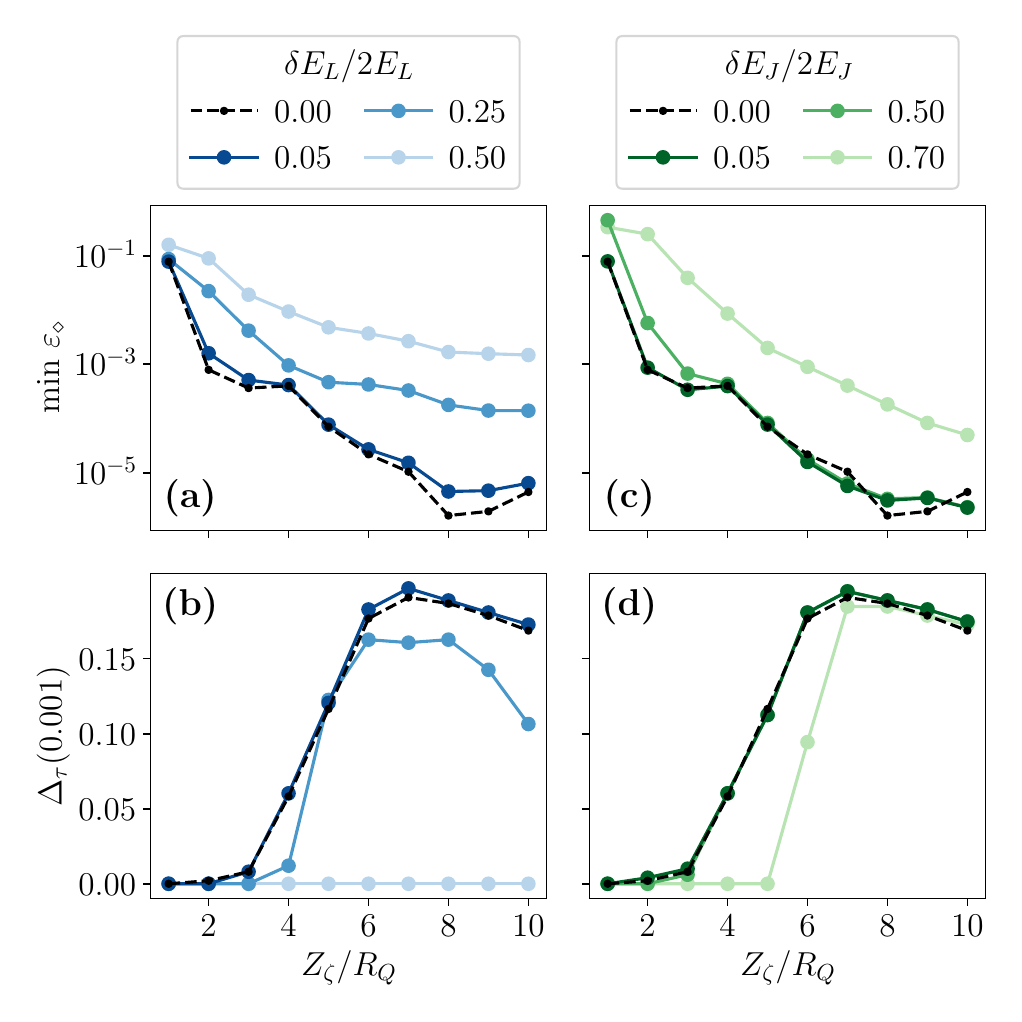}%
    \subfloat{\label{fig:imprecision-EL-disorder}}%
    \subfloat{\label{fig:robustness-EL-disorder}}%
    \subfloat{\label{fig:imprecision-EJ-disorder}}%
    \subfloat{\label{fig:robustness-EJ-disorder}}%
    \caption{
        Impact of circuit disorder on the protected phase gate for the $0$-$\pi$ qubit that uses the internal $\zeta$ mode.
        (a) Imprecision and (b) robustness as a function of the $\zeta$-mode impedance $Z_\zeta$, for different asymmetries in the inductors $\delta E_L$.
        (c) Imprecision and (d) robustness as a function of the $\zeta$-mode impedance $Z_\zeta$, for different asymmetries in the Josephson junctions $\delta E_J$.
        Throughout all simulations, the charging energy ratio is fixed to $E_{C_\varphi}/E_{C_\theta} = 100$ and the minimum and maximum Josephson couplings are $E_{J \minn} = 0$ and $E_{J \maxx}/E_{C_\zeta} = 100$.
        }
    \label{fig:circuit-disorder}
\end{figure}

\textit{Inductive asymmetry} leads to the flux-flux interaction term $\delta E_L \varphih \zetah$, where $\delta E_L = \phi_0^2/L_1 - \phi_0^2/L_2$. 
To simulate this, we add the term $\delta E_L \pi \n_\alpha \zetah$ to \cref{eqn:zero-pi-zeta-interaction-alpha} because $\varphih \to \pi\n_\alpha$ in the effective model.
In \cref{fig:imprecision-EL-disorder,fig:robustness-EL-disorder} we plot the gate error as a function of $Z_\zeta$ for different inductive asymmetries. 
We find that it is almost entirely insensitive to an inductive asymmetry of $\delta E_L / 2E_L = 5 \%$, and begins to lose its robustness at $\delta E_L / 2E_L = 25\%$.
The robustness of the gate to inductive asymmetries may be attributed to the $0$-$\pi$ qubit's insensitivity to flux noise.

\textit{Josephson asymmetry} leads to a nonlinear interaction term $\delta E_J \sin\thetah \sin\varphih$, where $\delta E_J = E_{J_1} - E_{J_2}$. 
This flattens the egg-carton-shaped potential of the $0$-$\pi$ qubit, and leads to an increase in $E_{J_\alpha}/E_{J_{2\alpha}}$ for the effective model in \cref{eqn:H-alpha}.
In \cref{fig:imprecision-EJ-disorder,fig:robustness-EJ-disorder} we plot the gate error as a function of $Z_\zeta$ for different Josephson asymmetries.
We find that the gate is almost entirely insensitive to asymmetries up to $\delta E_J / 2E_J = 50\%$, and only begins to lose its robustness once $\delta E_J / 2E_J = 70\%$.
This is due to the fact that the location of the minima in the potential are unaffected by the $\sin\thetah \sin\varphih$ term and tunnelling between the two qubit states remains exponentially suppressed in $E_J / E_{C_\theta}$ so long as the reduced barrier height remains larger than $E_{C_\theta}$.

\textit{Capacitive asymmetry} leads to charge-charge coupling between the $\theta$, $\varphi$ and $\zeta$ modes. 
Here we explain that these effects are less significant than the asymmetries in the inductors and Josephson junctions.
Additional terms that arise due to capacitive disorder may be obtained by substituting $\n_\theta \to \n_\theta - r_\varphi \n_\varphi - r_\zeta \n_\zeta$ into \cref{eqn:zero-pi-theta-phi}, where $r_\zeta = (C_1 - C_2)/(C_1 + C_2)$ and $r_\varphi = (C_{J_1} - C_{J_2})/(C_{J_1} + C_{J_2})$ quantify the relative disorder in the capacitances. 
We start by considering the effects of this disorder that are linear in the parameters $r_\zeta$ and $r_\varphi$.

Asymmetry in the junction capacitances leads to an interaction term $r_\varphi E_{C_\theta}  \n_\theta \n_\varphi$.
It is possible to repeat our numerical fitting of the effective one-dimensional model to \cref{eqn:zero-pi-theta-phi} with this additional contribution to the Hamiltonian.
However, we find that this term has a negligible impact on the spectrum of the $0$-$\pi$ qubit up to asymmetries of $|r_\varphi| = 1$.
This is consistent with the findings of Ref.~\cite{Dempster2014}. 

Asymmetry in the capacitors leads to an interaction term $r_\zeta E_{C_\theta}  \n_\theta \n_\zeta$. 
It is also possible to capture the effect of this noise in the tight-binding model of the $0$-$\pi$ qubit. 
The tight-binding eigenstates $\ket{n_\alpha}$ are approximately harmonic oscillator ground states localised at the minima of the potential of the Hamiltonian in \cref{eqn:zero-pi-theta-phi}. 
This makes it possible to approximate the matrix elements $\bra{ n'_\alpha}\n_\theta \ket{n_\alpha}$ for these states. 
When $n_\alpha \neq n'_\alpha$ these matrix elements couple distinct wells of the cosine potential and are exponentially suppressed since these states are highly localised.
On the other hand, the expectation values $\bra{n_\alpha} \n_\theta \ket{n_\alpha}$ vanish because of the $\thetah \rightarrow -\thetah$ symmetry of the potential. 
Note that this symmetry persists for all values of $E_L$ so that even in analysing corrections to the harmonic oscillator ground state, these mean values remain equal to zero.
The perturbative correction to the tight-binding model is of the form 
\begin{equation}
    r_\zeta E_{C_\theta} \n_\zeta \sum_{n_\alpha, n'_\alpha \in \Z} \bra{n'_\alpha} \n_\theta \ket{n_\alpha} \, \ketbra{n'_\alpha}{n_\alpha},
\end{equation}
and is therefore expected to be negligible compared to the inductive and Josephson disorder. 
This is consistent with the fact that transmons are insensitive to offset charge and the $0$-$\pi$ qubit is inheriting this insensitivity through the $\theta$ mode.

At the next order in $r_\zeta$, $r_\varphi$ there are three additional contributions to the Hamiltonian. 
The first two of these, $r_\varphi^2 E_{C_\theta} \n_\varphi^2$ and $ r_\zeta^2 E_{C_\theta} \n_\zeta^2$, can be absorbed as small adjustments to the charging energy of the $\varphi$ and $\zeta$ modes and we do not discuss them further. 
The final correction is $r_\zeta r_\varphi E_{C_\theta} \n_\varphi \n_\zeta$ and this can be analysed in the same way as the $\n_\theta \n_\zeta$ term.
However, here the inductive potential in \cref{eqn:zero-pi-theta-phi} does break the symmetry of the $\varphi$ wavefunctions for the $\ket{n_\alpha}$ states leading to nonzero mean values $\bra{n_\alpha} \n_\varphi \ket{n_\alpha}$. 
Nevertheless, this contribution is second order in the capacitive disorder parameters and will be suppressed by some power of $E_L/\sqrt{E_J E_{C_\varphi}}$. 
Therefore, we do not pursue a quantitative discussion of this contribution, which will be smaller than the effects of the inductive and Josephson junction asymmetries that we analysed above.

\subsection{Flux noise} \label{sec:flux-noise}

Flux noise from stray magnetic fields or noisy flux lines is a common source of infidelity for gates that rely on flux control, as well as a source of decoherence for qubits with inductive loops.
Whilst the $0$-$\pi$ qubit is designed to be robust against flux noise, the addition of the SQUID shunt adds two additional inductive loops to the circuit that may affect the performance of the gate, as well as the coherence time of the qubit while it is idling.

\begin{figure}
    \centering
    \includegraphics[width=\linewidth]{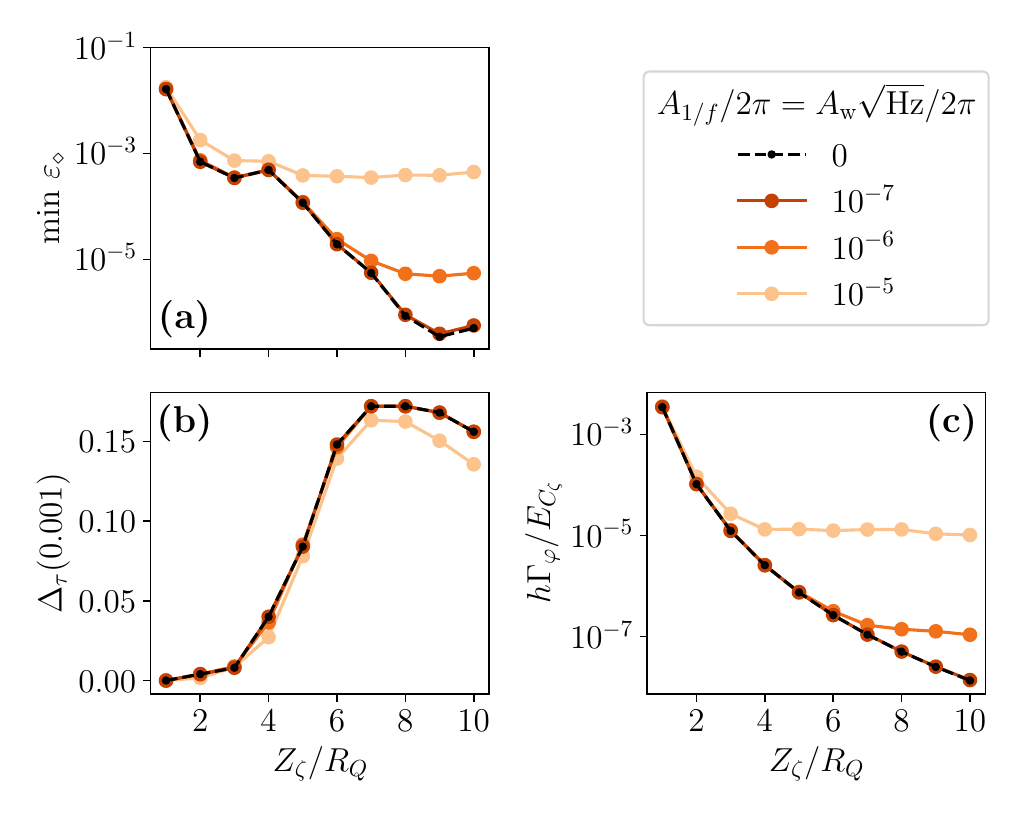}%
    \subfloat{\label{fig:imprecision-flux-noise}}%
    \subfloat{\label{fig:robustness-flux-noise}}%
    \subfloat{\label{fig:dephasing-rates}}%
    \caption{
        Impact of flux noise on the protected phase gate and coherence time of the $0$-$\pi$ qubit.
        (a) Imprecision and (b) robustness as a function of the $\zeta$-mode impedance $Z_\zeta$, for different flux noise amplitudes.
        (c) Dephasing rate $\Gamma_\varphi$ for the $0$-$\pi$ qubit as a function of the $\zeta$-mode impedance, for different flux noise amplitudes.
        Throughout all simulations, the charging energy ratio is fixed to $E_{C_\varphi}/E_{C_\theta} = 100$ and the minimum and maximum Josephson couplings (in the absence of flux noise) are $E_{J \minn} = 0$ and $E_{J \maxx}/E_{C_\zeta} = 50$.
        }
    \label{fig:flux-noise}
\end{figure}

\Cref{fig:imprecision-flux-noise,fig:robustness-flux-noise} show the imprecision and robustness of the gate as a function of the $\zeta$-mode impedance at different levels of flux noise.
These simulations account for flux noise in each of the three inductive loops of the circuit with a spectral density
\begin{equation}
    S(\omega) = A^2_{1/f} \frac{2\pi}{\abs{\omega}} + A^2_\mathrm{w},
\end{equation}
where $A_{1/f}$ and $A_\mathrm{w}$ represent flux noise amplitudes for $1/f$ and white noise, respectively. 
For simplicity, we fix $A_{1/f} = A_\mathrm{w}\sqrt{\mathrm{Hz}}$ in our simulations.
Experimentally measured flux noise amplitudes are on the order of $2\pi \times 10^{-6}$~\cite{Hutchings2017,Didier2019}.
In our simulations, we find that flux noise has negligible impact for amplitudes of $2\pi \times 10^{-7}$, but in the highly protected regime of the qubit there is an observable decrease in the precision of the gate for amplitudes of $2\pi \times 10^{-6}$ or greater.
Meanwhile, the robustness of the gate is largely unaffected at all flux noise amplitudes, a consequence of the gate's protection.
In \cref{app:flux-noise} we provide further details of our flux noise simulations.

Ideally the interaction term in \cref{eqn:zero-pi-zeta-interaction} is switched off when the qubit is idling, but inevitably stray magnetic fields will lead to a residual interaction.
This will cause dephasing since the interaction term effectively acts like a logical $\bar Z_\theta$ on the protected qubit.
However, in \cref{app:flux-noise} we show that the qubit is protected to first order in the strength of this interaction and only has a nonvanishing dephasing rate at second order.

\Cref{fig:dephasing-rates} shows the dephasing rate for an idling $0$-$\pi$ qubit.
We observe that for large flux noise amplitudes, the dephasing rate of the qubit begins to plateau with increasing impedance.
Fortunately, for flux noise amplitudes of $2\pi \times10^{-6}$ or less, the effect on the dephasing rate of the qubit is minimal.

\subsection{Tunable Josephson element}\label{sec:tunable-josephson-element}

In \cref{sec:BKP-simulations} we found that the tunable Josephson element requires a dynamic range of at least two orders of magnitude to obtain a protected gate, with larger dynamic ranges required at lower ancilla impedances. 
Whilst two orders of magnitude may be reliably achievable using a SQUID, where asymmetries in the Josephson energies of the junctions are usually at the percent level~\cite{Blais2021,Vanselow2025}, anything larger will be experimentally challenging.
Voltage-tunable Josephson junctions~\cite{Larsen2015,Banszerus2024} might provide an alternative as they have no such symmetry requirements. 
Instead, the dynamic range will be dictated by the sensitivity of the device to voltage fluctuations of the electrostatic gate that tunes the Josephson energy.

Another possible route to achieve a tunable Josephson element with a larger dynamic range is to use a SQUID with multiple inductive loops.
Similar devices have been used in Refs.~\cite{Miano2019,Lescanne2020,Miano2022}.
In \cref{app:tunable-josephson-element-dynamic-range} we calculate the effective Josephson energy of a multiloop SQUID.
We find that the dynamic range of the effective Josephson energy scales with the noise on the external magnetic fluxes, rather than the asymmetry in the Josephson energies of the junctions.
Since typical flux noise amplitudes are on the order of $2\pi \times 10^{-6}$~\cite{Hutchings2017,Didier2019}, this suggests that tunable Josephson elements with a larger dynamic range could be achieved for a protected gate.

We have also treated the tunable Josephson element as a purely inductive element. 
However, in reality it will have some nonzero capacitance $C \intt$. 
This reduces the effective charging energy ratio between the oscillator mode and qubit mode, and introduces undesired charge-charge coupling between the two modes.
In \cref{app:tunable-josephson-element-capacitance} we consider these effects for the case of the ideal qubit coupled to an oscillator.
Here, we find that the gate remains unaffected so long as $C \intt / C_\phi < 0.1$.

A smaller $C \intt$ also implies a smaller $E_{J \maxx}$ for a fixed plasma frequency $\omega_p = \sqrt{8 E_{J \maxx} E_{C \intt}}/\hbar$. 
This makes it very challenging to achieve the large values of $E^*_{J \maxx}/E_{C_\phi}$ required for the protected gate with the $0$-$\pi$ qubit using the external oscillator whilst maintaining a small ratio of $C \intt / C_\phi$.
For example, in \cref{sec:numerical-simulations-zero-pi} we found a minimum $\zeta$-mode impedance of $Z_\zeta / R_Q = 5$ at an oscillator impedance of $Z_\phi/R_Q = 10$ was necessary to achieve a gate with appreciable robustness. 
At this $\zeta$-mode impedance, a coupling strength of $E_{J \maxx} / E_{C_\phi} > 8 \times 10^4$ is required. 
Assuming the oscillator impedance may be obtained with $E_{C_\phi}/h = 1$~GHz and $E_{L_\phi}/h = 2$~MHz, the maximum Josephson coupling required for the protected gate would be greater than $E_{J \maxx}/h = 80$~THz. 
With a plasma frequency of the tunable Josephson element of $\omega_p/2\pi = 40$~GHz, this implies that $C \intt / C_\phi > 100$, which is well above the requirement for the gate to remain protected.

When using either of the internal modes of the $0$-$\pi$ qubit instead of the external oscillator, the capacitance of the tunable Josephson element simply adds to the capacitance of the $\zeta$ or $\varphi$ mode. In this way, so long as the capacitance in the device remains approximately symmetric, the capacitance of the tunable Josephson element has no effect on the performance of the gate, regardless of its value. This is another advantage of utilising the internal modes of the $0$-$\pi$ qubit to perform a gate.

\subsection{Charging energy ratio} \label{sec:charging-energy-ratio}
In all of the above numerical results we have fixed the largest ratio of the physical capacitances in the system to two orders of magnitude.
Ideally this quantity should be made as large as possible, but larger values pose an experimental challenge.
In the case of the $0$-$\pi$ qubit, a large $E_{C_\varphi}/E_{C_\theta}$ is achieved by minimising the capacitance of the Josephson junctions $C_J$ whilst simultaneously maximising the capacitance of the capacitors $C$.
In Ref.~\cite{Gyenis2021}, a charging energy ratio of $E_{C_\varphi}/E_{C_\theta} \approx 12$ was achieved for the soft $0$-$\pi$ qubit. 
This relatively small separation in charging energies is the reason for the partial protection of this device. 
Methods to increase this ratio include different capacitor geometries~\cite{Kim2024} as well using a silicon-on-insulator substrate for the device to reduce the effect of stray capacitances that decrease $E_{C_\varphi}$~\cite{Hassani2024}.
We choose to focus on a ratio of two orders of magnitude as a representative near-term value, but in \cref{app:varphi-theta-capacitance-ratio} we investigate the impact of different charging energy ratios on the performance of a protected gate.
Here we find that a protected gate can be recovered with smaller charging energy ratios, but that it requires a larger impedance of the $\zeta$ mode.

\subsection{Ancilla impedance}\label{sec:ancilla-impedance}

The results in \cref{sec:BKP-simulations} showed that an oscillator impedance of at least $Z_\phi/R_Q = 4$ was necessary to obtain a protected gate for an ideal qubit coupled to an oscillator. 
This value is achievable with current superinductors, where impedances up to $Z_\phi/R_Q = 4.8$ have been demonstrated using chains of Josephson junctions~\cite{Pechenezhskiy2020} or geometric coils~\cite{Peruzzo2020}.

In \cref{sec:numerical-simulations-zero-pi} we found that a $\zeta$-mode impedance of $Z_\zeta/R_Q \ge 5$ was necessary to obtain a protected gate. 
These results assumed a charging energy ratio of $E_{C_\varphi}/E_{C_\theta} = 100$ in the $0$-$\pi$ qubit, meaning that the $\varphi$-mode impedance $Z_\varphi = \sqrt{L/4C_J}$ is larger than $Z_\zeta$ by a factor of $10$. 
Therefore, the $\zeta$-mode impedance requirement of $Z_\zeta/R_Q = 5$ translates to a $\varphi$-mode impedance (which is the largest impedance in the device) of $Z_\varphi/R_Q = 50$.
In \cref{app:varphi-theta-capacitance-ratio} we simulate the gate at lower charging energy ratios, but find that this increases the required $\zeta$-mode impedance for the gate to be protected such that $Z_\varphi/R_Q = 50$ remains the lowest $\varphi$-mode impedance to achieve protection.
This is an order of magnitude larger than what has been achieved with current superinductors, suggesting that further improvements in superinductors are necessary to obtain a protected gate with the $0$-$\pi$ qubit. 
Increasing circuit impedance in superconducting qubits is an active area of experimental research~\cite{Pechenezhskiy2020,Peruzzo2020,Gruenhaupt2019,Junger2025,Manset2025}.

\subsection{Cooling} \label{sec:cooling}

As mentioned in \cref{sec:BKP}, the ancillary mode must be cooled between performing gates in order to extract entropy from the circuit. 
For an external oscillator, this is possible using well-established techniques for oscillator cooling~\cite{Blais2021}.
In the case of using one of the internal modes as an ancilla, this mode may be cooled using the scheme provided in Ref.~\cite{Paolo2019}, which involves capacitively coupling the $0$-$\pi$ qubit to a lossy oscillator, and activating sideband transitions between the ancillary mode and the oscillator. 
We note that this cooling oscillator does not require a superinductor, in contrast to the ancillary oscillator required to perform a protected gate. 
However, it does require a tunable inductance in order to modulate its frequency and coupling strength. 
Given that it is likely to be necessary to cool the $\zeta$ mode of the $0$-$\pi$ qubit to prevent coupling to the $\varphi$ mode during idling~\cite{Groszkowski2018,Paolo2019}, implementation of this cooling mechanism is likely a requirement regardless of whether or not an internal mode is used to perform the gate.

\subsection{Photon loss}\label{sec:photon-loss}

Throughout this paper we have used the gate's robustness to pulse mistiming as a proxy for its protection.
Another useful metric would be the error of the gate in the presence of photon loss in the ancillary mode. 
However, simulating this presents a theoretical and/or computational challenge due to the dynamic nonlinearity present in the interaction Hamiltonian. 

For example, let us consider the case of the ideal qubit coupled to an oscillator via \cref{eqn:BKP-interaction}. 
In the regime where $E_{J \intt}(t) \ll \sqrt{8 E_{L_\phi} E_{C_\phi}}$, photon loss may be efficiently simulated using a Lindblad master equation with collapse operators given by the annihilation operators for the oscillator mode $\phi$.
Likewise, photon loss may be efficiently simulated in the opposite regime, $E_{J \intt}(t) \gg \sqrt{8 E_{L_\phi} E_{C_\phi}}$, by using collapse operators that transition between tight-binding eigenstates~\cite{CohenTannoudji1992}. 

However, in the intermediate regime between these extremes, the situation is more complicated. 
Unlike qubit gates involving resonant driving, the Hamiltonian changes during the gate in ways that cannot be described perturbatively. 
This means that there are no consistent collapse operators for the full time-dependent gate simulation as $E_{J \intt}(t)$ crosses through the aforementioned regimes. 
For these general time-dependent systems, the master equation is much more computationally expensive since it involves diagonalising the system at each time step.
Moreover, the Markov approximation that leads to a master equation requires that the energy levels in the Hamiltonian are well-separated compared to the photon-loss rate. 
This requirement may not hold for intermediate values of $E_{J \intt}(t)$. 

Nevertheless, we expect the gate to be robust to photon loss for the following reasons.
In the large $E_{J \intt}(t)$ regime of the gate, the ancillary mode is encoded into the GKP code, which is is particularly resilient to photon loss~\cite{Albert2018}.
Therefore, photon loss errors that occur during this time should be correctable upon cooling the ancillary mode.
When this mode is not encoded in the GKP code during the small $E_{J \intt}(t)$ regime of the pulse, then photon loss benefits the gate by cooling the ancillary mode.

Finally, we note that our pulses are not optimised for pulse duration, and that pulse optimisation techniques such as DRAG~\cite{Motzoi2009} may yield faster gate times with smaller errors.

\section{Beyond the phase gate} \label{sec:other-gates}

In this paper, we have focused on the implementation of a protected single-qubit phase gate. In Refs.~\cite{Kitaev2006} and \cite{Brooks2013}, an extension of the single-qubit phase gate to a protected two-qubit phase gate is outlined.
Universal fault-tolerance is then obtained by supplementing these protected gates with unprotected single-qubit rotations and repeated noisy measurements in the logical $X$ and $Z$ bases. 
In \cref{sec:two-qubit-gate} we outline a way to use the internal modes of two coupled $0$-$\pi$ qubits to perform a protected two-qubit gate, and in \cref{sec:non-clifford-gate} we briefly discuss an extension of the protected $S$-gate to a $T$-gate.

\subsection{Two-qubit gate} \label{sec:two-qubit-gate}
The two-qubit gate proposed in Refs.~\cite{Kitaev2006,Brooks2013} involves connecting two $0$-$\pi$ qubits in series and shunting the entire circuit with a high-impedance oscillator. 
This is intended to implement an interaction term
\begin{equation}
    \hat H \intt
    =
    - E_{J\intt} (t) 
    \cos\big(\thetah_a + \thetah_b - \phih\big)
    ,
\end{equation}
where $\theta_a$ and $\theta_b$ label two protected qubit modes, and $\phi$ denotes the oscillator mode. 
This term means that a phase of $\pi/2$ is acquired in the oscillator when the logical states of the two qubits are $\ket{01}$ or $\ket{10}$ and no phase is accrued when the qubit states are in $\ket{00}$ or $\ket{11}$, enacting the logical unitary operation $\exp(-i \frac{\pi}{4} \hat{Z} \otimes \hat{Z})$. 
Similarly to the single-qubit case, we find that the envisaged circuit gives rise to coupling to internal $\zeta$ modes of both qubits, making the protected gate challenging to achieve due to the exponential suppression of the interaction strength. 

In \cref{app:series-zero-pi-circuit}, we show that it is possible to utilise these $\zeta$ modes to perform this gate by instead shunting the two qubits with a tunable Josephson element. 
For this gate to work, the two qubits must be brought into resonance with each other to utilise the symmetric superposition of the two internal $\zeta$ modes in place of the external oscillator. This suggests the need to make $0$-$\pi$ qubits tunable when scaling to a multi-qubit system based on $0$-$\pi$ qubits. 

\subsection{Non-Clifford gate} \label{sec:non-clifford-gate}
The protected $S$-gate relies on the fact that the quadratic potential enacts an encoded $S$-gate on the GKP code. In Refs.~\cite{Gottesman2001,Nguyen2025} higher-order polynomial potentials are shown to enact a $T$-gate on GKP codewords. 
However, the resulting gate is deemed to be non fault-tolerant due to the fact that higher-order polynomials distort GKP codewords in an uncorrectable way.
In \cref{app:protected-T-gate}, we investigate replacing the quadratic potential in \cref{eqn:BKP-oscillator} with a quartic potential in order to realise a potentially protected $T$-gate. 
A quartic potential may be generated using a superconducting nonlinear asymmetric inductive element (SNAIL) oscillator~\cite{Sivak2019,Hillmann2020,Eriksson2024} rather than a linear oscillator. 
We find that a $T$-gate with errors $< 10^{-3}$ may be obtained, but that the gate is less robust than the equivalent $S$-gate. 
This is a consequence of the encoded $T$-gate being non fault-tolerant for the GKP code. 
Nevertheless, this provides a way of performing a $T$-gate that does not explicitly break the protection of the qubit. 
We believe there is much room for exploring other protected gates based on different potentials or different bosonic codes.

\section{Conclusions}\label{sec:conclusions}

In conclusion, we have proposed and analysed a protected phase gate for the $0$-$\pi$ qubit that is compatible with the protected regime of the qubit. 
This opens up the possibility of performing a protected gate on a protected superconducting qubit.
Through numerical simulations, facilitated by a one-dimensional model for the $0$-$\pi$ qubit, we compared the performance of the gate using an external oscillator to the gate using the internal $\zeta$ mode.
The results revealed that a $\zeta$-mode impedance of $Z_\zeta / R_Q = 5$ is necessary for a protected gate in both cases, but that infeasibly large Josephson energies would be required to recover this protection when using an external oscillator.

Whilst our scheme substantially reduces the hardware requirements for the protected gate relative to using an external oscillator, we summarise three of the main experimental challenges.
First, a large dynamic range of the tunable Josephson element is required; we found a requirement of greater than two orders of magnitude.
This may be challenging with standard SQUIDs and necessitate a different coupling element, such as a SQUID with multiple inductive loops.
Second, a large charging energy ratio of $E_{C_\varphi}/E_{C_\theta} = 100$ is necessary. 
Whilst the soft $0$-$\pi$ qubit only achieved a charging energy ratio of $E_{C_\varphi}/E_{C_\theta} \approx 12$, there exist several approaches to substantially increase this value.
Third, owing to the large charging energy ratio, the $\zeta$-mode impedance requirement of $Z_\zeta/R_Q = 5$ translates to a $\varphi$-mode impedance of $Z_\varphi/R_Q = 50$.
This is roughly an order of magnitude larger than what has been achieved with current superinductors~\cite{Pechenezhskiy2020,Peruzzo2020}.
Increasing circuit impedance is an active experimental effort since this is a requirement of many types of superconducting qubits~\cite{Manucharyan2012,Kalashnikov2020,Smith2020,Pechenezhskiy2020,Hassani2023}.

Although improvements in current fabrication techniques are required to achieve a protected gate with the $0$-$\pi$ qubit, our results also show that unprotected gates with gate errors $<10^{-3}$ are possible with current parameters. 
Given the inherent challenge in manipulating protected qubits, our results provide valuable approaches to perform gates with these qubits in both the current and near term.
These techniques will be crucial for utilising protected qubits in near-term quantum computers.

Finally, we gave two possible extensions of the protected single-qubit phase gate to a protected non-Clifford and two-qubit gate.
We believe there are many opportunities for exploring other types of protected gates based on different bosonic codes, and tailoring protected gates to other types of protected qubits.

\section*{Acknowledgements}

We would like to thank Sean Barrett, Joshua Combes, Xanthe Croot, Farid Hassani and Mackenzie Shaw for insightful discussions.
We acknowledge support from the Australian Research Council via the Centre of Excellence in Engineered Quantum Systems (CE170100009), and the US Army Research Office (W911NF-23-10092). 
X. C. K. is supported by an Australian Government Research Training Program (RTP) Scholarship.
X. C. K. acknowledges access to the University of Sydney's high performance computing facility, Artemis, for obtaining numerical results.
F. T. is supported by the Sydney Quantum Academy.
We acknowledge the traditional owners of the land on which this work was undertaken at the University of Sydney, the Gadigal people of the Eora Nation.

\section*{Data availability}
The data that support the findings of this article are openly available at Ref.~\cite{data-repo}.

\appendix
\section{Circuit quantisation}\label{app:circuit-quantisation}

In this appendix we provide details for the quantisation of circuits discussed in the main text. 
After introducing the methodology used, we derive Hamiltonians for the following circuits: a $0$-$\pi$ qubit in \cref{app:zero-pi-circuit}, a $0$-$\pi$ qubit Josephson-coupled to an oscillator in \cref{app:zero-pi-oscillator-circuit}, modified $0$-$\pi$ qubits Josephson-coupled to their internal modes in \cref{app:zero-pi-zeta-circuit,app:zero-pi-phi-circuit}, and two $0$-$\pi$ qubits coupled in series in \cref{app:series-zero-pi-circuit}.
In \cref{app:tunable-josephson-element} we use the established quantisation procedure to analyse the effects of capacitance and extend the dynamic range of the tunable Josephson element.

We now provide an overview of our generic circuit quantisation procedure, which closely follows Refs.~\cite{Rajabzadeh2023,Vool2017}.
The classical Lagrangian of a circuit is
\begin{equation}\label{eqn:circuit-lagrangian}
    \mathcal L
    =
    T
    - 
    V
    ,
\end{equation}
where $T$ is the kinetic energy associated with capacitors and $V$ is the potential energy associated with inductors and Josephson junctions.
Each circuit element has a corresponding branch flux variable:
\begin{equation}\label{eqn:branch-flux}
    \Phi_j (t) 
    = 
    \int_{-\infty}^t \! \mathrm{d}\tau \: V_j (\tau)
    ,
\end{equation}
where $V_j$ is the voltage across the branch $j$.
Faraday's law and fluxoid quantisation leads to a constraint on branch fluxes that form a loop $\ell$:
\begin{equation}\label{eqn:constraint-1}
    \sum_{j \in \ell}
    \Phi_j
    =
    \phi \ext 
    ,
\end{equation}
where $\phi \ext$ is the external flux threading the loop $\ell$. 
This constraint provides a relationship between branch flux variables $\Phi_j$ and node flux variables $\phi_n$.
Specifically,
\begin{equation}\label{eqn:constraint-2}
    \Phi_j 
    =
    \phi_n - \phi_{n'}
    +
    \phi_{\mathrm{ext}_j} 
    ,
\end{equation}
where $\phi_{\mathrm{ext}_j}$ is the contribution of the external flux $\phi \ext$ to the branch $j$.
\Cref{eqn:constraint-1,eqn:constraint-2} may be written in a more compact vector form:
\begin{align}
    \label{eqn:constraint-1-vectorised}
    \vect G \vect \Phi 
    &= 
    \vect \phi \ext
    ,
    \\
    \label{eqn:constraint-2-vectorised}
    \vect \Phi
    &=
    \vect A \vect \phi
    +
    \vect B \vect \phi \ext 
    .
\end{align}
We define three column vectors: $\vect \Phi$ contains the $k$ branch fluxes, $\vect \phi$ contains the $n$ node fluxes, and $\vect \phi \ext$ contains the $l$ external fluxes.
The $l \times k$ matrix $\vect G$ ensures that the sum of branch fluxes in a loop equals the external flux threading that loop.
The $k \times n$ matrix $\vect A$ relates each branch flux to a phase difference between two node fluxes, which is determined by the choice of spanning tree.
The $k \times l$ matrix $\vect B$ assigns external fluxes to the closure branches of the circuit.

The kinetic energy of a circuit is given by
\begin{equation}\label{eqn:kinetic-energy-Phi}
    T (\dot{\vect \Phi})
    = 
    \frac12 
    \dot {\vect \Phi} \T
    \vect C
    \dot {\vect \Phi}
    ,
\end{equation}
where $\vect C$ is a diagonal matrix containing the capacitances of each branch.
Substituting \cref{eqn:constraint-2-vectorised} into \cref{eqn:kinetic-energy-Phi} gives
\begin{equation}\label{eqn:kinetic-energy-phiext}
    T (\dot{\vect \phi}, \dot{\vect \phi}\ext)
    = 
    \frac12
    \dot {\vect \phi} \T
    \vect C_{\vect \phi}
    \dot {\vect \phi}
    +
    \dot {\vect \phi} \ext \T
    \vect C \coup
    \dot {\vect \phi}
    ,
\end{equation}
where $\vect C_{\vect \phi} = \vect A \T \vect C \vect A$ and $\vect C \coup = \vect B \T \vect C \vect A$.
We omit terms that are quadratic in the classical external fluxes $\vect \phi \ext$, as these terms are discarded when the circuit Hamiltonian is quantised.
Importantly, the second term in \cref{eqn:kinetic-energy-phiext} indicates that time-dependence in the external fluxes $\vect \phi \ext$ generates an EMF acting on the node fluxes $\vect \phi$~\cite{You2019,Riwar2022}.
The irrotational gauge, where this coupling term vanishes, corresponds to the choice~\cite{Rajabzadeh2023}
\begin{equation}\label{eqn:irrotational-gauge}
    \vect B
    =
    \begin{pmatrix}
        \vect A \T \vect C
        \\
        \vect G
    \end{pmatrix}
    ^+
    \begin{pmatrix}
        \vect 0
        \\
        \vect I
    \end{pmatrix} 
    ,
\end{equation}
where the zero matrix $\vect 0$ has size $n \times l$, the identity matrix $\vect I$ has size $l \times l$, and $\vect O^+$ indicates the psuedoinverse of the matrix $\vect O$.
For this choice of $\vect B$, the kinetic energy has the simplified form
\begin{equation}\label{eqn:kinetic-energy-phi}
    T (\dot{\vect \phi})
    = 
    \frac12 
    \dot {\vect \phi} \T
    \vect C_{\vect \phi}
    \dot {\vect \phi}
    .
\end{equation}
The potential energy of a circuit is given by
\begin{equation}\label{eqn:potential-energy-Phi}
    V (\vect \Phi)
    = 
    \frac12 
    \vect \Phi \T
    \vect L \inv
    \vect \Phi
    -
    \sum_{j \in \mathcal S_J}
    E_{J_j}
    \cos{
        \left(
            \vect \Phi_j
            / \phi_0
        \right)
    }
    ,
\end{equation}
where $\vect L \inv$ is a diagonal matrix containing the inverse inductances of each branch, $\mathcal S_J$ is the set of branches that contain Josephson junctions, $E_{J_j}$ is the Josephson energy of the branch $\vect \Phi_j$, and $\phi_0 = \hbar/2e$ is the reduced flux quantum.
Substituting \cref{eqn:constraint-2-vectorised} into \cref{eqn:potential-energy-Phi} gives
\begin{equation}\label{eqn:potential-energy-phi}
    \begin{aligned}
        V (\vect \phi, \vect \phi \ext) 
        ={}&
        \frac12 
        \vect \phi \T
        \vect L_{\vect \phi} \inv
        \vect \phi  
        +
        \vect \phi \ext \T
        \vect L \coup \inv 
        \vect \phi 
        \\
        &-
        \sum_{j \in \mathcal S_J}
        E_{J_j}
        \cos{ \left[
            \left(
                \vect A_j \vect \phi 
                - \vect B_j \vect \phi \ext
            \right)
            / 
            \phi_0
        \right] }
        ,
    \end{aligned}
\end{equation}
where $\vect L_{\vect \phi} \inv = \vect A \T \vect L \inv \vect A$ and $\vect L \coup \inv = \vect B \T \vect L \inv \vect A$, and $\vect A_j$ and $\vect B_j$ are the $j$-th rows of $\vect A$ and $\vect B$ .
Again, we omit terms that are quadratic in $\vect \phi \ext$.

The Lagrangian for a circuit in the irrotational gauge is given by \cref{eqn:kinetic-energy-phi,eqn:potential-energy-phi}.
Taking the Legendre transform gives us the Hamiltonian:
\begin{equation}\label{eqn:legendre-transform}
    H 
    \equiv
    \vect q \T \dot {\vect \phi}
    -
    \mathcal L 
    ,
\end{equation}
where $\vect q$ is a column vector that contains the node charge variables that are canonically conjugate to the node flux variables $\vect \phi$.
By definition,
\begin{equation}\label{eqn:conjugate-charges-q}
    \vect q 
    \equiv 
    \frac{\partial \mathcal L}{\partial \dot{\vect \phi}} 
    = 
    \vect C_{\vect \phi} \dot {\vect \phi}
    ,
\end{equation}
where we use the fact that $\vect C_{\vect \phi} = \vect C_{\vect \phi} \T$.
Since \cref{eqn:kinetic-energy-phi} is quadratic in $\vect \phi$, the Hamiltonian for a circuit is given by
\begin{equation}\label{eqn:circuit-hamiltonian-q-phi}
    H 
    =
    T 
    +
    V 
    ,
\end{equation}
where the kinetic energy, expressed in terms of $\vect q$, is
\begin{equation}\label{eqn:kinetic-energy-q}
    T (\vect q)
    = 
    \frac12 
    \vect q \T
    \vect C_\mathrm{\vect \phi} \inv
    \vect q
    ,
\end{equation}
and the potential energy is given by \cref{eqn:potential-energy-phi}.

However, for many of the circuits that we consider, the node capacitance matrix $\vect C_\mathrm{\vect \phi}$ is singular, and a coordinate transformation is required to obtain a Hamiltonian.
We consider coordinate transformations of the form
\begin{equation}\label{eqn:coordinate-transformation-theta}
    \vect \theta
    =
    \frac{1}{\phi_0}
    \vect M \inv
    \vect \phi
    ,
\end{equation}
where $\vect \theta$ is a column vector containing (unitless) superconducting phase variables, and the coordinate transformation matrix $\vect M$ is invertible.
Substituting \cref{eqn:coordinate-transformation-theta} into \cref{eqn:kinetic-energy-phi,eqn:potential-energy-phi} gives
\begin{equation}\label{eqn:kinetic-energy-theta}
    T (\dot{\vect \theta})
    = 
    \frac{\phi_0^2}{2} 
    \dot {\vect \theta} \T
    \vect C_{\vect \theta}
    \dot {\vect \theta}
    ,
\end{equation}
where $\vect C_{\vect \theta} = \vect M \T \vect C_{\vect \phi} \vect M$, and 
\begin{equation}\label{eqn:potential-energy-theta}
    \begin{aligned}
        V (\vect \theta, \vect \varphi \ext) 
        ={}&
        \frac{\phi_0^2}{2} 
        \vect \theta \T
        \vect L_{\vect \theta} \inv
        \vect \theta  
        +
        \phi_0^2
        \vect \varphi \ext \T
        \vect L \coup \inv 
        \vect M
        \vect \theta
        \\
        &-
        \sum_{j \in \mathcal S_J}
        E_{J_j}
        \cos{
            \left(
                \vect A_j \vect M \vect \theta 
                - \vect B_j \vect \varphi \ext
            \right)
            }
        .
    \end{aligned}
\end{equation}
where $\vect L_{\vect \theta} \inv = \vect M \T \vect L_{\vect \phi} \inv \vect M$ and $\vect \varphi \ext = \vect \phi \ext / \phi_0$.
The variables canonically conjugate to $\vect \theta$ are given by
\begin{equation}\label{eqn:conjugate-charges-p}
    \vect p 
    \equiv 
    \frac{\partial \mathcal L}{\partial \dot{\vect \theta}} 
    =
    \phi_0^2 
    \vect C_{\vect \theta}
    \dot {\vect \theta}
    ,
\end{equation}
which have units of angular momentum.
It is conventional to define unitless charge number variables
\begin{equation}
    \vect n
    \equiv
    \frac{\vect p}{\hbar}   
    = 
    \frac{\phi_0}{2e} 
    \vect C_{\vect \theta}
    \dot {\vect \theta}
    .
\end{equation}
We can relate the new variables $\vect n$ to the old variables $\vect q$ by substituting in \cref{eqn:coordinate-transformation-theta,eqn:conjugate-charges-q}:
\begin{equation}\label{eqn:coordinate-transformation-n}
    \vect n
    = 
    \frac{1}{2e}
    \vect M \T
    \vect q
    .
\end{equation}
This allows us to rewrite \cref{eqn:circuit-hamiltonian-q-phi} in terms of $\vect \theta$ and $\vect n$:
\begin{equation}\label{eqn:circuit-hamiltonian-n-theta}
    H (\vect n, \vect \theta)
    = 
    T (\vect n)
    +
    V (\vect \theta, \vect \varphi \ext)
    ,
\end{equation}
where the kinetic energy, expressed in terms of $\vect n$, is
\begin{equation}\label{eqn:kinetic-energy-n}
    T (\vect n)
    = 
    \frac{4e^2}{2} 
    \vect n \T
    \vect C_{\vect \theta} \inv
    \vect n
    .
\end{equation}

Finally, the circuit Hamiltonian is quantised; classical variables are promoted to quantum operators that satisfy the commutation relations
\begin{equation}\label{eqn:commutation-relations-theta-n}
    \big[
        \hat {\vect \theta}_j
        ,
        \hat {\vect n}_{j'}
    \big]
    = 
    i \delta_{jj'} 
    .
\end{equation}
In the following sections, we use this procedure to derive Hamiltonians for the circuits presented in the main text.

\subsection{\texorpdfstring{$\vect0$}{0}-\texorpdfstring{$\vect\pi$}{π} qubit}\label{app:zero-pi-circuit}

The circuit for a $0$-$\pi$ qubit, shown in \cref{fig:zero-pi-circuit}, has $n = 4$ nodes, $k = 6$ branches, and $l = 1$ loops.
The first and second branches have Josephson energies $E_{J_{1,2}}$ and capacitances $C_{J_{1,2}}$, the third and fourth branches have capacitances $C_{1,2}$, and the fifth and sixth branches have inductances $L_{1,2}$. 
We ignore the capacitance of the inductors, capacitive coupling to ground, and capacitive coupling to any offset voltages.
In terms of the branch fluxes, the capacitance matrix is
\begin{equation}\label{eqn:zero-pi-capacitance-phi}
    \vect C
    =
    \mathrm{diag}
    \big(        
        C_{J_1},
        C_{J_2},
        C_1,
        C_2,
        0,
        0
    \big)
    ,
\end{equation}
and the inverse inductance matrix is
\begin{equation}\label{eqn:zero-pi-inductance-phi}
    \vect L \inv
    =
    \mathrm{diag}
    \big(
        0,
        0,
        0,
        0,
        L_1 \inv,
        L_2 \inv
    \big)
    .
\end{equation}
The branch fluxes $\vect \Phi$ are related to the node fluxes $\vect \phi$ according to \cref{eqn:constraint-2-vectorised}, where
\begin{equation}\label{eqn:zero-pi-A-matrix}
    \vect A 
    =
    \begin{pmatrix}
        1   & -1    & 0     & 0
        \\ 
        0   & 0     & 1     & -1
        \\
        1   & 0     & 0     & -1
        \\ 
        0   & 1     & -1    & 0
        \\ 
        1   & 0     & -1    & 0
        \\ 
        0   & 1     & 0     & -1
    \end{pmatrix}
    ,
\end{equation}
and $\vect B$ is given by \cref{eqn:irrotational-gauge}.
The external flux $\phi \ext$ is related to the branch fluxes $\vect \Phi$ by \cref{eqn:constraint-1-vectorised}, where
\begin{equation}\label{eqn:zero-pi-G-matrix}
    \vect G 
    =
    \begin{pmatrix}
        1   & -1     & 0     & 0    & -1     & 1
    \end{pmatrix}
    .
\end{equation}

The Hamiltonian for the circuit is $H = T + V$, where $T$ is given by \cref{eqn:kinetic-energy-q}, and $V$ is given by \cref{eqn:potential-energy-phi}.
However, the matrix $\vect C_{\vect \phi} = \vect A \T \vect C \vect A$ evaluates to
\begin{equation}
    \vect C_{\vect \phi} 
    =
    \begin{pmatrix}
        C_1 + C_{J_1}   & -C_{J_1}      & 0             & -C_1
        \\ 
        -C_{J_1}        & C_2 + C_{J_1} & -C_2          & 0
        \\ 
        0               & -C_2          & C_2 + C_{J_2} & -C_{J_2}
        \\ 
        -C_1            & 0             & -C_{J_2}      & C_1 + C_{J_2}
    \end{pmatrix}
    ,
\end{equation}
which is singular.
To proceed, we perform a coordinate transformation according to \cref{eqn:coordinate-transformation-theta,eqn:coordinate-transformation-n}.
We define new superconducting phase and number variables
\begin{equation}\label{eqn:zero-pi-variables-theta}
    \vect \theta
    = 
    \begin{pmatrix}
        \theta 
        \\ \varphi 
        \\ \zeta 
        \\ \Sigma
    \end{pmatrix}
    ,
    \quad \quad
    \vect n
    =
    \begin{pmatrix}
        n_\theta 
        \\ n_\varphi 
        \\ n_\zeta 
        \\ n_\Sigma
    \end{pmatrix}
    ,
\end{equation}
and the invertible transformation matrix~\cite{Dempster2014}
\begin{equation}\label{eqn:zero-pi-coordinate-transformation}
    \vect M \inv
    =
    \frac12
    \begin{pmatrix} 
        1       & -1    & 1     & -1 
        \\ 
        1       & -1    & -1    & 1 
        \\ 
        1       & 1     & -1    & -1 
        \\ 
        1       & 1     & 1     & 1 
    \end{pmatrix} 
    .
\end{equation}
The matrix $\vect M \inv$ is orthogonal, so $\vect M \inv = \vect M \T$.
The new variables $\vect \theta$ and $\vect n$ are quadrupole combinations of $\vect \phi$ and $\vect q$, respectively.
The quadrupoles are illustrated in \cref{fig:zero-pi-circuit}.
The transformed capacitance matrix is
\begin{equation}\label{eqn:zero-pi-capacitance-theta}
    \vect C_{\vect \theta}
    =
    \vect M \T
    \vect C_{\vect \phi}
    \vect M
    =
    \begin{pmatrix}
        C_\zeta + C_\varphi         & \delta C_\varphi    & \delta C_\zeta  & 0
        \\ 
        \delta C_\varphi      & C_\varphi           & 0         & 0
        \\ 
        \delta C_\zeta        & 0             & C_\zeta         & 0
        \\ 
        0               & 0             & 0         & 0
    \end{pmatrix}
    ,
\end{equation}
where $C_\zeta = C_1 + C_2$, $C_\varphi = C_{J_1} + C_{J_2}$, $\delta C_\zeta = C_1 - C_2$, and $\delta C_\varphi = C_{J_1} - C_{J_2}$.
Following a similar process, the transformed inverse inductance matrix is
\begin{equation}\label{eqn:zero-pi-inductance-theta}
    \vect L_{\vect \theta} \inv
    =
    \frac{1}{\phi_0^2}
    \begin{pmatrix}
        0       & 0             & 0             & 0
        \\ 
        0       & 2 \bar E_L           & \delta E_L    & 0
        \\ 
        0       & \delta E_L    & 2 \bar E_L           & 0
        \\ 
        0       & 0             & 0             & 0
    \end{pmatrix}
    ,
\end{equation}
where $\bar E_L = {\phi_0^2}/{2L_1} + {\phi_0^2}/{2L_2}$ and $\delta E_L = {\phi_0^2}/{L_1} - {\phi_0^2}/{L_2}$.

We can now write the circuit Hamiltonian in terms of the new variables, where $T$ is given by \cref{eqn:kinetic-energy-n} and $V$ is given by \cref{eqn:potential-energy-theta}.
The circuit Hamiltonian is obtained by inverting the capacitance matrix $\vect C_{\vect \theta}$.
Since the $\Sigma$ variable does appear in the Lagrangian, we can restrict our model to the three remaining variables.
It is also convenient to rewrite \cref{eqn:zero-pi-capacitance-theta} in the form
\begin{equation}\label{eqn:zero-pi-capacitance-rewritten}
    \vect C_{\vect \theta}
    =
    \begin{pmatrix}
        C_\theta + r_\varphi^2 C_\varphi + r_\zeta^2 C_\zeta    & r_\varphi C_\varphi   & r_\zeta C_\zeta 
        \\ 
        r_\varphi C_\varphi                         & C_\varphi       & 0
        \\ 
        r_\zeta C_\zeta                             & 0         & C_\zeta 
    \end{pmatrix}
    ,
\end{equation}
where $C_{\theta} = C_\zeta ( 1 - r_\zeta^2 ) + C_\varphi ( 1 - r_\varphi^2 )$, $r_\zeta = \delta C_\zeta / C_\zeta $, and $r_\varphi = \delta C_\varphi / C_\varphi$.
The inverse of \cref{eqn:zero-pi-capacitance-rewritten} is
\begin{equation}\label{eqn:zero-pi-capacitance-inverse}
    \vect C_{\vect \theta} \inv
    =
    \begin{pmatrix}
        \frac{1}{C_\theta}          & -\frac{r_\varphi}{C_\theta}                     & -\frac{r_\zeta}{C_\theta} 
        \\
        -\frac{r_\varphi}{C_\theta}       & \frac{1}{C_\varphi} + \frac{r_\varphi^2}{C_\theta}    & \frac{r_\zeta r_\varphi}{C_\theta}
        \\ 
        -\frac{r_\zeta}{C_\theta}         & \frac{r_\zeta r_\varphi}{C_\theta}                     & \frac{1}{C_\zeta } + \frac{r_\zeta^2}{C_\theta} 
    \end{pmatrix}
    .
\end{equation}

The irrotational gauge constraint \cref{eqn:irrotational-gauge} is particularly messy when $\delta C_\varphi \neq 0$.
For simplicity, we provide two sets of expressions for $T$ and $V$.
The first is for symmetric circuit elements, discussed in \cref{sec:zero-pi-hamiltonian},
\begin{equation}\label{eqn:zero-pi-hamiltonian-symmetric}
    \begin{aligned}
        \hat T_\mathrm{sym}
        ={}&
        4 E_{C_\theta} 
        \n_\theta^2
        +
        4 E_{C_\varphi} \n_\varphi^2
        +
        4 E_{C_\zeta} \n_\zeta^2
        ,
        \\
        \hat V_\mathrm{sym}
        ={}&
        E_L 
        \left[
            \left(
                \varphih
                -
                \frac{\varphi \ext}{2}
            \right)^2
            +
            \zetah^2
        \right]
        - 
        2E_J \cos \thetah \cos \varphih 
        ,
    \end{aligned}
\end{equation}
where $E_{C_\theta} = e^2/2C_\theta$, $E_{C_\varphi} = e^2/2C_\varphi$, $E_{C_\zeta} = e^2/2C_\zeta $, $E_L = \phi_0^2/L_1 = \phi_0/L_2$ and $E_J = E_{J_1} = E_{J_2}$.
The second is for asymmetric circuit elements and $\phi \ext = 0$,
\begin{equation}\label{eqn:zero-pi-hamiltonian-asymmetric}
    \begin{aligned}
        \hat T_\mathrm{asym}
        ={}&
        4 E_{C_\theta} 
        \big( 
            \n_\theta - r_\varphi \n_\varphi - r_\zeta \n_\zeta
        \big)^2
        +
        4 E_{C_\varphi} \n_\varphi^2
        +
        4 E_{C_\zeta} \n_\zeta^2
        ,
        \\
        \hat V_\mathrm{asym}
        ={}&
        \bar E_L 
        \big(
            \varphih^2 
            +
            \zetah^2
        \big)
        + 
        \delta E_L \varphih \zetah
        - 
        2 \bar E_J \cos \thetah \cos \varphih 
        \\ 
        &+ 
        \delta E_J \sin \thetah \sin \varphih
        ,
    \end{aligned}
\end{equation}
where $\bar E_J = (E_{J_1} + E_{J_2})/2$ and $\delta E_J = E_{J_1} - E_{J_2}$. 
This is discussed in \cref{sec:circuit-disorder}.

\subsection{\texorpdfstring{$\vect0$}{0}-\texorpdfstring{$\vect\pi$}{π} qubit coupled to an oscillator}\label{app:zero-pi-oscillator-circuit}

Next we consider the circuit for a $0$-$\pi$ qubit that is Josephson-coupled to a harmonic oscillator, as shown in \cref{fig:zero-pi-ancilla-circuit-intro}.
Here we treat the tunable Josephson element as a purely inductive element, consisting of two symmetric Josephson junctions with no capacitance threaded by an external flux.
In \cref{app:tunable-josephson-element-capacitance,app:tunable-josephson-element-dynamic-range} we consider the effects of nonzero capacitance and asymmetry in the tunable Josephson element, respectively.

This circuit has $n = 5$ nodes, $k = 10$ branches, and $l = 3$ loops.
The circuit is much the same as the bare $0$-$\pi$ qubit described in the previous section, with the addition of four new circuit elements.
We have added two Josephson junction with energies $E_{J_\text{s}}$, a capacitor with capacitance $C_\phi$ and inductor with inductance $L_\phi$.
This modifies the branch capacitance matrix
\begin{equation}\label{eqn:zero-pi-ancilla-capacitance-phi}
    \vect C
    =
    \mathrm{diag}
    \big(        
        C_{J_1},
        C_{J_2},
        0,
        0,
        C_1,
        C_2,
        0,
        0,
        C_\phi,
        0
    \big)
    ,
\end{equation}
and the inverse inductance matrix
\begin{equation}\label{eqn:zero-pi-ancilla-inductance-phi}
    \vect L \inv
    =
    \mathrm{diag}
    \big(        
        0,
        0,
        0,
        0,
        0,
        0,
        L_1 \inv,
        L_2 \inv,
        0,
        L_\phi \inv
    \big)
    .
\end{equation}
The vector of external fluxes for the three loops is
\begin{equation}
    \vect \phi \ext = 
    \begin{pmatrix}
        \phi^\mathrm{q} \ext \\
        \phi^\mathrm{s} \ext \\
        \phi^\mathrm{qs} \ext 
    \end{pmatrix},
\end{equation}
where $\phi^\mathrm{q} \ext$, $\phi^\mathrm{s} \ext$ are the external fluxes threading the $0$-$\pi$ qubit loop and the SQUID loop, and $\phi^\mathrm{qs} \ext$ is the external flux threading the loop between the $0$-$\pi$ qubit and the oscillator. 
We assume all external fluxes have the same orientation.
As before, we must specify the matrices $\vect A$ and $\vect G$ that relate the branch fluxes $\vect \Phi$ to the nodes fluxes $\vect \phi$ and external fluxes $\vect \phi \ext$, respectively.
The matrix $\vect A$ has four added rows (for the four added branches) and a single added column (for the single added node):
\begin{equation}\label{eqn:zero-pi-ancilla-A-matrix}
    \vect A 
    =
    \begin{pmatrix}
        1   & -1    & 0     & 0     & 0
        \\ 
        0   & 0     & 1     & -1    & 0
        \\
        1   & 0     & 0     & 0     & -1
        \\
        1   & 0     & 0     & 0     & -1
        \\
        1   & 0     & 0     & -1    & 0
        \\ 
        0   & 1     & -1    & 0     & 0
        \\ 
        1   & 0     & -1    & 0     & 0
        \\ 
        0   & 1     & 0     & -1    & 0
        \\
        0   & 0     & 0     & 1     & -1 \\
        0   & 0     & 0     & 1     & -1
    \end{pmatrix}
    .
\end{equation}
Likewise, the matrix $\vect G$ has two added rows (for the two added external fluxes) and four added columns (for the four added branches):
\begin{equation}\label{eqn:zero-pi-ancilla-G-matrix}
    \vect G 
    =
    \begin{pmatrix}
        1   & -1     & 0     & 0    & 0     & 0    & -1     & 1     & 0    & 0 
        \\ 
        0   & 0     & -1     & 1    & 0     & 0    & 0     & 0     & 0     & 0 
        \\
        -1   & 0     & 1     & 0    & 0     & 0    & 0     & -1      & 0     & -1   
    \end{pmatrix}
    .
\end{equation}

As before, we perform a coordinate transformation to a new set of variables, which we denote 
\begin{equation}\label{eqn:switch-variables-theta}
    \vect \theta
    = 
    \begin{pmatrix}
        \theta 
        \\ \varphi 
        \\ \zeta 
        \\ \Sigma
        \\ \phi
    \end{pmatrix}
    ,
    \quad \quad
    {\vect n} 
    =
    \begin{pmatrix}
        n_\theta 
        \\ n_\varphi 
        \\ n_\zeta 
        \\ n_\Sigma
        \\ n_\phi
    \end{pmatrix}
    ,
\end{equation}
where the invertible transformation matrix is~\cite{Paolo2019}
\begin{equation}\label{eqn:switch-coordinate-transformation}
    \vect M \inv
    =
    \frac12
    \begin{pmatrix} 
        1       & -1    & 1     & -1    & 0
        \\ 
        1       & -1    & -1    & 1     & 0
        \\ 
        1       & 1     & -1    & -1    & 0
        \\ 
        1       & 1     & 1     & 1     & 0
        \\
        0       & 0     & 0     & -2    & 2
    \end{pmatrix} 
    .
\end{equation}
The transformed capacitance matrix is
\begin{equation}\label{eqn:zero-pi-ancilla-capacitance-theta}
    \vect C_{\vect \theta}
    =
    \begin{pmatrix}
        C_\zeta  + C_\varphi         & \delta C_\varphi    & \delta C_\zeta   & 0 & 0
        \\ 
        \delta C_\varphi      & C_\varphi           & 0         & 0 & 0
        \\ 
        \delta C_\zeta         & 0             & C_\zeta          & 0 & 0
        \\  
        0               & 0             & 0         & 0 & 0
        \\ 
        0               & 0             & 0         & 0 & C_\phi
    \end{pmatrix}
    ,
\end{equation}
and the transformed inverse inductance matrix is
\begin{equation}\label{eqn:zero-pi-ancilla-inductance-theta}
    \vect L_{\vect \theta} \inv
    =
    \frac{1}{\phi_0^2}
    \begin{pmatrix}
        0       & 0             & 0             & 0 & 0
        \\ 
        0       & 2 \bar E_L           & \delta E_L    & 0 & 0
        \\ 
        0       & \delta E_L    & 2 \bar E_L           & 0 & 0
        \\ 
        0       & 0             & 0             & 0 & 0
        \\
        0       & 0             & 0             & 0 & E_{L_\phi}
    \end{pmatrix}
    ,
\end{equation}
where we have defined $E_{L_\phi} = {\phi_0^2}/{L_\phi}$.
In the regime of symmetric $0$-$\pi$ circuit elements, and after fixing the flux between the qubit and oscillator to be $\varphi^\mathrm{qs} \ext = -\varphi^\mathrm{q} \ext/2 - \varphi^\mathrm{s} \ext/2$, the full expressions for the kinetic and potential energies are
\begin{equation}
    \begin{aligned}
        \hat T
        &=
        \hat T_\mathrm{sym}
        + 
        4 E_{C_\phi} \hat n_\phi^2
        ,
        \\
        \hat V
        &=
        \hat V_\mathrm{sym}
        +
        \frac{E_{L_\phi}}{2} 
        \hat \phi^2
        - 2 E_{J_\mathrm{s}} \cos\Bigl(\frac{\varphi^\mathrm{s} \ext}{2}\Bigr)
        \cos{\big( \hat \phi - \hat \theta + \hat \zeta \big)}, 
    \end{aligned}
\end{equation}
where $\hat T_\mathrm{sym}$ and $\hat V_\mathrm{sym}$ are given in \cref{eqn:zero-pi-hamiltonian-symmetric} with $\varphi_\text{ext} = \varphi^\mathrm{q} \ext$. 
The last term in the potential gives rise to the interaction Hamiltonian in \cref{eqn:zero-pi-interaction} with $E_{J \intt}(t) = 2 E_{J_\mathrm{s}} \cos(\varphi^\mathrm{s} \ext/2)$, and the full Hamiltonian is therefore equal to $\H(t) = \H_\phi + \H_{\theta\varphi} + \H_\zeta + \H^\phi \intt(t)$, with the terms given by \cref{eqn:BKP-oscillator,eqn:zero-pi-theta-phi,eqn:zero-pi-zeta,eqn:zero-pi-interaction}, respectively.

\subsection{Coupling to the internal \texorpdfstring{$\zeta$}{ζ} mode}
\label{app:zero-pi-zeta-circuit}
\Cref{fig:zero-pi-zeta-circuit-intro} shows the circuit diagram for coupling to the internal $\zeta$ mode. 
Once again, we assume the tunable Josephson element to be a symmetric SQUID with Josephson energies $E_{J_\mathrm{s}}$ and no capacitance.
This circuit has $n=5$ nodes, $k=8$ branches and $l=3$ loops.
Similarly to the previous section we denote by $\phi^\mathrm{q} \ext$, $\phi^\mathrm{s} \ext$ and $\phi^\mathrm{qs} \ext$ the external fluxes threading the $0$-$\pi$ qubit, the tunable Josephson element, and the loop threading the qubit and tunable Josephson element, respectively.
As in the previous section, we fix the latter flux to be $\phi^\mathrm{qs} \ext = -\phi^\mathrm{q} \ext/2 - \phi^\mathrm{s} \ext/2$.
Then, following the same quantisation procedure and assuming symmetry in the $0$-$\pi$ qubit, the kinetic and potential energies become
\begin{equation} \label{eqn:zero-pi-zeta-kinetic-and-potential-energy}
    \begin{aligned}
        \hat T
        &=
        \hat T_\mathrm{sym}
        ,
        \\
        \hat V
        &=
        \hat V_\mathrm{sym}
        - 2 E_{J_\mathrm{s}} \cos\Bigl(\frac{\varphi^\mathrm{s} \ext}{2}\Bigr)
        \cos{\big(\hat \zeta + \hat \theta \big)}, 
    \end{aligned}
\end{equation}
where $\hat T_\mathrm{sym}$ and $\hat V_\mathrm{sym}$ are given in \cref{eqn:zero-pi-hamiltonian-symmetric} with $\varphi_\text{ext} = \varphi^\mathrm{q} \ext$. 
The last term in the potential gives rise to the interaction Hamiltonian in \cref{eqn:zero-pi-zeta-interaction} with $E_{J \intt}(t) = 2 E_{J_\mathrm{s}} \cos(\varphi^\mathrm{s} \ext/2)$, and the full Hamiltonian is therefore equal to $\H(t) = \H_\phi + \H_{\theta\varphi} + \H_\zeta + \H^\phi \intt(t)$, with the terms given by \cref{eqn:BKP-oscillator,eqn:zero-pi-theta-phi,eqn:zero-pi-zeta,eqn:zero-pi-zeta-interaction}, respectively.

Relaxing the symmetry requirement in the $0$-$\pi$ qubit leads to the same Hamiltonian, but with $\hat T_\mathrm{sym}$ and $\hat V_\mathrm{sym}$ in \cref{eqn:zero-pi-zeta-kinetic-and-potential-energy} replaced with $\hat T_\mathrm{asym}$ and $\hat V_\mathrm{asym}$ from \cref{eqn:zero-pi-hamiltonian-asymmetric}. 
However, the irrotational gauge constraint requires $\phi^\text{qs} \ext$ to be given by a different linear combination of $\phi^\text{q} \ext$ and $\phi^\text{s} \ext$, which may be found by solving a linear system. 
We omit its expression here for conciseness.

Finally, we note that nonzero capacitance in the tunable Josephson element may be accounted for by simply redefining the capacitive disorder in $\hat T_\mathrm{asym}$ to be $\delta C_\zeta  = C_1 + C \intt - C_2$, where $C \intt$ is the capacitance of the tunable Josephson element.

\subsection{Coupling to the internal \texorpdfstring{$\varphi$}{φ} mode}
\label{app:zero-pi-phi-circuit}
The inset of \cref{{fig:max-r-vs-EL-phi}} shows the circuit diagram for coupling to the internal $\varphi$ mode for the gate considered in \cref{sec:phi,app:phi-mode-gate}. 
We assume both tunable Josephson elements are symmetric SQUIDS with Josephson energies $E_{J_\mathrm{s}}$ and capacitances $C_J/2$. 
Furthermore, we assume the flux threading both tunable Josephson elements are identical.
This circuit has $n=5$ nodes, $k=8$ branches and $l=3$ loops.
Here we denote by $\phi^\mathrm{q} \ext$ and $\phi^{\mathrm{s}} \ext$ the external fluxes threading the qubit and the two tunable Josephson elements, respectively.
Following the same procedure as in the previous sections, and assuming symmetry in the $0$-$\pi$ qubit, the kinetic and potential energies become
\begin{equation} \label{eqn:zero-pi-phi-kinetic-and-potential-energy}
    \begin{aligned}
        \hat T
        ={}&
        \hat T_\mathrm{sym}
        ,
        \\
        \hat V
        ={}&
        E_L 
        \left[
            \left(
                \varphih
                -
                \frac{\varphi^\mathrm{q} \ext}{2}
                -
                \frac{\varphi^\mathrm{s} \ext}{2}
            \right)^2
            +
            \zetah^2
        \right] \\
        {}&- 
        4 E_{J_\mathrm{s}} \cos\Bigl(\frac{\varphi^\mathrm{s} \ext}{2}\Bigr) \cos \thetah \cos \varphih 
        ,
    \end{aligned}
\end{equation}
where $\hat T_\mathrm{sym}$ is given in \cref{eqn:zero-pi-hamiltonian-symmetric}.
Upon identifying $\varphi^\mathrm{q} \ext + \varphi^\mathrm{s} \ext$ with $\varphi \ext$ and $4 E_{J_\mathrm{s}} \cos(\varphi^\mathrm{s} \ext/2)$ with $2 E_J$, this becomes identical to \cref{eqn:zero-pi-hamiltonian-symmetric}.
The last term in the potential becomes the interaction Hamiltonian, which we use in \cref{eqn:zero-pi-phi-interaction} with $E_{J \intt}(t) = 4 E_{J_\mathrm{s}} \cos(\varphi^\mathrm{s} \ext/2)$. 

\subsection{Series \texorpdfstring{$\vect0$}{0}-\texorpdfstring{$\vect\pi$}{π} qubits}
\label{app:series-zero-pi-circuit}

\begin{figure}
    \includegraphics[scale=1]{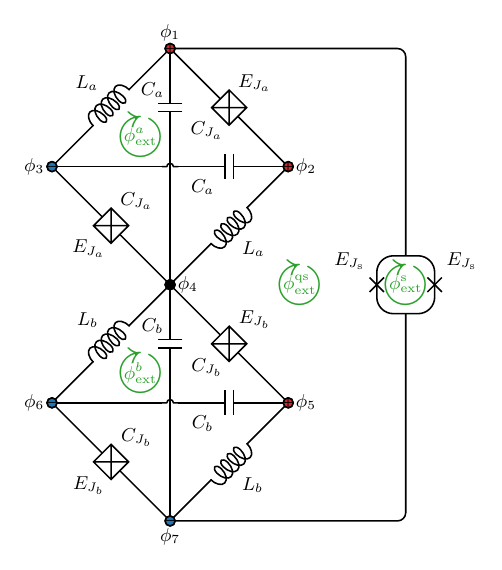}%
    \caption{Circuit diagram for a protected two-qubit gate using the internal $\zeta$ modes of two $0$-$\pi$ qubits connected in series. The node colours and signs correspond to the symmetric coupled $\zeta$ mode, $\zeta_+$, which is used to peform the gate. Each $0$-$\pi$ qubit is assumed to be symmetric, and the gate is performed by bringing the qubits onto resonance ($C_a = C_b$, $L_a = L_b$).}
    \label{fig:series-zero-pis-circuit}%
\end{figure}

\Cref{fig:series-zero-pis-circuit} shows the circuit diagram for two $0$-$\pi$ qubits connected in series and shunted by a SQUID for the two-qubit gate discussed in \cref{sec:two-qubit-gate}.
This circuit consists of $n=7$ nodes, $k=14$ branches and $l=4$ loops.
To separate out the symmetric and antisymmetric $\zeta$ modes of the coupled system, we define the following new set of variables
\begin{equation}\label{eqn:series-zero-pi-variables-theta}
    \vect \theta
    = 
    \begin{pmatrix}
        \theta_a \\
        \varphi_a \\
        \theta_b \\
        \varphi_b \\
        \zeta_+ \\
        \zeta_- \\
        \Sigma
    \end{pmatrix}
    ,
    \quad \quad
    {\vect n} 
    =
    \begin{pmatrix}
        n_{\theta_a} \\
        n_{\varphi_a} \\
        n_{\theta_b} \\
        n_{\varphi_b} \\
        n_{\zeta_+} \\
        n_{\zeta_-} \\
        n_{\Sigma}
    \end{pmatrix}
    ,
\end{equation}
which are related to the seven nodes of the circuit shown in \cref{fig:series-zero-pis-circuit} via the following coordinate transformation
\begin{equation}\label{eqn:series-zero-pi-coordinate-transformation}
    \vect M \inv
    =
    \frac12
    \begin{pmatrix} 
        1 & -1 & 1 & -1 & 0 & 0 & 0 \\
        1 & -1 & -1 & 1 & 0 & 0 & 0 \\
        0 & 0 & 0 & 1 & -1 & 1 & -1 \\
        0 & 0 & 0 & 1 & -1 & -1 & 1 \\
        1 & 1 & -1 & 0 & 1 & -1 & -1 \\
        1 & 1 & -1 & -2 & -1 & 1 & 1 \\
        1 & 1 & 1 & 1 & 1 & 1 & 1 
    \end{pmatrix}
    .
\end{equation}

For simplicity, we assume each $0$-$\pi$ qubit to be symmetric, and treat the tunable Josephson element as a symmetric SQUID with no capacitance. 
We denote the external fluxes in each $0$-$\pi$ qubit by $\phi^a \ext$ and $\phi^b \ext$, the external flux threading the tunable Josephson element by $\phi^\mathrm{s} \ext$, and the external flux threading the loop between the qubits and the tunable Josephson element by $\phi^\mathrm{qs} \ext$.

Using the well-established quantisation procedure, and fixing $\phi^\mathrm{qs} \ext = -(\phi^a \ext + \phi^b \ext + \phi^\mathrm{s} \ext)/2$, the kinetic and potential energies for the circuit are
\begin{equation}
\label{eqn:H-series-zero-pis}
\begin{aligned}
    \hat T ={}& \sum_{i=a,b} \Bigl[ 4 E_{C_{\theta_i}} \n_{\theta_i}^2 + 4 E_{C_{\varphi_i}} \n_{\varphi_i}^2 \Bigr] \\
    &+ 4E_{C_\zeta} (\n_{\zeta_+}^2 + \n_{\zeta_-}^2 - 2r_\zeta \n_{\zeta_+} \n_{\zeta_-}), \\
    \hat V = {}& \sum_{i = a,b} \Bigl[ E_{L_i} \left( \varphih_i + \frac{\varphi^i \ext}{2}\right)^2 - 2 E_{J_i} \cos(\thetah_i)\cos(\varphih_i) \Bigr] \\
    &+ E_{L_\zeta} (\zetah_+^2 + \zetah_-^2) - \delta E_L \zetah_+ \zetah_- \\
    &- 2E_{J_\mathrm{s}} \cos\left(\frac{\varphi^\mathrm{s} \ext}{2}\right)\cos(\thetah_a + \thetah_b + \zetah_+).
    \end{aligned}
\end{equation}

Here, $E_{C_{\theta_i}} = e^2/4( C_i + C_{J_i})$, $E_{C_{\varphi_i}} = e^2/4 C_{J_i}$, $E_{C_{\zeta}} = e^2/4E_{C_a} + e^2/4C_b$, $r_\zeta = (C_a - C_b)/(C_a + C_b)$, $E_{L_i} = \phi_0^2/L_i$, $E_{L_\zeta} = (E_{L_a} + E_{L_b})/2$ and $\delta E_L = (E_{L_a} - E_{L_b})/2$. Crucially, when the two $0$-$\pi$ qubits are brought on resonance ($C_a = C_b$ and $L_a = L_b$), then $r_\zeta = \delta E_L = 0$ and the symmetric and antisymmetric $\zeta$-modes decouple. 
In this case, the final term facilitates a conditional phase of $\pi/2$ acquired by the $\zeta_+$ mode when $\theta_a = \theta_b = \pi \bmod 2\pi$, thus enacting the logical unitary operation $\exp(-i\frac{\pi}{4} \hat{Z} \otimes \hat{Z})$. 

If the SQUID is replaced by an ancillary oscillator, as originally proposed in Ref.~\cite{Kitaev2006,Brooks2013}, then the resulting Hamiltonian is the same as \cref{eqn:H-series-zero-pis} with the addition of the bare-oscillator Hamiltonian and with the replacement $\cos(\thetah_a + \thetah_b + \zetah_+) \to \cos(\thetah_a + \thetah_b + \zetah_+ - \phih)$ in the final term in \cref{eqn:H-series-zero-pis}, where $\phih$ represents the degree of freedom for the ancillary oscillator. 
Therefore, the coupling to the $\zeta$-mode persists when coupling two qubits to an external oscillator, leading to the same exponential suppression of the interaction strength as for the single-qubit gate.

\subsection{Tunable Josephson element} \label{app:tunable-josephson-element}
Here we give supporting analysis for the implementation of the tunable Josephson element.
In \cref{app:tunable-josephson-element-capacitance} we discuss the effect of its capacitance and in \cref{app:tunable-josephson-element-dynamic-range} we discuss ways to obtain a large dynamic range.

\subsubsection{Capacitance of the tunable Josephson element}\label{app:tunable-josephson-element-capacitance}
 
For simplicity, we consider the case of the ideal qubit coupled to an external oscillator.
Here, a nonzero capacitance in the tunable Josephson element leads to the following kinetic energy for the coupled system
\begin{equation}
    \hat T = 4 E'_C \n_\theta^2 + 4 E'_{C_\phi} \n_\phi^2 + 4 E_{C_{\theta\phi}} \n_\theta \n_\phi,
\end{equation}
where $E'_C = e^2(C_\phi + C \intt)/2\tilde{C}$ and $E'_{C_\phi} = e^2(C_\theta + C \intt)/2\tilde{C}$ are the dressed charging energies for the qubit and oscillator, respectively, and $E_{C_{\theta\phi}} = e^2 C \intt / 2\tilde{C}$ is an effective charging energy for the charge-charge coupling term. Here, $C_\theta = C_J + C$, $C_\phi$ and $C \intt $ are the capacitances of the qubit, oscillator and the tunable Josephson element, respectively, and $\tilde{C} = C_\theta C_\phi + C_\theta C \intt + C_\phi C \intt$. In the limit that $C \intt \to 0$, we recover $E'_C = E_C$, $E'_{C_\phi} = E_{C_\phi}$ and $E_{C \intt} = 0$, which corresponds to the kinetic energy considered in \cref{sec:BKP}. 
The dressed charging energies may be expressed in terms of the ratios $r \intt = C \intt / C_\phi$ and $r_{\theta\phi} = C_\theta / C_\phi$ as
\begin{align}
    \frac{E_{C \intt}}{E'_{C_\phi}} &= \frac{r \intt}{r_{\theta\phi} + r \intt} \approx \frac{r \intt}{r_{\theta\phi}}, \\
    \frac{E'_{C_\theta}}{E'_{C_\phi}} &= \frac{1 + r \intt}{r_{\theta\phi} + r \intt} \approx \frac{1}{r_{\theta\phi}} + \frac{r \intt}{r_{\theta\phi}},
\end{align}
where the approximations hold for $r \intt \ll 1 \ll r_{\theta \phi}$. This reveals that the spurious charge-charge coupling is approximately proportional to $r \intt/r_{\theta\phi} = C \intt/C_\theta$ and that the charging energy ratio, $E'_{C_\theta}/E'_{C_\phi}$ is increased by this same quantity. 

\begin{figure}
    \centering%
    \includegraphics[width=\linewidth]{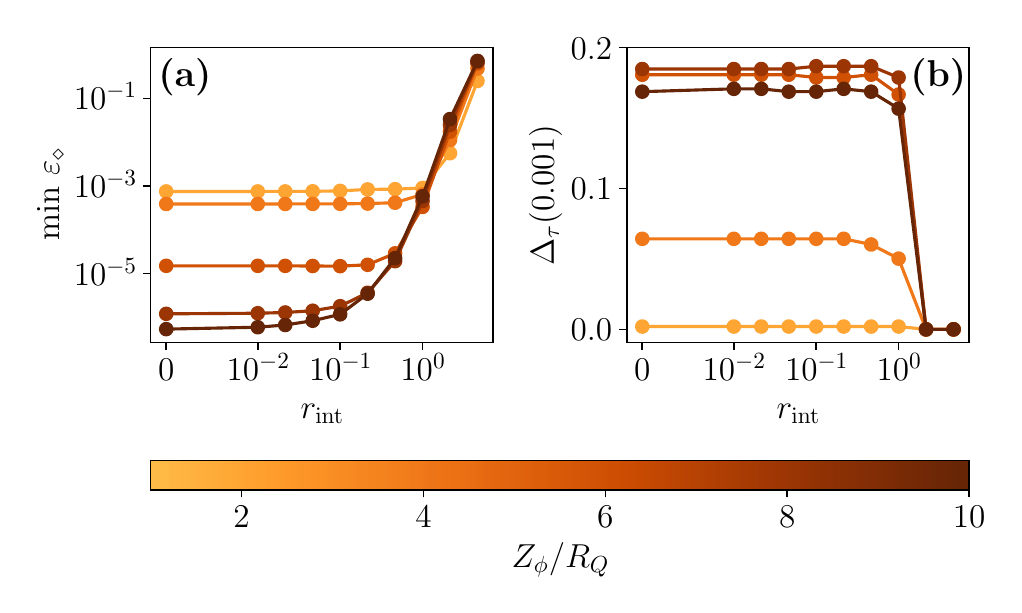}%
    \subfloat{\label{fig:precision-Cint}}%
    \subfloat{\label{fig:robustness-Cint}}%
    \caption{Gate errors for an ideal qubit coupled to an external oscillator with nonzero Josephson element capacitance. Imprecision (a) and robustness (b) of the gate as a function of the capacitance ratio $r \intt = C \intt / C_\phi$, for different impedances of the oscillator $Z_\phi$. Throughout, $r_{\theta\phi} = C_\theta/C_\phi = 100$, $E_{J \minn} = 0$, $E_{J \maxx}/E'_{C_\phi} = 100$ and $E_J/E'_{C_\phi} = 1$.}
    \label{fig:gate-erorrs-Cint}%
\end{figure}

\Cref{fig:gate-erorrs-Cint} shows the effect of $r \intt$ on the performance of a gate with an ideal qubit coupled to an external oscillator.
The results show that the gate is nearly unaffected up until $r \intt = 0.1$ and begins to lose its protection once $r \intt > 1$.
For a fixed plasma frequency, $\hbar\omega_p = \sqrt{8 E_{J \maxx} E_{C \intt}}$, where $E_{C \intt} = e^2/2 C \intt$ is the charging energy of the tunable Josephson element, the maximal interaction strength is given by 
\begin{align}
    \frac{E_{J \maxx}}{E'_{C_\phi}} &= \frac{r_{\theta\phi} r \intt + r_{\theta\phi} r \intt^2 + r \intt^2}{8(r_{\theta\phi} + r \intt)} \left( \frac{\hbar\omega_p}{E_{C_\phi}} \right)^2 \label{eqn:EJ-max-Cint} \\
    &\approx \frac{r \intt}{8} \left( \frac{\hbar\omega_p}{E_{C_\phi}} \right)^2,
\end{align}
where the approximation holds for $r \intt \ll 1 \ll r_{\theta \phi}$.

The above analysis may be taken as a best-case scenario for a $0$-$\pi$ qubit coupled to an external oscillator. 
Assuming the oscillator impedance may be obtained with $E_{C_\phi}/h = 1$~GHz and $E_{L_\phi}/h = 2$~MHz, the maximum Josephson coupling required for the protected gate would be greater than $E_{J \maxx}/h = 8 \times 10^4$~GHz. 
With a plasma frequency of the tunable Josephson element of $\omega_p/2\pi = 40$~GHz, this implies that $C \intt / C_\phi > 100$, which is well above the requirement for the gate to remain protected. Assuming a plasma frequency of $\omega_p/2\pi = 40$~GHz and a bare oscillator charging energy of $E_{C_\phi}/h = 1$~GHz yields a maximal coupling strength of $E_{J \maxx}/E'_{C_\phi} \approx 200 r \intt$.
Requiring $r \intt <1$ then restricts the dynamic range to values $E_{J \maxx}/E'_{C_\phi} < 200$.
This upper bound is larger than the values considered in \cref{sec:BKP}.
However, taking this bound as a best-case scenario for the $0$-$\pi$ qubit coupled to an external oscillator, its value is too small to achieve a protected gate, where we found a value of $E_{J \maxx}/E_{C_\phi} > 8 \times 10^4 $ was necessary in \cref{sec:numerical-simulations-zero-pi}.
As mentioned in \cref{app:zero-pi-zeta-circuit}, the capacitance in the tunable Josephson element may be lumped into the capacitance for the $\zeta$ mode, which avoids these spurious effects so long as the $0$-$\pi$ qubit remains symmetric.

\subsubsection{Extending the dynamic range of a flux-tunable Josephson element}\label{app:tunable-josephson-element-dynamic-range}

Consider the potential energy for a SQUID with Josephson junction energies $E_{J_1}$ and $E_{J_2}$, 
\begin{equation}
    \begin{aligned}
    \hat V = {}& -(E_{J_1} + E_{J_2}) \cos(\varphi_\text{ext})\cos(\phih) \\ &- (E_{J_1} - E_{J_2}) \sin(\varphi_\text{ext})\sin(\phih) \\
    = {}& -E_{J_\mathrm{eff}}(\varphi \ext)\cos(\phih - \varphi_0),
    \end{aligned}
\end{equation}
where 
\begin{equation}
    E_{J_\mathrm{eff}}(\varphi \ext) = (E_{J_1} + E_{J_2}) \sqrt{1 + d^2 \tan^2(\varphi_\text{ext})},
\end{equation}
$d = (E_{J_1} - E_{J_2})/(E_{J_1} + E_{J_2})$ is the Josephson asymmetry and $\tan \varphi_0 = d\tan\varphi \ext $. 
Its dynamic range is then given by $1/d$, and is therefore determined by the asymmetry of the two Josephson junctions. 

\begin{figure}
    \centering%
    \includegraphics[scale=1]{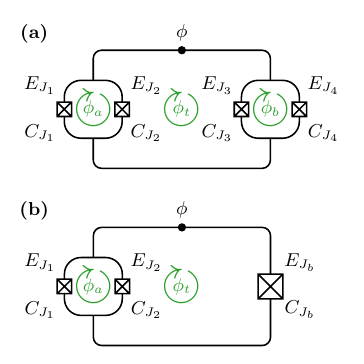}%
    \subfloat{\label{fig:three-loop-squid}}%
    \subfloat{\label{fig:two-loop-squid}}%
    \caption{(a) Three-loop and (b) two-loop SQUIDs for realising a tunable Josephson element. $\phi_a$ and $\phi_b$ are time-independent external fluxes, whereas $\phi_t$ is a time-dependent external flux. $\phi$ denotes the single degree of freedom in the circuit.}
    \label{fig:multi-loop-squids}%
\end{figure}

Now consider the case where each Josephson junction is replaced by a SQUID as shown in \cref{fig:three-loop-squid}. The basic idea is that the two time-independent fluxes in each of the smaller loops $\phi_a$ and $\phi_b$ are used to bring the effective Josephson energies of each loop closer to each other, whilst the time-dependent flux in the central loop $\phi_t$ is used to tune the overall potential energy. Here we show that this results in a dynamic range that is limited by the flux noise in the external fluxes rather than the static Josephson asymmetries.

Following the procedure outlined in \cref{app:circuit-quantisation}, the potential energy for the circuit is
\begin{equation} \label{eqn:tunable-JJ-potential-1}
    \hat V = - \sum_{i=1}^4 E_{J_i} \cos( \phih + \vect C_{J_i} \vect \varphi),
\end{equation}
where 
\begin{equation}
    \vect \varphi = 
    \begin{pmatrix}
        \varphi_a \\
        \varphi_t \\
        \varphi_b
    \end{pmatrix}
\end{equation}
and $\vect C_{J_i}$ denotes the $i$-th row of the matrix 
\begin{equation}
    \vect C_J
    =
    \frac{1}{C_\Sigma}
    \begin{pmatrix} 
        C_{J_1} - C_\Sigma & -C_{J_3} - C_{J_4} & -C_{J_4} \\
        C_{J_1} & -C_{J_3} - C_{J_4} & -C_{J_4} \\
        C_{J_1} & C_{J_1} + C_{J_2} & -C_{J_4} \\
        C_{J_1} & -C_{J_1} - C_{J_2} & C_{J_4} - C_\Sigma
    \end{pmatrix},
\end{equation}
with $C_\Sigma = C_{J_1} + C_{J_2} + C_{J_3} + C_{J_4}$ the total capacitance.

Taking all Josephson junctions to be equal up to the following asymmetries
\begin{align}
    d_a &= \frac{C_{J_1} - C_{J_2}}{C_{J_1} + C_{J_2}} \nonumber\\
    &= \frac{E_{J_1} - E_{J_2}}{E_{J_1} + E_{J_2}}, \\
    d_b &= \frac{C_{J_3} - C_{J_4}}{C_{J_3} + C_{J_4}} \nonumber\\
    &= \frac{E_{J_3} - E_{J_4}}{E_{J_3} + E_{J_4}}, \\
    d_{ab} &= \frac{C_{J_1} + C_{J_2} - C_{J_3} - C_{J_4}}{C_{J_1} + C_{J_2} + C_{J_3} + C_{J_4}} \nonumber\\
    &= \frac{E_{J_1} + E_{J_2} - E_{J_3} - E_{J_4}}{E_{J_1} + E_{J_2} + E_{J_3} + E_{J_4}},
\end{align}
\cref{eqn:tunable-JJ-potential-1} may be rewritten as
\begin{equation}
\begin{aligned} \label{eqn:tunable-JJ-potential-2}
    \hat V = &{} -E_{J_a^\mathrm{eff}}(\varphi_a') \cos(\phih - \varphi_t' - \varphi_0 + d_{ab}\varphi_t') \\
    &{} - E_{J_b^\mathrm{eff}}(\varphi_b') \cos(\phih + \varphi_t' - \varphi_0 + d_{ab}\varphi_t'), 
\end{aligned}
\end{equation}
where 
\begin{equation} \label{eqn:EJeff}
    E_{J_i^\mathrm{eff}}(\varphi'_i) = E_{J_i}\sqrt{1 + d_i^2 \tan^2(\varphi'_i)}\cos(\varphi'_i),
\end{equation}
are the effective tunable Josephson energies for the left ($i=a$) and right ($i=b$) loops with $E_{J_a} = E_{J_1} + E_{J_2}$ and $E_{J_b} = E_{J_3} + E_{J_4}$. In these equations, the transformed fluxes are 
\begin{align}
    \varphi_a' &= \frac{\varphi_a}{2} \\
    \varphi_b' &= \frac{\varphi_b}{2} \\
    \varphi_t' &= \frac{1}{2}\left( \varphi_t - \varphi'_a - \varphi'_b + \alpha_a - \alpha_b \right)\\
    \varphi_0 &= \frac{1}{2}\left[(1 + d_{ab})(d_a \varphi_a' + \alpha_a) + (1 - d_{ab})(d_b \varphi_b' + \alpha_b)\right],
\end{align}
with $\tan\alpha_a = d_a\tan\varphi_a'$ and $\tan\alpha_b = d_b\tan\varphi_b'$. Importantly, we note that $\varphi_t'$ is the only time-dependent quantity in these equations. \Cref{eqn:tunable-JJ-potential-2} can now be rewritten in terms of a single cosine potential as
\begin{equation} \label{eqn:EJeff-final}
    \hat V = -E_{J_t^\mathrm{eff}}(\varphi'_t)\cos(\phih - \varphi_0 + d_{ab}\varphi_t' - \alpha_t), 
\end{equation}
where $E_{J_t^\mathrm{eff}}$ is given by \cref{eqn:EJeff} with $i=t$ and
\begin{align}
    E_{J_t} &= E_{J_a^\mathrm{eff}}(\varphi'_a) + E_{J_b^\mathrm{eff}}(\varphi'_b), \\
    d_t &= \frac{E_{J_a^\mathrm{eff}}(\varphi'_a) - E_{J_b^\mathrm{eff}}(\varphi'_b)}{E_{J_a^\mathrm{eff}}(\varphi'_a) + E_{J_b^\mathrm{eff}}(\varphi'_b)}, \\
    \tan\alpha_t &= d_t \tan\varphi'_t.
\end{align}
Therefore, by fixing the time-independent fluxes in the left and right loops such that 
\begin{equation} \label{eqn:time-independent-flux-condition}
    E_{J_a^\mathrm{eff}}(\varphi_a') = E_{J_b^\mathrm{eff}}(\varphi_b'),
\end{equation}
then $d_t$ and $\alpha_t$ vanish in \cref{eqn:EJeff-final}. Since $\varphi_0$ is a time-independent phase, it may be ignored, leaving us with the potential
\begin{equation}
    \hat V = -E_{J_t^\mathrm{eff}} \cos(\varphi'_t) \cos(\phih + d_{ab}\varphi_t').
\end{equation}
Crucially, the dynamic range is not affected by the static asymmetries in the Josephson energies and the time-dependent phase $d_{ab}\varphi_t'$ only leads to a small deformation of the cosine potential when the pulse is being turned on and off.

One such choice of the time-independent fluxes that satisfies \cref{eqn:time-independent-flux-condition} is 
\begin{align}
    \varphi_a &= 2\arctan \sqrt{\frac{1 - r_{E_J}^2}{r_{E_J}^2 - d_a^2}}, \\
    \varphi_b &= 0,
\end{align}
where $r_{E_J} = E_{J_b}/E_{J_a}$ and we have assumed $E_{J_a} > E_{J_b}$ without loss of generality. Since the right loop is now superfluous, this choice may be realised with three Josephson junctions and two external fluxes as shown in \cref{fig:two-loop-squid}.  

Assuming there is noise in the external flux so that $\varphi_a \to \varphi_a + \delta \varphi$, and expanding $d_\text{eff}$ to first order in $\delta\varphi$, yields
\begin{align}
   d_\text{eff} &= \delta \varphi \frac{\sqrt{(1 - r_{E_J}^2)(r_{E_J}^2 - d_a^2)}}{(r_{E_J}^2 + 1)^2} \\
   &\approx \delta\varphi\frac{\sqrt{d_{ab}}}{2}
\end{align}
where we have expanded to leading order in the asymmetries in the final line. This shows that the dynamic range is limited by the flux noise rather than the static Josephson energy asymmetries.

\section{Numerical simulations}\label{app:numerical-simulations}
In this appendix we explain the numerical methods used to simulate the gate. 
Our simulations make use of the split-operator method~\cite{Fleck1976} to perform time evolution and find ground states via imaginary time evolution.
For ease of notation we set $\hbar =1$ in this section.

\subsection{Time evolution}
We exploit the fact that there are no terms that couple position and momentum of the same mode. Considering the simple case of a single-mode Hamiltonian to begin with, the time-evolution unitary may be decomposed via a second-order Trotterisation,
\begin{equation}
    \U(dt) =  \U_x(dt/2) \U_p(dt) \U_x(dt/2) + O(dt^3),
\end{equation}
where $\U(t) = e^{-i (\H_x + \H_p) t}$ and $\U_{x/p}(dt) = e^{-i \H_{x/p} t}$, are unitary operators, with $\H_{x/p}$ denoting the parts of the Hamiltonian that are functions of only position/momentum operators. Time evolution may then be simulated through alternating multiplications of the unitaries $\U_x(dt/2)$ and $\U_p(dt)$ onto a state's wavefunction interleaved with Fourier and inverse Fourier transforms to alternate between the position and momentum bases:
\begin{equation}
    \psi(t + dt) =  \U_x(dt/2) \, \mathcal{F}^{-1} \, \U_p(dt) \, \mathcal{F} \, U_x(dt/2) \, \psi(t), \label{eqn:split-op-time-step}
\end{equation}
where $\psi(t)$ denotes the position wavefunction at time $t$, and $\mathcal{F}$ and $\mathcal{F}^{-1}$ denote the Fourier and inverse Fourier transform operators, respectively. That is, by repeated applications of the time step in \cref{eqn:split-op-time-step} the time-evolved state $\psi(t)$ may be computed from an initial state $\psi(0)$. Crucially, owing to the Fourier transforms, $\U_x$ and $\U_p$ are diagonal in the basis on which they act, and therefore may be represented by vectors with the same dimensions as $\psi(t)$. Moreover, by concatenating \cref{eqn:split-op-time-step}, the half time-steps may be combined at intermediate times so that $N-1$ applications of the sequence of operators  
\begin{equation}
    \U_x(dt) \mathcal{F}^{-1} \U_p(dt) \mathcal{F} \label{eqn:split-op-update-rule}
\end{equation}
onto the state $\U_x(dt/2)\psi(0)$, with a final application of $\U_x(dt/2) \mathcal{F}^{-1} \U_p(dt) \mathcal{F}$, yields the time-evolved state $\psi(t)$ with an error of $O(dt^3) = O((t/N)^3)$.\\

All of the Hamiltonians considered in this manuscript are two-mode Hamiltonians with no coupling between position and momentum of the same mode. Furthermore, the Hamiltonian for the gate with the ideal protected qubit involves no cross terms that couple position of one mode to momentum of another mode. In this case, the update rule in \cref{eqn:split-op-update-rule} may be applied by casting the vectors $\psi$, $\U_x$ and $\U_p$ into matrices, with each dimension representing position or momentum in each mode, and using a two-dimensional Fourier transform. In contrast, the interaction Hamiltonians for simulating a gate with the $0$-$\pi$ qubit [\cref{eqn:zero-pi-interaction-exp-suppression-alpha,eqn:zero-pi-zeta-interaction-alpha}] involve a term that couples position of one mode to momentum of the other. In this case, the full Hamiltonian may be written as
 \begin{equation}
    \H = \sum_{i=1,2} \H_{x_i} + \H_{p_i} + \H_{x_1, p_2}, \label{eqn:H-x-p}
\end{equation}
where $\H_{x_i/p_i}$ denotes the part of the Hamiltonian that is a function of position/momentum operators acting solely on the $i$-th mode, and $\H_{x_1, p_2}$ couples the position of one mode to the momentum of the other. The Fourier transforms in each dimension must now be staggered, yielding the following modified update rule
\begin{equation}
   \mathcal{F}_2^{-1} \, U_{x_1,p_2}(dt) \, \mathcal{F}_2 \, U_x(dt/2) \, \mathcal{F}^{-1} \, U_p(dt) \mathcal{F} \, U_x(dt/2),
\end{equation}
for all $N-1$ intermediate updates, with the initial and final updates given by
\begin{subequations}
\begin{align}
    & \mathcal{F}^{-1}_2 \, \U_{x_1,p_2}(dt/2) \, \mathcal{F}_2, \\
    & \mathcal{F}_2^{-1} \, \U_{x_1,p_2}(dt/2) \, \mathcal{F}_2 \, \U_x(dt/2) \, \mathcal{F}^{-1} \, \U_p(dt) \, \mathcal{F} \, \U_x(dt/2),
\end{align}
\end{subequations}
respectively. 
Here, $\U_{x/p}(t) = e^{-i(\H_{x_1/p_1} + \H_{x_2/p_2})t}$, $\U_{x_1,p_2}(t) = e^{-i \H_{x_1,p_2}t}$, $\mathcal{F}$ denotes the two-dimensional Fourier transform and $\mathcal{F}_2$ denotes the partial Fourier transform on the second mode.
Similarly, this method yields the time-evolved state $\psi(t)$ with an error of $O(dt^3)$.\\

\subsection{Imaginary time evolution}
For each gate simulation, we first find the ground states of the two-mode Hamiltonian by applying the split-operator method with an imaginary time-step, $dt \to -idt$, to an initial ansatz wavefunction $\psi(0)$. For the gate with the ideal protected qubit, this ansatz wavefunction is chosen to be 
\begin{equation}
    \psi(0) = \frac{1}{\sqrt{2}}[\mathcal{N}_{\sigma_1,\sigma_2}(0,0) + \mathcal{N}_{\sigma_1,\sigma_2}(0, \pi)], \label{eqn:initial-wavefunction}
\end{equation}
where $\mathcal{N}_{\sigma_1,\sigma_2}(\mu_1,\mu_2)$ denotes a two-dimensional Gaussian with standard deviations $\sigma_1$, $\sigma_2$ and means $\mu_1$, $\mu_2$. After evolving under the Hamiltonian for sufficiently long, each Gaussian in \cref{eqn:initial-wavefunction} converges to a ground-state wavefunction that is localised near $x_2 = 0$ or $x_2 = \pi$, and therefore the equal superposition defines a $+1$ eigenstate of $\bar X_\mathrm{eff}$ as per our subsystem decomposition [see \cref{eqn:X-eff-theta} in \cref{app:subsystem-decomp}]. Whilst the true ground state of the Hamiltonian may correspond to a different superposition of the two localised wavefunctions, for the Hamiltonians we consider, tunnelling in the $x_2$ direction occurs on a much longer time scale than tunnelling in the $x_1$ direction due to the large charging energy separation, e.g., $E_{C_\theta} \ll E_{C_\phi}$.
Therefore, we choose a total (imaginary) evolution time that is on the order of the harmonic oscillator period in the $x_1$ direction. 

When simulating the gate with the $0$-$\pi$ qubit, the ansatz wavefunction must be modified since position in $\theta$ corresponds to momentum in the effective mode $\alpha$. 
We choose
\begin{equation}
    \psi(0) = \mathcal{N}_{\sigma_1,\sigma_2}(0,0),
\end{equation}
which approximately corresponds to a $+1$ eigenstate of $\bar X_\text{eff}$ when $x_2 = \alpha$ [see \cref{eqn:X-eff-alpha} in \cref{app:subsystem-decomp}].

\section{Quantifying the gate error}\label{app:subsystem-decomp-and-dnorm}

In \cref{app:subsystem-decomp} we introduce subsystem decompositions for the full Hilbert space of the protected qubit systems to determine the performance of the various gates discussed in the main text. 
In \cref{app:dnorm}, we use these subsystem decompositions to derive expressions for the diamond norm error of these gates.

\subsection{Subsystem decompositions} \label{app:subsystem-decomp}
To analyse the performance of our gates while taking advantage of the error-correcting properties of the qubits considered in this paper, we utilise the subsystem decompositions in Refs.~\cite{Pantaleoni2020,Shaw2024}, which are developed for GKP codes. 
These subsystem decompositions break the full Hilbert space $\mathcal{H}$ into a tensor product of a \textit{logical} subsystem $\mathcal{H}_L$, which contains the logical information, and a \textit{stabiliser} subsystem $\mathcal{H}_S$, which contains the information from stabiliser measurements:
\begin{equation}
    \mathcal{H} = \mathcal{H}_L \otimes \mathcal{H}_S.
\end{equation}
The choice of tensor product structure is not unique and can be thought of as a choice of decoder and correction protocol for the error-corrected qubit. 

For any given subsystem decomposition one may define \textit{effective qubit Pauli operators} that act trivially on the stabiliser subsystem:
\begin{subequations}%
    \begin{align}
        \bar{X}_\mathrm{eff} &= \hat{X} \otimes \hat{I}, \\
        \bar{Y}_\mathrm{eff} &= \hat{Y} \otimes \hat{I}, \\
        \bar{Z}_\mathrm{eff} &= \hat{Z} \otimes \hat{I}, 
    \end{align}
\end{subequations}
where $\hat{X}, \hat{Y}, \hat{Z}$ are the usual $2\times 2$ Pauli matrices acting on the logical subspace and $\hat{I}$ denotes the identity operator. Note that these operators are distinct from the logical operators of the stabiliser code [e.g., \cref{eqn:GKP-stabilisers}].
Using the effective qubit operators, we can define a \textit{logical} density matrix $\rhoh \in \mathcal{H}_L$ via
\begin{equation}\label{eqn:logical-dm}
    \rhoh = \frac{1}{2} \left( \hat{I} + \expval{\bar{X}_\text{eff}} \hat{X} + \expval{\bar{Y}_\text{eff}} \hat{Y} + \expval{\bar{Z}_\text{eff}} \hat{Z} \right), 
\end{equation}
where the expectation values are taken with respect to the state in the full Hilbert space. 
Conceptually, $\rhoh$ is the state of the logical qubit after the error correction procedure captured by the choice of subsystem decomposition. 
We will use these logical states before and after a gate operation to determine the quantum operation that is performed on the logical qubit. 
Gate performance is then characterised by a diamond norm error associated with this quantum operation.

In the following, we introduce three separate subsystem decompositions reflecting three different decoders.
These are used for the following three gate scenarios: the gate with the ideal protected qubit; the gate with the $0$-$\pi$ qubit using either its internal $\zeta$ mode or an external oscillator; and the gate with the $0$-$\pi$ qubit using its internal $\varphi$ mode.

First, we consider the case of the ideal qubit coupled to an external oscillator.
The full Hilbert space is decomposed as follows
\begin{equation} \label{eqn:Hilbert-space-decomp}
    \mathcal{H} = \mathcal{H}_\theta \otimes \mathcal{H}_\phi = (\mathcal{H}_L \otimes \mathcal{H}_{S}) \otimes \mathcal{H}_\phi,
\end{equation}
where $\mathcal{H}_\theta$ and $\mathcal{H}_\phi$ represent the subsystems for the qubit and oscillator modes, respectively.
We choose a subsystem decomposition of the qubit subsystem that allows us to define the qubit state based on its modular position in $\theta$.
To this end, the qubit subsystem $\mathcal{H}_\theta$ is decomposed into the logical subsystem $\mathcal{H}_L$ and the stabiliser subsystem $\mathcal{H}_S$ via the following map:
\begin{equation} \label{eqn:subsystem-decomp-mapping}
    \ket{\theta} \to \ket{\mu}\ket*{\tilde{\theta}}, 
\end{equation}
such that $\theta = \mu\pi + \tilde{\theta}$, where $\mu = 0,1$ and $\tilde{\theta} \in [-\pi/2, \pi/2)$. 
In this way, we define states whose support lies in $[-\pi/2, \pi/2)$ to be $+1$ eigenstates of the Pauli-$Z$ effective qubit operator $\bar{Z}_\text{eff}$, and states whose support lies in $[\pi/2, 3\pi/2)$ to be the $-1$ eigenstates of $\bar{Z}_\text{eff}$. 
This facilitates the following effective qubit operators
\begin{subequations} \label{eqn:effective-qubit-operators-theta}
    \begin{align}
        \bar{X}_\text{eff} &= \int_{-\pi/2}^{\pi/2} d\theta \, \ketbasisbra{\theta}{\theta}{\theta+\pi} + \ketbasisbra{\theta+\pi}{\theta}{\theta}, \label{eqn:X-eff-theta}\\
        \bar{Y}_\text{eff} &= \int_{-\pi/2}^{\pi/2} d\theta \, -i\ketbasisbra{\theta}{\theta}{\theta+\pi} +i \ketbasisbra{\theta+\pi}{\theta}{\theta}, \\
        \bar{Z}_\text{eff} &= \int_{-\pi/2}^{\pi/2} d\theta \, \ketbasisbra{\theta}{\theta}{\theta} - \ketbasisbra{\theta+\pi}{\theta}{\theta+\pi},
    \end{align}
\end{subequations} 
which implicitly act as the identity on the oscillator mode. 
Here and in what follows we use the notation $\ket{x}_x$ to represent an eigenstate of the operator $\hat{x}$ with eigenvalue $x$.
It is easily verified that these operators are unitary and Hermitian, commute with both the stabiliser generators in \cref{eqn:GKP-stabilisers}, anticommute with each other, and satisfy $\bar Y_\mathrm{eff} = i \bar X_\mathrm{eff} \bar Z_\mathrm{eff}$.

Next, we consider the gate with the $0$-$\pi$ qubit using either its internal $\zeta$ mode or an external oscillator.
In this case, the Hamiltonian for the $0$-$\pi$ qubit is described by the effective Hamiltonian in \cref{eqn:H-alpha}.
Here, the Hilbert space may be decomposed using \cref{eqn:Hilbert-space-decomp} with $\theta \to \alpha$. 
Additionally, $\phi \to \zeta$ when the internal $\zeta$ mode is used as the ancillary mode for the gate.
Since $\theta = 0$ corresponds to even $n_\alpha$ and $\theta = \pi$ corresponds to odd $n_\alpha$ in the effective model, we define a logical $\ket{0}$ state to be one with support on only even $n_\alpha$ and a logical $\ket{1}$ state to be one with support on only odd $n_\alpha$.
This means that a logical state localised at $\theta = 0$ now corresponds to a wavefunction over $\alpha$ that is a symmetric superposition of states localised at $\alpha = 0$ and $\alpha = \pi$, and a logical state localised at $\theta = \pi$ corresponds to the antisymmetric superposition. 
The resulting effective qubit operators are
\begin{subequations} \label{eqn:effective-qubit-operators-alpha}
    \begin{align}
        \bar{X}_\text{eff} &= \int_{-\pi/2}^{\pi/2} d\alpha \, \ketbasisbra{\alpha}{\alpha}{\alpha} - \ketbasisbra{\alpha+\pi}{\alpha}{\alpha+\pi}, \label{eqn:X-eff-alpha}\\
        \bar{Y}_\text{eff} &= \int_{-\pi/2}^{\pi/2} d\alpha \, i\ketbasisbra{\alpha}{\alpha}{\alpha+\pi} -i \ketbasisbra{\alpha+\pi}{\alpha}{\alpha}, \\
        \bar{Z}_\text{eff} &= \int_{-\pi/2}^{\pi/2} d\alpha \, \ketbasisbra{\alpha}{\alpha}{\alpha+\pi} + \ketbasisbra{\alpha+\pi}{\alpha}{\alpha}.
    \end{align}
\end{subequations}
If we relabelled $\alpha \to \theta$ these correspond to the effective qubit operators in \cref{eqn:effective-qubit-operators-theta} after a logical Hadamard.
This is a consequence of the fact that flux gets mapped to charge under the effective model, and that the Fourier transform corresponds to a logical Hadamard for the GKP code.

Finally, in the case of using the internal $\varphi$ mode to perform a gate with the $0$-$\pi$ qubit, we choose to decompose the Hilbert space via \cref{eqn:Hilbert-space-decomp} with $\phi \to \varphi$ and the effective qubit operators are the same as in \cref{eqn:effective-qubit-operators-theta}.
This corresponds to using the entire $\varphi$ mode as the ancillary mode and disregarding any stabiliser information contained in this mode.
We do not believe this is the optimal choice, particularly in the large $E_J/E_{C_\varphi}$ regime.
An improved decoder should make use of $\theta$ and $\varphi$ simultaneously.
However, due to the large size of the $\varphi$-mode Hilbert space, this approach is computationally expensive because the measurement operators have support on both dimensions in the numerics. 
While it would be of interest to study the optimal performance of this gate, we do not believe that our choice of decoder here affects the qualitative conclusion that the $\zeta$ mode is a more promising avenue to a protected gate.

\subsection{Diamond norm deviation}\label{app:dnorm}

To quantify the performance of the gate we use the diamond norm deviation. This is used as our primary metric instead of the average gate fidelity due to the fact that it is a worst-case estimate for the gate performance~\cite{Kueng2016}, and coincides with the metric used in threshold theorems~\cite{Kitaev1997,Aharonov1997,Knill1998}. In particular, for the case of coherent errors, the average gate infidelity potentially underestimates the gate error~\cite{Sanders2015}.
Here we derive the expression for the diamond norm deviation in terms of the effective qubit operators defined in the previous section, and compare with the average gate infidelity.

We assume that no logical bit-flips occur during the gate, which is a valid assumption so long as $E_J \gg E_C$, in the case of the ideal qubit, or $E_J \gg E_{C_\theta}$ in the case of the $0$-$\pi$ qubit.
In other words, the value of $\expval*{\bar{Z}_\mathrm{eff}}$ remains unchanged throughout the gate. This allows us to write an arbitrary pure state before and after the gate as 
\begin{equation}\label{eqn:initial-and-final-states}
    \ket{\psi_{\text{i}/\text{f}}} = a\ket{0}\ket{\psi_{0,\text{i}/\text{f}}} + b\ket{1}\ket{\psi_{1,\text{i}/\text{f}}}, 
\end{equation}
where $a$ and $b$ are two complex amplitudes satisfying $\abs{a}^2 + \abs{b}^2 = 1$, and subscripts ``$\text{i}$" and ``$\text{f}$" denote initial and final states, respectively.
Here, we have divided the Hilbert space in two, where $\ket{\mu} \in \mathcal{H}_L$ is the logical state, with $\mu = 0,1$, and $\ket{\psi_{\mu, \text{i}/\text{f}}} \in \mathcal{H}_{S} \otimes \mathcal{H}_\xi$ is the joint state on the stabiliser and ancillary modes, with $\xi = \phi,\zeta$ depending on the ancillary mode used to perform the gate. 
Note that we have absorbed the phase of $-i$ from the gate into the definition for the state $\ket{\psi_{1,\mathrm{f}}}$. 
The corresponding logical density matrices in the $\ket{\mu}$ basis are
\begin{equation} \label{eqn:initial-and-final-dms}
    \rhoh_{\text{i}/\text{f}} = \begin{pmatrix} \abs{a}^2 & a b^* \braket{\psi_{1,\text{i}/\text{f}}}{\psi_{0,\text{i}/\text{f}}} \\ a^* b \braket{\psi_{0,\text{i}/\text{f}}}{\psi_{1,\text{i}/\text{f}}} & \abs{b}^2 \end{pmatrix}. 
\end{equation} 

The map $\rhoh_\text{i} \to \rhoh_\text{f}$ corresponds to a CPTP quantum channel so long as the initial logical and stabiliser-ancilla states are not too entangled, which is satisfied in the case of a good decoder.
Specifically, $\rhoh_\text{i} \to \rhoh_\text{f}$ describes a miscalibration and dephasing noise channel~\cite{Kueng2016}
\begin{equation} \label{eqn:noisy-channel}
\begin{aligned}
    \mathcal{U}(\rhoh) = {}& p \hat{Z} e^{i(\pi/4 - \delta) \hat{Z}} \rhoh e^{-i(\pi/4 - \delta) \hat{Z}} \hat{Z}\\
    & + (1-p) e^{i(\pi/4 - \delta) \hat{Z}} \rhoh e^{-i(\pi/4 - \delta) \hat{Z}},
\end{aligned}
\end{equation}
where $p$ is the dephasing rate and $\delta$ is the unitary overrotation angle. These are given in terms of the overlap between the initial and final stabiliser-ancilla states by
\begin{subequations}
    \begin{align}
        p &= \frac{1}{2}\left(1 
        - \frac{
        \abs{\braket{\psi_{0,\text{f}}}{\psi_{1,\text{f}}}}}{\abs{\braket{\psi_{0,\text{i}}}{\psi_{1,\text{i}}}}}
        \right), \\
        \delta &= \frac{1}{2} 
        \left( 
        \arg\braket{\psi_{0,\text{f}}}{\psi_{1,\text{f}}} 
        - \arg \braket{\psi_{0,\text{i}}}{\psi_{1,\text{i}}}
        + \frac{\pi}{2}
        \right),
    \end{align}
\end{subequations}
where $|\braket{\cdot}|$ and $\arg \braket{\cdot}$ denote the magnitude and phase of the inner products, respectively.
Crucially, if the initial logical and stabiliser-ancilla states are not too entangled then $\abs{\braket{\psi_{0,\text{i}}}{\psi_{1,\text{i}}}} \approx 1$ and $\abs{\braket{\psi_{0,\text{i}}}{\psi_{1,\text{i}}}} \ge \abs{\braket{\psi_{0,\text{f}}}{\psi_{1,\text{f}}}}$, ensuring $p$ is positive and the map is CPTP.
In the parameter regimes we study, this is satisfied by all the decoders described in the previous section.

The diamond norm deviation of \cref{eqn:noisy-channel} from the ideal channel $\rhoh \to e^{i\pi\hat Z/4} \rhoh e^{-i\pi\hat Z/4}$ is~\cite{Kueng2016}
\begin{equation} \label{eqn:dnorm-Kueng}
    \eps_\diamond = \frac{1}{2} \abs{1 - (1-2p) e^{2i \delta}}.
\end{equation}
Using \cref{eqn:logical-dm,eqn:noisy-channel}, the dephasing rate $p$ and overrotation angle $\delta$ may be found in terms of the expectation values of the effective qubit operators:
\begin{subequations} \label{eqn:p-delta-eff-qubit-operators}
\begin{align}
    p &= \frac{1}{2} \left( 
    1 - 
    \sqrt{\frac{
    \expval*{\bar{X}_\mathrm{eff}}_\mathrm{f}^2 
    + \expval*{\bar{Y}_\mathrm{eff}}_\mathrm{f}^2}
    {\expval*{\bar{X}_\mathrm{eff}}_\mathrm{i}^2 
    + \expval*{\bar{Y}_\mathrm{eff}}_\mathrm{i}^2}}
    \right), \\
    \tan(2\delta) &= \frac{
    \expval*{\bar{X}_\mathrm{eff}}_\mathrm{i} 
    \expval*{\bar{X}_\mathrm{eff}}_\mathrm{f}
    + \expval*{\bar{Y}_\mathrm{eff}}_\mathrm{i} 
    \expval*{\bar{Y}_\mathrm{eff}}_\mathrm{f}}
    {\expval*{\bar{Y}_\mathrm{eff}}_\mathrm{i} 
    \expval*{\bar{X}_\mathrm{eff}}_\mathrm{f}
    - \expval*{\bar{X}_\mathrm{eff}}_\mathrm{i} 
    \expval*{\bar{Y}_\mathrm{eff}}_\mathrm{f}}.
\end{align}
\end{subequations}
Here, and in what follows we use the notation $\expval*{\cdot}_{\mathrm{i}/\mathrm{f}}$ to denote an expectation value taken with respect to the inital/final state $\rhoh_{\mathrm{i}/\mathrm{f}}$.
Substituting into \cref{eqn:dnorm-Kueng} leads to the following expression for the diamond norm deviation in terms of the effective qubit operators
\begin{equation} \label{eqn:dnorm-1}
    \eps_\diamond = \frac{1}{2} \sqrt{\frac{(\expval*{\bar{X}_\text{eff}}_\text{i} + \expval*{\bar{Y}_\text{eff}}_\text{f})^2 + (\expval*{\bar{Y}_\text{eff}}_\text{i} - \expval*{\bar{X}_\text{eff}}_\text{f})^2}{\expval*{\bar{X}_\text{eff}}_\text{i}^2 + \expval*{\bar{Y}_\text{eff}}_\text{i}^2}}.
\end{equation}
For an initial state with $\expval*{\bar{X}_\text{eff}}_\text{i} = 1$ and $\expval*{\bar{Y}_\text{eff}}_\text{i} = 0$ [equivalently, $a = b = 1/\sqrt{2}$ in \cref{eqn:initial-and-final-states}], which we may choose with a good enough decoder, the diamond norm deviation reduces to 
\begin{equation}
	\eps_\diamond = \frac{1}{2} \sqrt{(1+ \expval*{\bar{Y}_\text{eff}}_\text{f})^2 + \expval*{\bar{X}_\text{eff}}_\text{f}^2}. \label{eqn:dnorm-2}
\end{equation}
For clarity, we state this expression in the main text [\cref{eqn:dnorm-0}], but use the more accurate expression in \cref{eqn:dnorm-1} for the numerical results to account for the small deviations of $\expval*{\bar{X}_\text{eff}}_\text{i}$ from 1 and $\expval*{\bar{Y}_\text{eff}}_\text{i}$ from 0.

\subsubsection{Comparison to average gate infidelity}
To compare with the average gate infidelity, we compute the latter in terms of the effective qubit operator expectation values.
The average gate infidelity for the noisy channel in \cref{eqn:noisy-channel} is~\cite{Kueng2016}
\begin{equation}
    1 - \mathcal{F} = \frac{1}{3}\left[ 1 - (1 - 2p) \cos 2 \delta \right].
\end{equation}
Substituting the expressions for the dephasing rate and overrotation angle in \cref{eqn:p-delta-eff-qubit-operators} results in the following expression for the average gate infidelity
\begin{equation}
    1 - \mathcal{F} = \frac{1}{3} \left( 1 - \frac{ 
    \expval*{\bar{Y}_\mathrm{eff}}_\mathrm{i} \expval*{\bar{X}_\mathrm{eff}}_\mathrm{f}
    - \expval*{\bar{X}_\mathrm{eff}}_\mathrm{i} \expval*{\bar{Y}_\mathrm{eff}}_\mathrm{f}}
    {\expval*{\bar{X}_\mathrm{eff}}_\mathrm{i}^2
    + \expval*{\bar{Y}_\mathrm{eff}}_\mathrm{i}^2} \right).
\end{equation}
Upon setting $\expval*{\bar X _\mathrm{eff}}_\mathrm{i} = 1$ and $\expval*{\bar Y _\mathrm{eff}}_\mathrm{i} = 0$, this reduces to
\begin{equation}
    1 - \bar{\mathcal{F}} = \frac{1}{3}\left(1 + \expval*{\bar{Y}_\text{eff}}_\text{f}\right).
\end{equation}
Comparing with \cref{eqn:dnorm-2}, we see that the gate infidelity differs from the diamond norm deviation in that it does not account for nonzero $\expval{\bar{X}_\text{eff}}_\text{f}$, and it is smaller by a factor of $2/3$.

\subsubsection{Extension to arbitrary \texorpdfstring{$Z$}{Z}-rotations}

The expression for the diamond norm deviation may be easily generalised to the case where the ideal operation corresponds to an arbitrary rotation, $\vartheta$, about the $Z$-axis by replacing $\pi/4$ with $\vartheta/2$ in \cref{eqn:noisy-channel}. 
In particular, in the case of a $T$-gate, as considered in \cref{sec:non-clifford-gate,app:protected-T-gate}, we have $\vartheta = \pi/4$.
The resulting expression for the diamond norm deviation is then the same as \cref{eqn:dnorm-1} after the replacements, $\expval*{\bar{X}_\mathrm{eff}}_\mathrm{i} \to( \expval*{\bar{X}_\mathrm{eff}}_\mathrm{i} - \expval*{\bar{Y}_\mathrm{eff}}_\mathrm{i} )/\sqrt{2}$ and $\expval*{\bar{Y}_\mathrm{eff}}_\mathrm{i} \to( \expval*{\bar{X}_\mathrm{eff}}_\mathrm{i} + \expval*{\bar{Y}_\mathrm{eff}}_\mathrm{i} )/\sqrt{2}$, and reduces to
\begin{equation}
    \eps_\diamond = \frac{1}{2} \abs*{1 + e^{-i\pi/4} \expval*{\bar{X}_\text{eff}}_\text{f} + e^{i\pi/4} \expval*{\bar{Y}_\text{eff}}_\text{f}} \label{eqn:T-gate-dnorm}
\end{equation}
in the case that $\expval*{\bar{X}_\text{eff}}_\text{i} = 1$ and $\expval*{\bar{Y}_\text{eff}}_\text{i} = 0$.

\section{Sources of gate error}\label{app:more-gate-simulations}

In this appendix we analyse the impact of different noise sources on the protected gate.
In \cref{app:phase-latching} we consider over-rotation due to pulse wait-time variation, and give analytical justification for the gate's protection and the optimal impedance of the ancillary oscillator.
In \cref{app:ramp-time} we quantify the effects of pulse ramp-time variation. 
In \cref{app:charging-energy-ratio} we study the impact of the charging energy ratio on gate performance.
In \cref{app:flux-noise} we provide details on simulating the gate in the presence of external flux noise, and derive a dephasing rate for the qubit due to flux noise.

\subsection{Pulse wait-time variation}\label{app:phase-latching}

In this section we give analytical justification for the protection of the gate against pulse wait-time variation studied in \cref{sec:BKP-simulations}.
This analysis also explains the optimal ancilla impedance to maximise the robustness for a given gate error. 
Our approach follows a similar line of analysis to Refs.~\cite{Brooks2013,Preskill2007}.
We consider the case of an ideal qubit coupled to an external oscillator. 
To simplify the analysis, we neglect the errors that occur due to the finite ramp-time in the pulse and only consider the part of the gate when the pulse is on [points (ii) to (iii) in \cref{fig:BKP-gate}].  Furthermore, we consider only the finite width of the Gaussian envelope in the GKP states of the oscillator and disregard the nonzero width of each peak. To this end, the states in the oscillator may be approximated as a superposition of $\phi$-eigenstates at the centre of each cosine well, with amplitudes weighted by an envelope parameter $\kappa$:
\begin{equation}
    \ket{\tilde{\mu}} = \frac{1}{\mathcal{N}_\mu} \sum_{n \in \Z} e^{-\kappa \pi (n + \mu/2)^2} \ket{(n+\mu/2)2\pi}_\phi, \label{eqn:approx-gkp-states}
\end{equation}
where $\mathcal{N}_\mu$ are normalisation factors given by
\begin{equation}
    \mathcal{N}_\mu^2 = \sum_{n \in \Z} e^{-2\pi \kappa (n + \mu/2)^2},
\end{equation}
and $\mu = 0,1$ labels the qubit state.
In the limit that $\kappa \to \infty$ these become $\pm 1$ eigenstates of the logical operator $\bar{Z}_\phi$ and $+1$ eigenstates of the stabilisers $\hat{S}_{X_\phi}$ and $\hat{S}_{Z_\phi}$.
We now consider the evolution of these states under a noisy version of the ideal unitary,
\begin{equation}
    \hat{S}_\eta = e^{i(1 + \eta)\phih^2/2\pi},
\end{equation}
where $\eta$ corresponds to a dimensionless overrotation in the gate. Under this operation, the states evolve to
\begin{equation}
    \hat{S}_\eta \ket{\tilde{\mu}} = \frac{i^\mu}{\mathcal{N}_\mu}\sum_{n \in \Z} e^{-2\pi(\kappa/2 -i\eta)(n + \mu/2)^2} \ket{2\pi(n+\mu/2)}_\phi.
\end{equation}
Since the states $\hat{S}_\eta \ket*{\tilde{\mu}}$ are orthogonal, we cannot use their overlap to estimate the gate error. Instead, we directly estimate their relative phase,
\begin{equation}
\begin{aligned}
    \frac{\bra{\tilde{1}}\hat{S}_\eta \ket{\tilde{1}}}{\bra{\tilde{0}}\hat{S}_\eta \ket{\tilde{0}}} ={}& i \left(\frac{ \displaystyle \sum_{n \in \Z} e^{-2\pi \kappa n^2}}{\displaystyle \sum_{n \in \Z} e^{-2\pi \kappa (n+1/2)^2}}\right) \\
    {}&\times\left(\frac{\displaystyle \sum_{n \in \Z} e^{-2\pi(\kappa - i\eta)(n+1/2)^2}}{\displaystyle \sum_{n \in \Z} e^{-2\pi(\kappa - i\eta)n^2}} \right).
\end{aligned}
\end{equation}
Using the Poisson summation formula, the sums may be rewritten over their frequency components, resulting in 
\begin{equation}
\begin{aligned}
    \frac{\bra{\tilde{1}}\hat{S}_\eta \ket{\tilde{1}}}{\bra{\tilde{0}}\hat{S}_\eta \ket{\tilde{0}}} = {}&i \left(\frac{\displaystyle \sum_{k \in \Z} e^{-\frac{\pi}{2\kappa}k^2}}{\displaystyle \sum_{k \in \Z} (-1)^k e^{-\frac{\pi}{2\kappa}k^2}}\right)\\
    {}&\times \left(\frac{\displaystyle \sum_{k \in \Z} (-1)^k e^{-\frac{\pi}{2(\kappa - i\eta)}k^2}}{\displaystyle \sum_{k \in \Z} e^{-\frac{\pi}{2(\kappa - i\eta)}k^2}}\right).
\end{aligned}
\end{equation}
Assuming both the overrotation error $\eta$ and the envelope parameter $\kappa$ to be small facilitates the truncation of each sum to the leading order corrections $k = \pm 1$:
\begin{subequations}
    \begin{align}
        \frac{\bra{\tilde{1}}\hat{S}_\eta \ket{\tilde{1}}}{\bra{\tilde{0}}\hat{S}_\eta \ket{\tilde{0}}} &\approx i\left( \frac{1 + 2e^{-\frac{\pi}{2\kappa}}}{1 - 2e^{-\frac{\pi}{2\kappa}}} \right) \left(\frac{1 - 2e^{-\frac{\pi}{2(\kappa -i \eta)}}}{1 + 2e^{-\frac{\pi}{2(\kappa -i \eta)}}}\right) \\
        &\approx i(1 + 4e^{-\frac{\pi}{2\kappa}} - 4e^{-\frac{\pi}{2(\kappa - i \eta)}}),
    \end{align}
\end{subequations}
where we have expanded to first order in $e^{-1/\kappa}$ and $e^{-1/\eta}$ in the final line. 

\begin{figure}
    \centering
    \includegraphics[width=\linewidth]{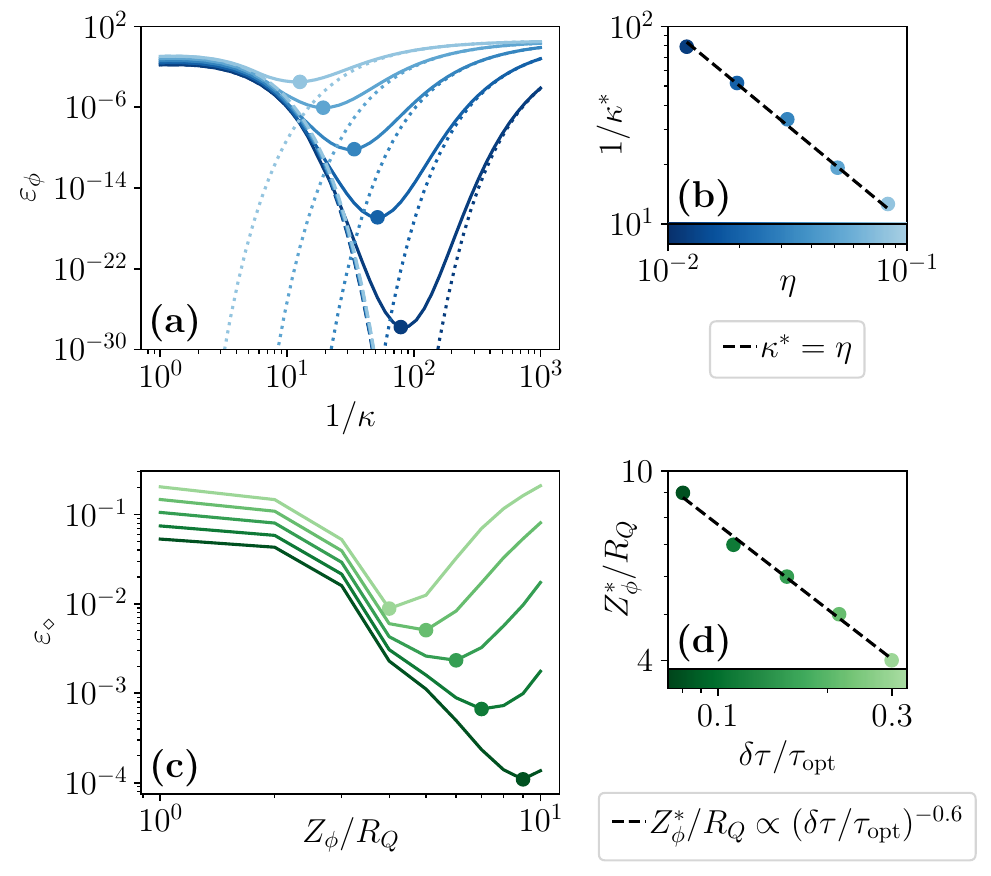}
    \subfloat{\label{fig:phase-error-vs-envelope}}
    \subfloat{\label{fig:optimal-envelope-vs-overrotation}}
    \subfloat{\label{fig:gate-error-vs-impedance}}
    \subfloat{\label{fig:optimal-impedance-vs-mistiming}}

    \caption{
        Optimal oscillator impedance. 
        (a) Relative phase error defined in \cref{eqn:phase-error} as a function of the envelope parameter $\kappa$ for different overrotation errors $\eta$. The circles denote the optimal envelope parameter $\kappa^*$ and the dotted (dashed) curves show the asymptotic scalings given in \cref{eqn:phase-latching-asymptotic-scalings} for $\kappa \ll \eta \ll 1$ ($\eta \ll \kappa \ll 1$). 
        (b) The optimal envelope parameter as a function of the overrotation error $\eta$. 
        The black dashed line denotes $\kappa^* = \eta$. 
        (c) Gate errors as a function of oscillator impedance for different mistiming errors $\delta \tau$. The circles denote the optimal oscillator impedance $Z_\phi^*$.
        (d) Optimal oscillator impedance as a function of the mistiming error.
        The black dashed line denotes a power-law fit to the data.
        }
    \label{fig:optimal-Z}
\end{figure}

The gate error may then be approximated by the deviation of this relative phase from its ideal value of $i$,
\begin{subequations}
    \begin{align}
        \eps_\phi &= \abs{i - \frac{\bra{\tilde{1}}\hat{S}_\eta \ket{\tilde{1}}}{\bra{\tilde{0}}\hat{S}_\eta \ket{\tilde{0}}}} \label{eqn:phase-error} \\
        &\approx \abs{ 4e^{-\frac{\pi}{2\kappa}} - 4 e^{-\frac{\pi}{2(\kappa -i\eta)}}} \\
        &\sim \begin{cases} 4e^{-\frac{\pi\kappa}{2\eta^2}} & \text{for } \kappa \ll \eta \ll 1 \\ \frac{2 \pi \eta}{\kappa^2} e^{-\frac{\pi}{2\kappa}} & \text{for } \eta \ll \kappa \ll 1 \end{cases}, \label{eqn:phase-latching-asymptotic-scalings}
    \end{align}
\end{subequations}
where $\sim$ in the final line denotes the asymptotic scaling of this error as a function of $\kappa$.
Since the first of these exponentials decays with increasing $\kappa$, whilst the second grows with increasing $\kappa$, these scalings imply an optimal envelope parameter of $\kappa = \eta$ for a given overrotation $\eta$. 
In \cref{fig:phase-error-vs-envelope,fig:optimal-envelope-vs-overrotation} we confirm these scalings and the existence of this optimal envelope parameter by plotting this error as a function of the envelope parameter $\kappa$ for different overrotations $\eta$. Here we find that the optimal envelope is given by $\kappa = \eta$.
Crucially, \cref{eqn:phase-latching-asymptotic-scalings} and \cref{fig:phase-error-vs-envelope} also show that the phase error is exponentially suppressed in $1/\kappa$ when $\eta \ll \kappa \ll 1$, owing to the protection of the gate.

To compare this approximate analysis with the full behaviour of the gate we plot the gate errors from \cref{fig:gate-errors} as a function of the oscillator impedance at different pulse mistimings in \cref{fig:gate-error-vs-impedance}. 
Similarly to \cref{fig:phase-error-vs-envelope} we observe an optimal impedance for a given pulse misting. 
This is because the envelope of the GKP states is inherited from the ground state of the oscillator, whose variance is controlled by the impedance; if the envelope parameter of the GKP states is exactly given by the variance of the ground state, then $\kappa = R_Q/Z_\phi$.
\Cref{fig:optimal-impedance-vs-mistiming} shows the optimal oscillator impedance as a function of the pulse mistiming. 
A fit to the data reveals that $Z^*_\phi/R_Q \propto (\delta\tau/\tau_\mathrm{opt})^{-0.6}$, which is similar to the scaling expected from the envelope parameter inherited from the ground state of the oscillator, $\kappa = R_Q/Z_\phi$.
The deviation from the scaling $Z^*_\phi/R_Q \propto (\delta\tau/\tau_\mathrm{opt})^{-1}$ is due to the finite ramp-time of the pulse and the nonzero width of the peaks in the GKP state wavefunctions neglected in the previous analysis.
The optimal impedance for a given mistiming error implies an optimal impedance for a given gate error threshold, explaining the maxima in the gate robustness as a function of the oscillator impedance observed in \cref{fig:robustness-Z}.

\subsection{Pulse ramp-time variation} 
\label{app:ramp-time}

In this section we examine the impact of varying the ramp-time $\tau_J$ on the performance of the gate. 
As discussed in the main text, the ramp-time is chosen to be approximately equal to an oscillation period $\tau_J \approx \hbar/\sqrt{8 E_{C_\phi} E_{L_\phi}}$.
This ensures that the evolution remains adiabatic with respect to the envelope inherited from the oscillator, but is fast enough to prevent the wavefunction from collapsing into a single well at the centre of the cosine potential.
However, just as the gate is insensitive to variations in the precise wait-time $\tau$, it is also expected to be insensitive to variations in this ramp-time.
We verify this numerically below.

\begin{figure}
    \centering
    \includegraphics[width=\linewidth]{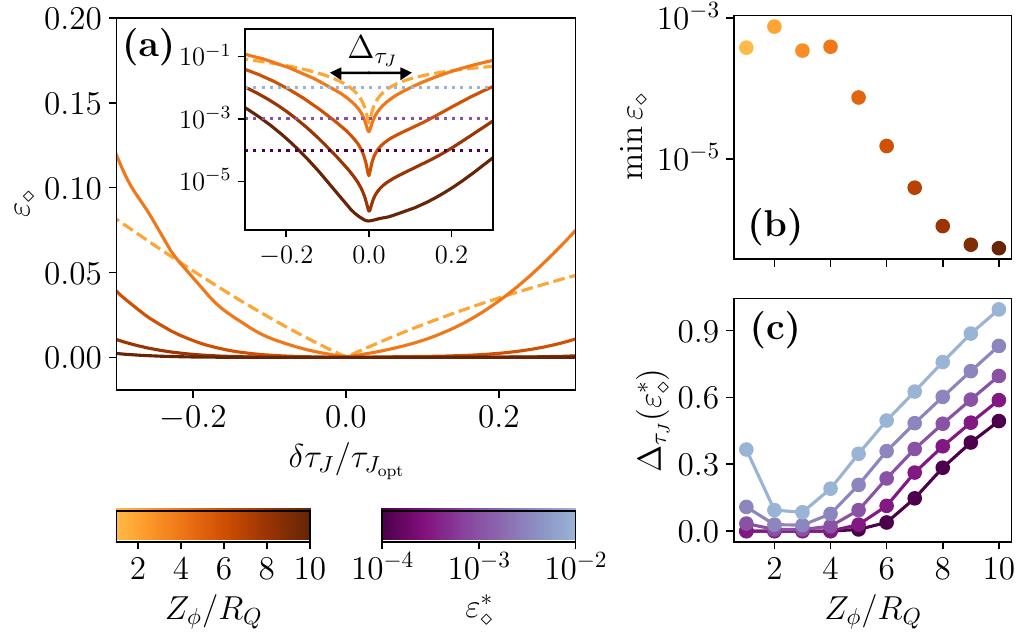}
    \subfloat{\label{fig:gate-error-curves-tau-J}}%
    \subfloat{\label{fig:imprecision-tau-J}}%
    \subfloat{\label{fig:robustness-tau-J}}%
    \caption{
        Gate errors as a function of ramp-time. 
        (a) Gate error $\varepsilon_\diamond$ as a function of deviations in the pulse ramp-time $\delta \tau_J$, where $\delta \tau_J$ is given by \cref{eqn:delta-tau-J}, calculated for different values of the oscillator impedance $Z_\phi$.
        Protected gates are marked in solid lines whereas the dashed line denotes an unprotected gate. 
        Inset: identical data on a log-linear plot. 
        Dotted lines correspond to threshold gate errors $\eps_\diamond^*$ for (c).
        (b) Gate imprecision as a function of $Z_\phi$. 
        (c) Gate robustness as a function of $Z_\phi$.
        For these simulations, $E_{C_\phi}/E_C = 100$, $E_J/E_{C_\phi} = 1$, $E_{J \minn} = 0$, $E_{J \maxx} / E_{C_\phi} = 100$ and $\tau = \tau_\mathrm{opt}$.
        }
    \label{fig:gate-errors-tau-J}
\end{figure}

\Cref{fig:gate-errors-tau-J} shows analogous results to \cref{fig:gate-errors,fig:imprecision-Z,fig:robustness-Z}, where we have defined the ramp-time variation parameter
\begin{equation} \label{eqn:delta-tau-J}
    \delta_{\tau_J} = \tau_J - \tau_{J_\text{opt}}.
\end{equation}
The error curves in \cref{fig:gate-error-curves-tau-J} are very similar in appearance to those in \cref{fig:gate-errors}.
Specifically, the gate error increase linearly with $\delta \tau_J$ for low oscillator impedances (dashed line), and are flat near $\delta \tau_J = 0$ for higher oscillator impedances (solid lines).
The imprecision shown in \cref{fig:imprecision-tau-J} is identical to that shown in \cref{fig:imprecision-Z} since the pulse parameters at the minimal gate error are the same in both cases.
The most notable difference is in the robustness, where we observe the gate to be more robust to relative imperfections in the ramp-time than the wait-time.
In particular, for the range of impedances considered here, the robustness parameter $\Delta_{\tau_J}$ does not reach its maximum value for the same thresholds $\eps_\diamond^*$ considered in \cref{fig:robustness-Z}, with values as large as $\Delta_{\tau_J}(0.01) = 0.995$ achieved.
This is in contrast to \cref{fig:robustness-Z} where the optimal impedances for the same error thresholds were $Z_\phi/R_Q \le 10$, with a maximum robustness of $\Delta_\tau(0.01) = 0.55$.

\subsection{Charging energy ratio} \label{app:charging-energy-ratio}

\begin{figure}
    \centering
    \includegraphics[scale=1]{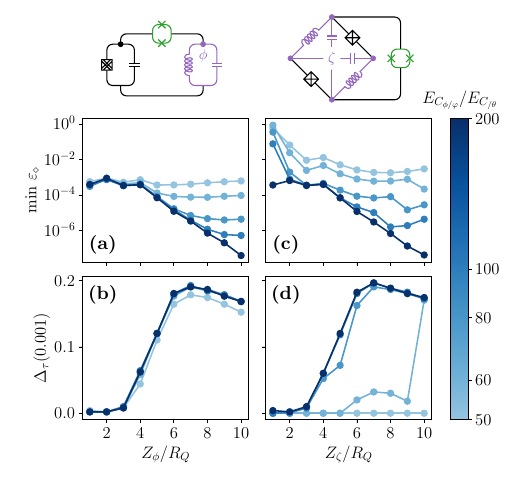}%
    \subfloat{\label{fig:theta-ancilla-imprecision-mass-ratio}}%
    \subfloat{\label{fig:theta-ancilla-robustness-mass-ratio}}%
    \subfloat{\label{fig:theta-zeta-imprecision-mass-ratio}}%
    \subfloat{\label{fig:theta-zeta-robustness-mass-ratio}}%
    \caption{
        Impact of charging energy ratio on the performance of protected phase gates.
        (a) Imprecision and (b) robustness of a gate for an ideal qubit coupled to an external oscillator as a function of the oscillator impedance $Z_\phi$, at different values of the charging energy ratio $E_{C_\phi}/E_C$.
        (c) Imprecision and (d) robustness of a gate for a $0$-$\pi$ qubit using its internal $\zeta$ mode as a function of the $\zeta$-mode impedance $Z_\zeta$, at different values of the charging energy ratio $E_{C_\varphi}/E_{C_\theta}$.
        Throughout all simulations, the minimal and maximal coupling strengths have been fixed to $E_{J \minn} = 0$ and $E_{J \maxx}/E_{C_{\phi/\varphi}} = 100$, and in panels (a) and (b) $E_J / E_{C_\phi} = 1$.
        The circuit diagrams corresponding to each version of the gate are shown above each column.
        }
    \label{fig:gate-errors-EL-zeta}
\end{figure}

Here we analyse the impact of the charging energy ratio on the performance of the gate. 
Specifically, in \cref{app:phi-mode-gate-theta-capacitance-ratio} we analyse the impact of varying $E_{C_\phi}/E_{C}$ on the performance of the gate for the ideal qubit coupled to an external oscillator and in \cref{app:varphi-theta-capacitance-ratio} we analyse the impact of varying $E_{C_\varphi}/E_{C_\theta}$ on the performance of the gate for the $0$-$\pi$ qubit using its internal $\zeta$ mode.

\subsubsection{Ideal qubit coupled to an external oscillator}
\label{app:phi-mode-gate-theta-capacitance-ratio}
\Cref{fig:theta-ancilla-imprecision-mass-ratio,fig:theta-ancilla-robustness-mass-ratio} show the imprecision and robustness for a gate with an ideal qubit coupled to an external oscillator as a function of the oscillator impedance at different values of the charging energy ratio $E_{C_\phi}/E_C$. 
The data at $E_{C_\phi}/E_C = 100$ correspond to the same data as in \cref{fig:imprecision-Z,fig:robustness-Z} at $\eps_\diamond^* = 10^{-3}$, as a point of comparison.
\Cref{fig:theta-ancilla-imprecision-mass-ratio} shows that the value of the oscillator impedance at which the imprecision plateaus increases with the charging energy ratio $E_{C_\phi}/E_C$.
In particular, for $E_{C_\phi}/E_C=200$ this critical value is larger than $Z_\phi/R_Q = 10$, the largest value considered here.
Therefore, $E_{C_\phi}/E_C$ sets the error floor for the gate assuming an arbitrarily large oscillator impedance can be obtained.
\Cref{fig:theta-ancilla-robustness-mass-ratio} shows that despite the decrease in precision with smaller $E_{C_\phi}/E_C$, the robustness remains largely unaffected for all $E_{C_\phi}/E_C \ge 50$.

\subsubsection{\texorpdfstring{$0$}{0}-\texorpdfstring{$\pi$}{π} qubit using its internal harmonic mode}
\label{app:varphi-theta-capacitance-ratio}

Here we investigate the impact of varying the largest charging energy ratio in the $0$-$\pi$ qubit, $E_{C_\varphi}/E_{C_\theta}$, on the performance of the gate when using its internal $\zeta$ mode. \Cref{fig:theta-zeta-imprecision-mass-ratio,fig:theta-zeta-robustness-mass-ratio} show the imprecision and robustness as a function of the impedance of the $\zeta$ mode for different charging energy ratios $E_{C_\varphi}/E_{C_\theta}$. 
Similarly to the case of the ideal qubit coupled to an oscillator, we find that larger charging energy ratios improve the performance of the gate. 
This is due to the fact that the nearest-neighbour tunnelling term in \cref{eqn:H-alpha} is exponentially suppressed relative to the next-nearest-neighbour tunnelling term in $E_{C_\varphi}/E_{C_\theta}$ [see \cref{fig:tunnelling-rates-mass-ratio}].

Comparing to \cref{fig:theta-ancilla-imprecision-mass-ratio,fig:theta-ancilla-robustness-mass-ratio} shows that nearly identical performance to the case of the ideal qubit coupled to an oscillator can be recovered at a charging energy ratio of $E_{C_\varphi}/E_{C_\theta} = 200$.
The results also reveal that a minimum value of $E_{C_\varphi}/E_{C_\theta} = 60$ is required for a protected gate.
However, at this charging energy ratio, a $\zeta$-mode impedance of $Z_\zeta/R_Q = 10$ is required to obtain appreciable robustness, which corresponds to a $\varphi$-mode impedance of $Z_\varphi/R_Q = 77$, where $Z_\varphi = \sqrt{L/4C_J}$. This exceeds the minimal $\varphi$-mode impedance of $Z_\varphi/R_Q = 50$ required for a protected gate at a charging energy ratio of $E_{C_\varphi}/E_{C_\theta} = 100$. Therefore, reducing the charging energy ratio from $E_{C_\varphi} / E_{C_\theta} = 100$ to $E_{C_\varphi} / E_{C_\theta} = 60$ does not lead to a smaller critical $Z_\varphi$ to obtain a protected gate.

\subsection{Flux noise} \label{app:flux-noise}
In this section we give more details on the impact of flux noise on the protected gate and the coherence time of the qubit.
In \cref{app:flux-noise-model} we explain how we model flux noise for the gate using the internal $\zeta$ mode, and in \cref{app:dephasing-rate} we explain the origin of the dephasing.

\subsubsection{Flux noise model} \label{app:flux-noise-model}
We start by deriving the Hamiltonian for the $0$-$\pi$ qubit shunted by a tunable Josephson element [\cref{fig:zero-pi-zeta-circuit-intro}] in the presence of noisy external fluxes threading each of the three inductive loops.
This may be achieved by following the same procedure as in \cref{app:zero-pi-zeta-circuit}, but with $\varphi^i_\mathrm{ext} \to \varphi^i_\mathrm{ext} + \delta \varphi^i_\mathrm{ext}$, with $i = \mathrm{q}, \mathrm{s}, \mathrm{qs}$.
In the irrotational gauge, the kinetic energy remains the same as in \cref{eqn:zero-pi-hamiltonian-symmetric} and the potential energy becomes
\begin{equation} \label{eqn:V-flux-noise}
\begin{aligned} 
    \V = &\;E_L\left( \varphih + \frac{\varphi_\mathrm{ext}^\mathrm{q} + \delta \varphi_\mathrm{ext}^\mathrm{q}}{2} \right)^2 \\
    &+ 
    E_L\left( \zetah + \frac{C^2+C_J^2}{6C^2 + 4 C_J^2} \delta \varphi_\mathrm{ext} \right)^2 \\
    &- 2 E_J\cos\left( \thetah + \frac{C^2}{6 C^2 + 4 C_J^2} \delta\varphi_\mathrm{ext}\right) \cos\varphih \\
    &- 2E_{J_\mathrm{s}} \cos\left( \frac{\varphi^\mathrm{s}_\mathrm{ext} + \delta \varphi^\mathrm{s}_\mathrm{ext}}{2}\right) \\
    &\times\cos\left( \zetah + \thetah - \frac{C^2 + C_J^2}{6 C^2 + 4 C_J^2} \delta \varphi_\mathrm{ext} \right),
\end{aligned}
\end{equation}
where $\delta \varphi_\mathrm{ext} = \delta \varphi_\mathrm{ext}^\mathrm{q} + \delta \varphi_\mathrm{ext}^\mathrm{s} + 2\delta \varphi_\mathrm{ext}^\mathrm{qs}$. 
Here we have treated the tunable Josephson element as a purely inductive element.
To leading order, introducing a nonzero capacitance $C_\mathrm{s}$ modifies the magnitude of the noisy inductive term by a term of order $C_\mathrm{s}/C$ and contributes a $\sin\thetah\cos\varphih$ and $\sin(\zetah + \thetah)$ potential, both of which are suppressed by a factor $|\delta \varphi_\mathrm{ext}|C_\mathrm{s}/C_\Sigma$. 
Since the term of order $C_\mathrm{s}/C$ reduces the magnitude of the inductive noise, and the latter two terms are negligibly small for $|\delta\varphi| \ll 1$, $C_\mathrm{s}/C  \ll 1$, we treat the tunable Josephson element as a purely inductive element in the following analysis.
We also fix $\varphi_\mathrm{ext}^\mathrm{q} = 0$ for the remainder of this section so that $\delta \varphi_\mathrm{ext}^\mathrm{q}$ represents stray magnetic field fluctuations threading the loop of the $0$-$\pi$ qubit.

The first two terms in \cref{eqn:V-flux-noise} represent linear flux noise in the two unbounded variables $\varphi$ and $\zeta$ of the $0$-$\pi$ qubit.
The qubit is expected to be robust to the former, and the latter has no effect in the limit of no coupling to the $\zeta$ mode.
Note that inductive flux noise in $\varphi$ leads to an ``effective charge noise" since flux gets mapped to charge in the effective model.

The effect of the third term in \cref{eqn:V-flux-noise} is most easily seen by moving away from the irrotational gauge via making a time-dependent unitary transformation given by the unitary operator $e^{i C^2 \delta\varphi\ext \n_\theta/(6C^2 + 4C_J^2)}$.
This returns the third term to the ideal form $\cos\thetah\cos\varphih$, but introduces charge noise on the $\theta$ mode with a magnitude of $C^2 \dot{\delta\varphi}\ext/16E_{C_\theta}(3C^2 + 2C_J^2)$, where $\dot{\delta\varphi}\ext$ represents the time-derivative of the external magnetic flux noise (a time-dependent magnetic flux leads to a fluctuating EMF via the Faraday effect). 
Since charge noise in the $\theta$ mode is exponentially suppressed in $E_J/E_{C_\theta}$, this is expected to have negligible effect on the coherence time of the $0$-$\pi$ qubit.
Making this frame transformation also increases the magnitude of the shift in the final term in \cref{eqn:V-flux-noise} by an amount $C^2/(6C^2 + 4C_J^2)$. 

The final term in \cref{eqn:V-flux-noise} represents a nonlinear coupling between the qubit and ancillary oscillator degree of freedom.
Whilst the $0$-$\pi$ qubit is designed to be protected against noise operators that are polynomials in its charge or flux degrees of freedom, this nonlinear interaction is not of this form.
Nevertheless, for the same reason that the interaction strength is exponentially suppressed in the impedance of the $\zeta$ mode, to first order the effect of this term on the coherence time of the $0$-$\pi$ qubit is also exponentially suppressed.
However, to higher order in perturbation theory, this term leads to dephasing of the qubit as explained in \cref{app:dephasing-rate}.

To model the gate with flux noise we map the changes discussed above to the effective model. 
This is achieved by simulating the Hamiltonian $\H(t) = \H_\zeta + \H_\alpha + \H \intt(t)$, where the Hamiltonian terms are given by \cref{eqn:zero-pi-zeta,eqn:H-alpha,eqn:zero-pi-zeta-interaction-alpha} with the following modifications:
\begin{itemize}
    \item $\zetah \to \zetah + (C^2 + C_J^2)\delta\varphi_\mathrm{ext}/(6 C^2 + 4 C_J^2)$ in \cref{eqn:zero-pi-zeta},
    \item $\n_\alpha \to \n_\alpha + \delta \varphi_\mathrm{ext}^\mathrm{q}/2\pi$ in \cref{eqn:H-alpha},
    \item $\zetah + \pi \n_\alpha \to \zetah + \pi \n_\alpha - (2C^2 + C_J^2)\delta\varphi_\mathrm{ext}/(6C^2 + 4 C_J^2)$ in \cref{eqn:zero-pi-zeta-interaction-alpha},
    \item $E_{J_\mathrm{int}}(t) \to E_{J_\mathrm{int}}(t) - E_{J\maxx}\delta\varphi_\mathrm{ext}^\mathrm{s}$ in \cref{eqn:zero-pi-zeta-interaction-alpha}.
\end{itemize}
Note that we have used a small-angle approximation in the final substitution, which is accurate for realistic flux noise amplitudes.

Flux noise may be characterised by its spectral density $S(\omega)$, which is defined to be
\begin{equation}
    S(\omega) = \int_{-\infty}^\infty d\tau \, \expval{\delta \varphi\ext(t)\delta\varphi\ext(t+\tau)} e^{-i\omega\tau},
\end{equation}
where $\expval{\delta \varphi\ext(t)\delta\varphi\ext(t+\tau)}$ is the two-correlation function for $\delta \varphi\ext$.
In superconducting circuits, flux noise has been observed to have the following spectral density~\cite{Didier2019,Hutchings2017}
\begin{equation}
    S(\omega) = \frac{2\pi}{\abs{\omega}} A^2_{1/f} + A^2_\mathrm{w},
\end{equation}
which represents a combination of $1/f$ noise and white noise with amplitudes $A_{1/f}$ and $A_\mathrm{w}$, respectively.
For our simulations of the gate with flux noise we generate noise vectors with this spectral density using the algorithm in Ref.~\cite{Timmer1995} and the approach outlined in Ref.~\cite{Kolesnikow2024}.

\subsubsection{Dephasing rate} \label{app:dephasing-rate}
Here we provide a derivation for the dephasing rate of an ideal protected qubit, in the presence of a weak coupling to an oscillator through a cosine nonlinearity. 
For simplicity, we take the limit of $E_J/E_C \to \infty$ in \cref{eqn:BKP-qubit} so that the qubit may be treated by a classical binary variable $\mu$.
The Hamiltonian for the qubit-oscillator system is then given by $\H(t) = \H_\phi + \H\intt(t)$, where $\H_\phi$ is the Hamiltonian for an oscillator given in \cref{eqn:BKP-oscillator}, and the interaction Hamiltonian is
\begin{equation}
    \H\intt(t) = - E_J(t) \cos(\pi\mu - \phih).
\end{equation}
Let $\{\ket{\mu,m}\}_{m=0}^\infty$ be the pairwise degenerate eigenstates of $\H_\phi$ with energies $m\hbar\omega_\phi$, where $\hbar\omega_\phi = \sqrt{8 E_{C_\phi}E_{L_\phi}}$.

To first order, the nonlinear interaction splits the degeneracy of the ground states by
\begin{subequations}
\begin{align}
    \Delta E(t) &= \bra{1,0}\H\intt(t)\ket{1,0} - \bra{0,0}\H\intt(t)\ket{0,0} \\
    &= 2E_J(t) e^{-\pi Z_\phi / 2 R_Q},
\end{align}
\end{subequations}
Therefore, to first order, the ground states become degenerate in the limit of large oscillator impedance. 

Now we compute the time-evolved ground states using second-order time-dependent perturbation theory.
The first- and second-order corrections to the time-evolved ground state $\ket{\psi_\mu(t)} = \hat{\mathcal{T}}e^{-i \int_0^t dt' \,\H(t') } \ket{\mu, 0}$ are
\begin{widetext}
    \begin{subequations} \label{eqn:time-dependent-state-corrections}
    \begin{align}
        \ket*{\psi_\mu^{(1)}(t)} &= i (-1)^\mu  \sum_{m=0}^\infty \int_0^t dt' \, E_J(t')e^{im\omega_\phi t'} \bra{m}\cos\phi\ket{0} \ket{\mu,m}, \\ 
        \ket*{\psi_\mu^{(2)}(t)} &=- \sum_{m,m'=0}^\infty \int_0^t dt' \int_0^{t'}dt'' \, e^{im\omega_\phi t'} e^{im'\omega_\phi(t'' - t')} E_J(t') E_J(t'') \bra{m}\cos\phi\ket{m'} \bra{m'} \cos\phi \ket{0} \ket{\mu,m}.
    \end{align}
    \end{subequations}
\end{widetext}
Now consider the initial unentangled state $\ket{+,0} = (\ket{0,0} + \ket{1,0})/\sqrt{2}$, which is a $+1$ eigenstate of the effective Pauli-X operator $\bar{X}_\mathrm{eff} = (\ketbra{0}{1} + \ketbra{1}{0}) \otimes I$.
The nonlinear perturbation leads to the qubit and oscillator becoming entangled and a reduction in $\expval{\bar{X}_\mathrm{eff}}$. 
To see this, we compute the expectation value of the state $\hat{\mathcal{T}}e^{-i \int_0^t dt' \,\H(t') } \ket{+,0}$ with the effective qubit operators using \cref{eqn:time-dependent-state-corrections}, keeping terms up to second order in $E_J(t)$.
After taking the ensemble average, these are
\begin{subequations}
\begin{align}
    \langle\langle\bar{X}_\mathrm{eff}\rangle\rangle &\approx 1 - 2\sum_{m=0}^\infty  I^X_m \abs*{\bra{m}\cos\phih\ket{0}}^2, \\
    \langle\langle\bar{Y}_\mathrm{eff}\rangle\rangle &\approx -2 \bar{E}_J t \bra{0}\cos\phih\ket{0} + \sum_{m=0}^\infty I^Y_m \abs*{\bra{m}\cos\phih\ket{0}}^2, \\
    \langle\langle\bar{Z}_\mathrm{eff}\rangle\rangle &\approx 0, 
\end{align}
\end{subequations}
where 
\begin{subequations}
\begin{align}
    I^X_m &= \int_0^t dt' \int_0^t dt'' \, \expval{E_J(t')E_J(t'')} \cos[m\omega_\phi(t'-t'')], \\
    I^Y_m &= \int_0^t dt' \int_0^t dt'' \, \expval{E_J(t')E_J(t'')} \sin[m\omega_\phi(t'-t'')].
\end{align}
\end{subequations}
In the case of white noise with mean $0$ and standard deviation $\sigma$, we have that $\bar{E}_J = 0$ and $\expval{E_J(t') E_J(t'')} = \sigma^2 \delta(t'-t'')$.
Thus, the effective-qubit operator expectation values become
\begin{subequations}
\begin{align}
    \langle\langle\bar{X}_\mathrm{eff}\rangle\rangle &= 1 - 2\sigma^2t \sum_{m=0}^\infty \abs*{\bra{m}\cos\phih\ket{0}}^2, \\
    \langle\langle\bar{Y}_\mathrm{eff}\rangle\rangle &\approx 0, \\
    \langle\langle\bar{Z}_\mathrm{eff}\rangle\rangle &\approx 0,
\end{align}
\end{subequations}
which shows that the qubit dephases at a rate given by
\begin{subequations}
\begin{align}
    \Gamma_\varphi &= 2\sigma^2 \sum_{m=0}^\infty \abs*{\bra{m}\cos\phih\ket{0}}^2 \\
    &= 2 \sigma^2 e^{-\pi Z_\phi/R_Q} \sum_{m=0}^\infty \frac{(\pi Z_\phi/R_Q)^{2k}}{(2k)!} \label{eqn:dephasing-rate-mat-elts}\\
    &\sim \sigma^2 \label{eqn:dephasing-rate-asymptotics}
\end{align}
\end{subequations}
Here we used the following matrix element expressions for $\cos\phih$ in \cref{eqn:dephasing-rate-mat-elts}
\begin{subequations} \label{eqn:cos-phi-mat-elts}
\begin{align} 
    \bra{2k}\cos\phi\ket{0} &= \frac{-e^{-\pi Z_\phi/2R_Q} (\pi Z_\phi/R_Q)^{k}}{\sqrt{(2k)!}}, \\
    \bra{2k+1}\cos\phi\ket{0} &= 0,
\end{align}
\end{subequations}
and used the following asymptotic equivalence in \cref{eqn:dephasing-rate-asymptotics}
\begin{equation} \label{eqn:asymptotic-equivalence}
    e^{-x} \sum_{k=1}^\infty \frac{x^{2k}}{(2k)^\alpha (2k)!} \sim \frac{1}{2x^\alpha}, \qquad \alpha\ge0,  x \to \infty.
\end{equation}
Thus, in the large impedance limit the qubit dephases with a rate given by the variance of the noise. 
The contribution of any given harmonic oscillator level to the dephasing is exponentially suppressed in the limit of high impedance. 
To see the correct dependence of the dephasing rate, it is important to sum over the infinite number of levels in the harmonic oscillator. 

\section{Additional gates}\label{app:gate-extensions}

In this section we describe two extensions to the protected phase gates analysed in the main text.
In \cref{app:phi-mode-gate} we investigate the possibility of using the internal $\varphi$ mode of the $0$-$\pi$ qubit to perform a protected gate.
In \cref{app:protected-T-gate} we propose a protected $T$-gate.

\subsection{Phase gate via the \texorpdfstring{$\vect\varphi$}{ϕ} mode}\label{app:phi-mode-gate}

\begin{figure}
    \centering%
    \includegraphics[scale=1]{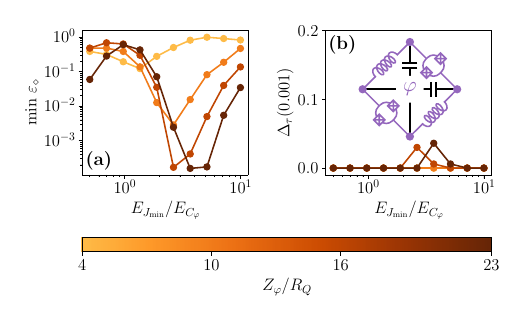}%
    \subfloat{\label{fig:min-dnorm-vs-EL-phi}}%
    \subfloat{\label{fig:max-r-vs-EL-phi}}%
    \caption{
        Gate errors for a phase gate with the $0$-$\pi$ qubit using its internal $\varphi$ mode. 
        (a) Imprecision and 
        (b) robustness of the gate as a function of the impedance of the $\varphi$ mode $Z_\varphi$, for different minimal coupling strengths $E_{J \minn}/E_{C_\varphi}$. 
        The charging energy ratio is fixed to $E_{C_\varphi}/E_{C_\theta} = 100$, and the maximal coupling strength to $E_{J \maxx}/E_{C_\varphi} = 100$. The inset of panel (b) shows the circuit diagram for the gate.
        }     
    \label{fig:gate-errors-varphi}
\end{figure}

Here we investigate the possibility of using the internal $\varphi$ mode of the $0$-$\pi$ qubit to perform a protected gate.
Unlike the $\zeta$ mode, the $\varphi$ mode is already coupled to the $\theta$ mode. 
Therefore, in order to utilise the $\varphi$ mode in place of the ancillary oscillator, this coupling must be made tunable. 
This can be achieved by replacing each of the static Josephson junctions in the $0$-$\pi$ qubit with a tunable Josephson element, and threading each with equal external fluxes. 
The inset of \cref{fig:max-r-vs-EL-phi} shows the circuit diagram for the gate and in \cref{app:zero-pi-phi-circuit} we provide the circuit quantisation.

To simulate the gate using this internal mode, we use \cref{eqn:zero-pi-theta-phi} with $2E_J \to E_{J \intt}(t)$. This means the interaction term now becomes
\begin{equation} \label{eqn:zero-pi-phi-interaction}
    \H \intt(t) = -E_{J \intt}(t) \cos\thetah \cos\varphih,
\end{equation}
which differs from the interaction term for the ancillary oscillator [\cref{eqn:BKP-interaction}] by $\phih \to \varphih$ and an additional $-E_{J \intt}(t) \sin\thetah \sin \varphih$ term. 
As explained in \cref{sec:circuit-disorder}, this term does not change the location of the potential energy minima in $\theta$-$\varphi$ space, resulting in a similar potential.
However, the interaction strength now also affects the qubit, whereby a small value of $E_{J \intt}(t)$ leads to increased tunnelling between the two qubit states. 
On the other hand, the results of \cref{sec:BKP-dynamic-range} showed that a small initial coupling strength is necessary for protection of the gate. 
In the following we investigate the performance of the gate as a function of the minimal coupling strength $E_{J \minn}$.

\Cref{fig:gate-errors-varphi} shows the performance of the gate as a function of $E_{J \minn}$. 
In \cref{fig:min-dnorm-vs-EL-phi} we find optimal values of $E_{J \minn}$, which minimise the imprecision. These optimal values increase with increasing impedance, but the minimal imprecision begins to saturate for $Z_\varphi/R_Q \ge 16$. \Cref{fig:max-r-vs-EL-phi} shows that the gate remains nearly unprotected, with a maximal robustness of $\Delta_\tau(0.001)<4$\% observed over all impedances and $E_{J \minn}$ values considered here. 
The lack of protection is because the optimal values of $E_{J \minn}/E_{C_\varphi}$ observed in \cref{fig:min-dnorm-vs-EL-phi} are greater than one, which exceeds the maximum value of $E_{J \minn}/E_{C_\phi}$ required for a protected gate found in \cref{sec:BKP-dynamic-range}; there we found that $E_{J \minn}/E_{C_\phi} < 0.08$ for a protected gate.

One way to decrease the tunnelling between the two qubit states whilst maintaining a sufficiently low $E_{J \minn}/E_{C_\varphi}$ for a protected gate is to increase the ratio of $E_{C_\varphi}/E_{C_\theta}$. However, given that current experiments~\cite{Gyenis2021,Kim2024,Hassani2024} have yet to reach the value of $E_{C_\varphi}/E_{C_\theta}$ considered here ($E_{C_\varphi}/E_{C_\theta}=100$), obtaining a protected gate in this way is likely to be challenging.

Whilst we have considered using the $\varphi$ mode analogously to the oscillator or $\zeta$ mode, an alternative way of utilising the $\varphi$ mode to perform a gate is to tune a single Josephson junction in the $0$-$\pi$ qubit, as proposed in Ref.~\cite{Hassani2024}. This tunes the barrier height in the $\theta + \varphi$ direction, allowing tunnelling to occur between the two qubit states and implementing a rotation about the $X$ axis. The benefit of this approach is that it eliminates the need to tune two Josephson elements symmetrically, but the gate is manifestly unprotected due to removing the tunnel barrier entirely. A comprehensive simulation of the gate in Ref.~\cite{Hassani2024} and comparison to the unprotected gate we describe in this section would be an interesting direction for future work.

\subsection{Protected \texorpdfstring{$\boldsymbol T$}{T}-gate} \label{app:protected-T-gate}

Here we investigate an extension of the protected $S$-gate to a protected $T$-gate. 
The basic idea is that the quadratic potential in \cref{eqn:BKP-oscillator} is replaced with a different potential such that a $T$-gate is performed on the GKP codewords. 

In Ref.~\cite{Gottesman2001}, the unitary gate 
\begin{equation} \label{eqn:GKP-T-gate-unitary}
    \U = \exp\left[-i 2\pi \left(\frac{ \x^3}{4} + \frac{\x^2}{8} - \frac{\x}{4}\right)\right],
\end{equation}
where $\x = \phih/\pi$ is suggested as a way to perform a logical $T$-gate on GKP codewords. 
Therefore, by replacing the quadratic potential in \cref{eqn:BKP-oscillator} with the cubic potential 
\begin{equation} \label{eqn:GKP-T-gate-potential}
    V(x) = V_0 \left(\frac{x^3}{4} + \frac{x^2}{8} - \frac{x}{4}\right)
\end{equation}
and turning on the interaction for a time $\tau \approx 2\pi\hbar / V_0$, a $T$-gate will be enacted on the qubit states. 
In our context, odd-ordered potentials are problematic since they are nonconfining, meaning that there will be no well-defined ground state of the ancillary mode. 
However, the polynomial in \cref{eqn:GKP-T-gate-potential} is not unique for enacting a logical $T$-gate. 
In particular, recent work~\cite{Nguyen2025} has shown that the following quartic polynomial potential also realises a $T$-gate on the GKP code
\begin{equation} \label{eqn:T-gate-quartic-potential}
    V(x) = V_0 \left( \frac{x^4}{24} - \frac{x^2}{6} \right).
\end{equation}
This can be seen to enact a phase of $e^{i\mu\pi/4}$ on the GKP codestates $\ket{\bar{\mu}}$, with $\mu=0,1$ by letting $x = 2n + \mu$, with $n$ an integer, and observing that $4x^2 - x^4 = 3\mu \bmod{24}$. 
Thus, $\exp[i 2\pi \hat V(\x)/V_0] \ket{\bar{\mu}} = e^{i\mu\pi/4}\ket{\bar{\mu}}$. 

Higher-order potentials such as the quartic potential in \cref{eqn:T-gate-quartic-potential} may be realised by flux-pumping a SNAIL oscillator~\cite{Sivak2019,Hillmann2020,Eriksson2024}. 
Therefore, in order to engineer this protected $T$-gate, the linear oscillator in \cref{fig:BKP-circuit} would be replaced with a SNAIL oscillator.
We do not give a detailed analysis of the experimental requirements here, but note that engineering the precise coefficients in \cref{eqn:T-gate-quartic-potential} would be an additional challenge.

\begin{figure}
    \centering%
    \includegraphics[width=\linewidth]{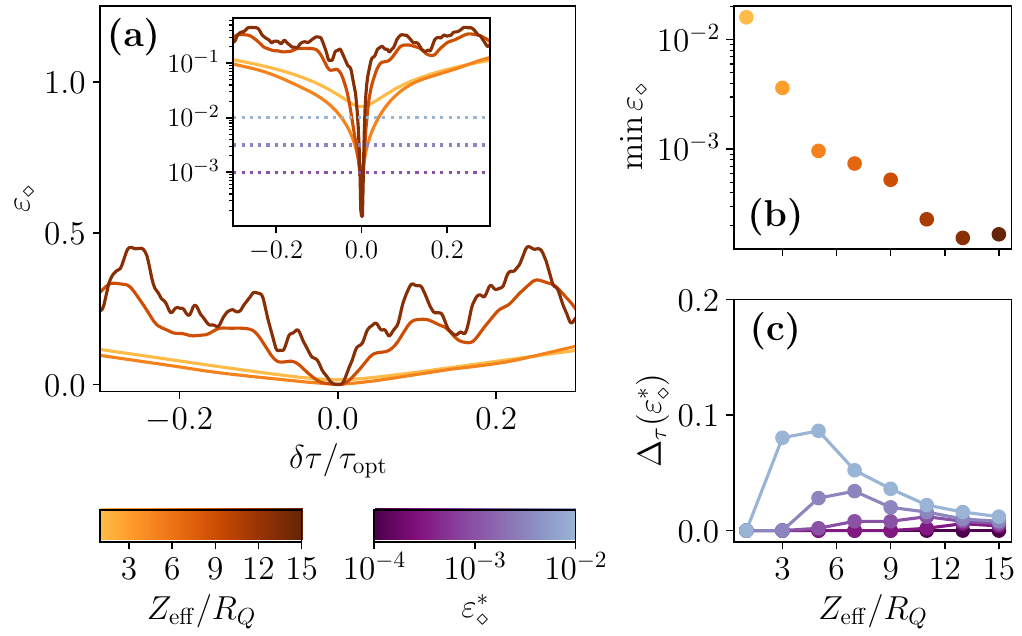}%
    \subfloat{\label{fig:gate-errors-T-gate}}%
    \subfloat{\label{fig:imprecision-T-gate}}%
    \subfloat{\label{fig:robustness-T-gate}}%
    \caption{
    Gate errors for a $T$-gate using a quartic potential. 
    (a) Gate error $\varepsilon_\diamond$ as a function of deviations in the pulse wait-time $\delta \tau$, calculated for different values of the effective impedance $Z_\mathrm{eff}$, where $Z_\mathrm{eff}$ is defined in \cref{eqn:effective-impedance}.
    Inset: identical data on a log-linear plot. 
    Dotted lines correspond to threshold gate errors $\eps_\diamond^*$ for (c).
    (b) Gate imprecision as a function of $Z_\mathrm{eff}$. 
    (c) Gate robustness as a function of $Z_\mathrm{eff}$.
    For these simulations, $E_{C_\phi}/E_C = 100$, $E_J/E_{C_\phi} = 1$, $E_{J \minn} = 0$ and $E_{J \maxx} / E_{C_\phi} = 100$
    }
    \label{fig:T-gate-numerics}
\end{figure}

To simulate this gate for an ideal qubit coupled to the SNAIL oscillator, we use the Hamiltonian $\H(t) = \H_\theta + \H_\phi + \H \intt(t)$, where the first and last terms are given by \cref{eqn:BKP-qubit,eqn:BKP-interaction} respectively, and the second term is given by
\begin{equation} \label{eqn:T-gate-oscillator}
    \H_\phi = 4 E_{C_\phi} \n_\phi^2 + V_0 \left( \frac{\phih^4}{24 \pi^4} -\frac{\phih^2}{6\pi^2} \right).
\end{equation}
We define an effective impedance $Z_\mathrm{eff}$ as function of $V_0$ as follows
\begin{equation} \label{eqn:effective-impedance}
    \frac{Z_\mathrm{eff}}{R_Q} =\left( \frac{24 E_{C_\phi}}{V_0} \right)^{1/4},
\end{equation}
which ensures that $V(\phi) = E_{C_\phi}$ at approximately the same value of $\phi$ as for the inductive potential with an inductance $L = Z_\mathrm{eff}^2 C_\phi$. 
This means that the ground-state wavefunction of \cref{eqn:T-gate-oscillator} will have a similar expectation value with the GKP stabiliser $\hat S_{X_\phi}$ as the ground-state of a linear oscillator with an impedance of $Z_\mathrm{eff}$. The definition of the gate error must also be modified due to the different logical gate. 
In \cref{app:dnorm}, we give the modification.

\Cref{fig:T-gate-numerics} shows the gate errors for a range of different effective impedances. 
As in \cref{sec:BKP-impedance}, we use a charging energy ratio of $E_{C_\phi}/E_{C_\theta} = 100$, a protected qubit Josephson energy of $E_J/E_{C_\phi}=1$, and minimum and maximum Josephson couplings of $E_{J \minn} = 0$ and $E_{J \maxx}/E_{C_\phi} = 100$. 
The results show that the imprecision saturates just above $10^{-4}$, and that the robustness does not exceed $10$\% over all effective impedances and threshold gate errors. 
These values are roughly 2 orders of magnitude and a factor of 8 worse than for the analogous $S$-gate results in \cref{fig:gate-errors,fig:imprecision-Z,fig:robustness-Z}, respectively.
The worse performance of the $T$-gate may be attributed to the fact that unitary operations constructed from quadratic potentials (such as the inductive potential used for the $S$-gate) are fault-tolerant for the GKP code, whereas higher-order potentials (such as the quartic potential used for the $T$-gate) are not~\cite{Gottesman2001}. 
Nonetheless, this provides a way to obtain an unprotected non-Clifford gate that does not explicitly break the protection of the qubit, since the quartic potential in \cref{eqn:T-gate-quartic-potential} commutes with the stabilisers for the qubit.

\section{Additional analysis of the \texorpdfstring{$\vect0$}{0}-\texorpdfstring{$\vect\pi$}{π} qubit}

In this appendix we give supplementary information for the $0$-$\pi$ qubit.
In \cref{app:optimal-EJ} we analyse the optimal Josephson energy for the $0$-$\pi$ qubit, and in \cref{app:effective-model} we give details on the effective model for the $0$-$\pi$ qubit.

\subsection{Optimal Josephson energy}\label{app:optimal-EJ}

Here we analyse the optimal Josephson energy for a $0$-$\pi$ qubit. 
This was also examined in Ref.~\cite{Dempster2014} but here we perform our own analysis to be consistent with the parameters and notation used in this work.
For fixed $C$, $C_J$ and $L$ (equivalently fixed $E_{C_\varphi}/E_{C_\theta}$ and  $Z_\zeta/R_Q$) in the $0$-$\pi$ qubit, there is an optimal Josephson energy $E_J$ that maximises the protection of the $0$-$\pi$ qubit.
This is for the following reason.
On the one hand, a larger $E_J$ raises the tunnel barrier in the $\theta$ direction, suppressing the tunnelling between the two qubit states localised at $\theta = 0$ and $\theta = \pi$, resulting in a longer lifetime.
On the other hand, increasing $E_J$ simultaneously raises the tunnel barrier in the $\varphi$ direction, leading to more localised qubit state wavefunctions that are more susceptible to flux-noise induced dephasing. 
Thus, there is an optimal $E_J$, which balances out these two competing effects. 

To quantify the protection of the qubit, we compute its degeneracy, defined as~\cite{Dempster2014}
\begin{equation}
    D = \log_{10}\frac{E_2 - E_0}{E_1 - E_0},
\end{equation}
where $E_i$ denotes the $i$-th energy of the $0$-$\pi$ qubit. 
Note that at the flux sweetspot $\varphi_\text{ext} = \pi$, the degeneracy of the qubit is maximised in the $E_J \to \infty$ limit since the potential energy minima at $\theta =0$ and $\theta = \pi$ take the same value. 
However, away from this exact value, the coherence time suffers from very large $E_J$ due to highly localised wavefunctions. 
Therefore, we fix $\varphi_\text{ext} = 0$ when finding the optimal $E_J$ in order to maximise the qubit's protection to dephasing. 

\begin{figure}
    \centering%
    \includegraphics[width=\linewidth]{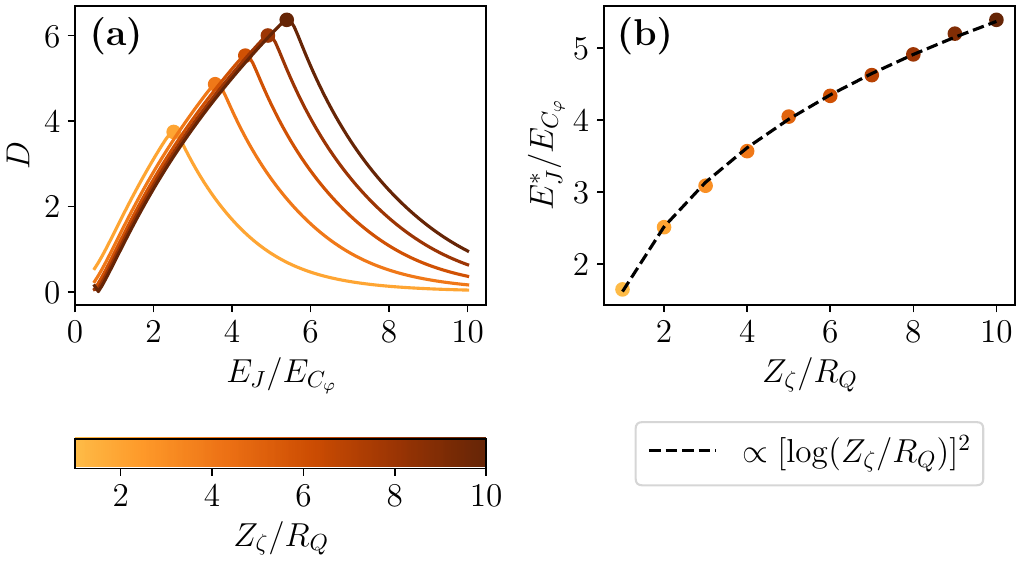}%
    \subfloat{\label{fig:D-vs-EJ-large-mass-ratio}}%
    \subfloat{\label{fig:EJ-opt-vs-Z-large-mass-ratio}}%
    \caption{
        Optimal Josephson energy for the $0$-$\pi$ qubit. 
        (a) Degeneracy of the $0$-$\pi$ qubit as a function of $E_J$ and $Z_\zeta$ for a fixed charging energy ratio of $E_{C_\varphi}/E_{C_\theta} = 100$ and $\varphi_\text{ext} = 0$.
        (b) The value of $E_J$ that maximises the degeneracy as a function of $Z_\zeta$. 
        The black dashed line shows a fit to the data via \cref{eqn:EJ-opt-scaling}. 
        }
    \label{fig:optimal-EJ}
\end{figure}

\Cref{fig:D-vs-EJ-large-mass-ratio} shows the degeneracy of the qubit as a function of $E_J$ and $Z_\zeta/R_Q$ with a fixed charging energy ratio of $E_{C_\varphi}/E_{C_\theta} = 100$. 
Here we see that there are clear optimal values of $E_J$ that maximise the degeneracy of the qubit.
In \cref{fig:EJ-opt-vs-Z-large-mass-ratio}, we plot these optimal values $E_J^*$ as a function of $Z_\zeta$. We find that the square root of these optimal values scales logarithmically with the $\zeta$-mode impedance, 
\begin{equation} \label{eqn:EJ-opt-scaling}
    \sqrt{E_J^*/E_{C_\varphi}} \propto \log(Z_\zeta/R_Q),
\end{equation}
which is consistent with the results in Ref.~\cite{Dempster2014}. 
For our simulations of the gate with the $0$-$\pi$ qubit, we choose $E_J = E_J^*$.

\subsection{Effective model}\label{app:effective-model}

Here we provide a derivation and supporting explanations for the effective one-dimensional model for the $0$-$\pi$ Hamiltonian. 
A heuristic effective model for the $0$-$\pi$ qubit is
\begin{equation}\label{eqn:H-heuristic}
	\H = 4 E_{C_\alpha} \n_\alpha^2 - E_{J_\alpha} \cos\alphah - E_{J_{2\alpha}} \cos 2\alphah, 
\end{equation}
with $[\alphah, \n_\alpha] = i$. In the limit of $E_{J_\alpha} \ll E_{C_\alpha} \ll E_{J_{2\alpha}}$, this represents a $4e$-tunnelling element perturbed by a $2e$-tunnelling term. 
This heuristic model was originally used to explain the behaviour of the current mirror qubit~\cite{Kitaev2006}, which the $0$-$\pi$ qubit is based on, and was later used in Refs.~\cite{Shen2015,Paolo2019,Hassani2024} for numerical analyses of the $0$-$\pi$ qubit. 
In Ref.~\cite{Shen2015}, expressions for the parameters $E_{C_\alpha}$, $E_{J_\alpha}$ and $E_{J_{2\alpha}}$ are derived in terms of the parameters for the $0$-$\pi$ qubit in a particular parameter regime. 
This derivation makes use of a Born-Oppenheimer approximation, motivated by the ``mass separation" given by the energy restriction $E_{C_\theta} \ll E_{C_\varphi}$. 
Here we explain that the Born-Oppenheimer approximation is only valid in a restricted parameter regime where, in addition, $E_{C_\theta} < 2E_L$ and $2E_J \ll E_{C_\varphi}$. 
Note that neither of these latter inequalities are a requirement for topological protection of the $0$-$\pi$ qubit [\cref{eqn:zero-pi-energy-constraint}]. 
Moreover, we find that this is not the relevant parameter regime for performing a gate with the $0$-$\pi$ qubit. 
In the optimal operating regime we find that a tight-binding model is a much more accurate description of the qubit. 

In \cref{app:born-oppenheimer-approx} we closely follow the approach taken in Ref.~\cite{Shen2015} to derive an effective model using the Born-Oppenheimer approximation in the regime where $E_{C_\theta} < 2E_L$, and in \cref{app:tight-binding-regime} we derive an effective model using a tight-binding approximation in the regime where $E_{C_\theta} \ge 2E_L$. The latter is the one-dimensional model that is used and discussed in the main text.
In \cref{app:tunnelling-rates} we compute the exponential scalings for the tunnelling rates in the tight-binding model, which are given in \cref{eqn:EJ1,eqn:EJ2}.

\subsubsection{Born-Oppenheimer approximation}\label{app:born-oppenheimer-approx}

The Born-Oppenheimer approximation makes use of the fact that $E_{C_\theta} \ll E_{C_\varphi}$ to separate the dynamics of the ``heavy" degree of freedom ($\theta$) from the ``light" degree of freedom ($\varphi$) by assuming the latter is in its ground state. To this end, we treat $\thetah$ as a classical variable and take $4 E_{C_\theta} \n_\theta \to 0$ in \cref{eqn:zero-pi-theta-phi}, yielding the following Hamiltonian for the $\varphi$ degree of freedom
\begin{equation}\label{eqn:Hphi}
    \H_\varphi = 4 E_{C_\varphi} \n_\varphi^2 + E_L \varphih^2 - 2E_J\cos\theta \cos\varphih, 
\end{equation}
where we have fixed $\varphi_\text{ext} = 0$ for simplicity. 
The effective one-dimensional Hamiltonian in the Born-Oppenheimer approximation is then given by
\begin{equation}
    \H_\text{BO} = 4 E_{C_\theta} \n_\theta^2 + E_0(\thetah), \label{eqn:HBO}
\end{equation}
where $E_0(\thetah)$ is the ground-state energy of \cref{eqn:Hphi}, and $\theta$ has been promoted back to an operator. 

\begin{figure}
    \centering
    \includegraphics[width=\linewidth]{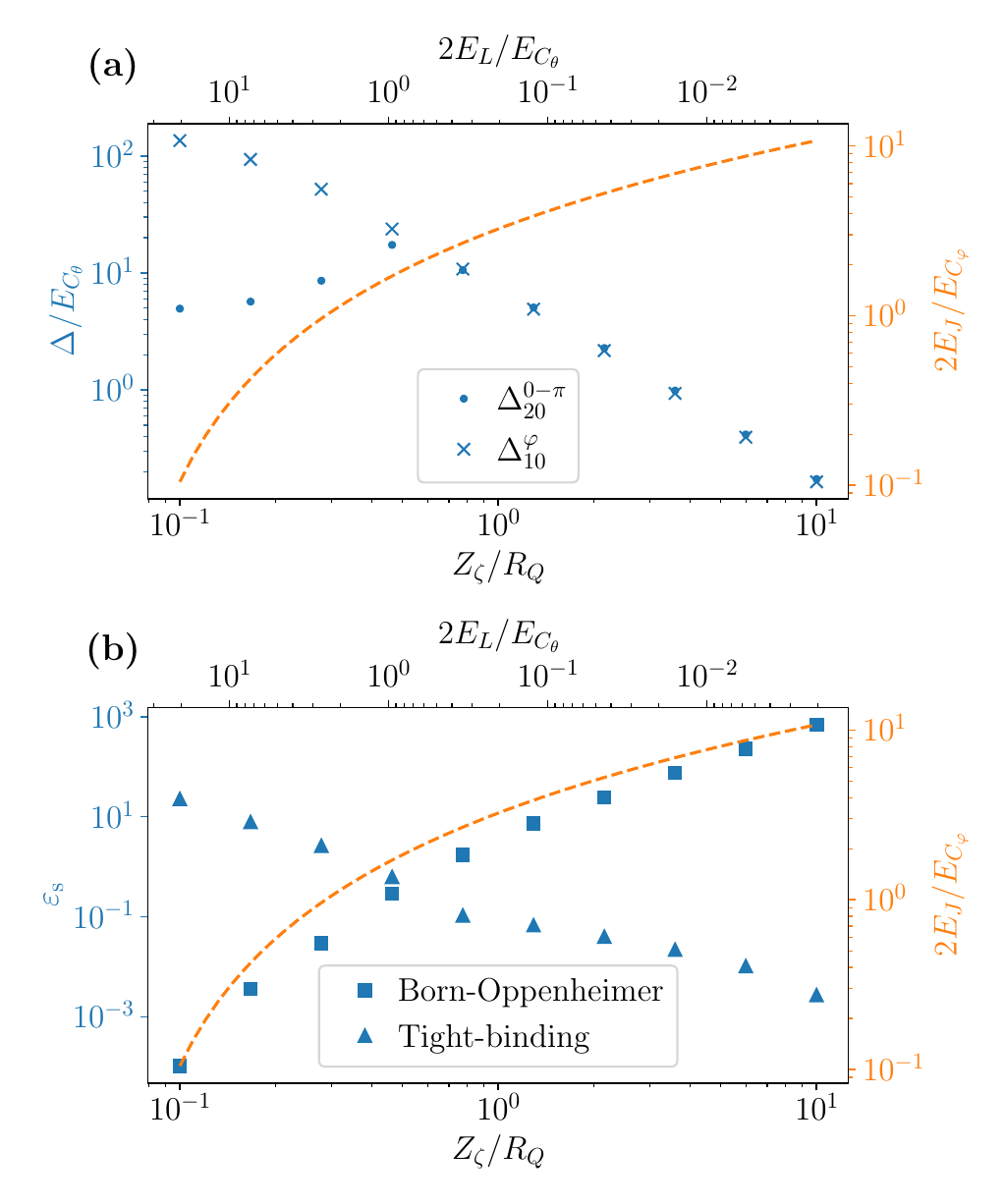}%
    \subfloat{\label{fig:energy-gaps}}%
    \subfloat{\label{fig:spectral-errors}}%
    \caption{Effective models. 
    (a) Energy gaps for the two-dimensional $0$-$\pi$ Hamiltonian [\cref{eqn:zero-pi-theta-phi}] and the Hamiltonian for the $\varphi$ mode [\cref{eqn:Hphi}] as a function of the $\zeta$-mode impedance $Z_\zeta$. Energy gaps for the $0$-$\pi$ Hamiltonian, $\Delta_{02}^{0\mathrm{-}\pi}$, are shown by the dots and energy gaps for the $\varphi$ mode Hamiltonian, $\Delta_{01}^\varphi$ are shown by crosses. $\Delta_{i0}$ denotes the energy gap between the $i$-th excited state and the ground state. (b) Spectral error $\eps_\mathrm{s}$ [\cref{eqn:spectral-error}] for the effective models as a function of the $\zeta$-mode impedance $Z_\zeta$. Spectral errors for the Born-Oppenheimer model [\cref{eqn:HBO}] are shown with squares and spectral errors for the tight-binding model [\cref{eqn:HTB}] are shown with triangles. In (a) and (b) the charging energy ratio is fixed to $E_{C_\varphi}/E_{C_\theta} = 100$, and the corresponding values of $E_J$ and $E_L$ are shown by the orange dashed line and the upper axis, respectively. 
    }
\end{figure}

Crucially, owing to the assumption that the $\varphi$ mode is in its ground state, the spectrum of \cref{eqn:HBO} is only accurate up to the $i$-th energy level of the $0$-$\pi$ Hamiltonian if $\Delta_{i0}^{0\mathrm{-}\pi} < \Delta_{10}^{\varphi}$, where $\Delta^{0\mathrm{-}\pi}_{i0}$ and $\Delta^{\varphi}_{i0}$ denote the energy gap from the ground state to the $i$-th excited state of the $0$-$\pi$ Hamiltonian [\cref{eqn:zero-pi-theta-phi}] and the Hamiltonian for the $\varphi$ mode [\cref{eqn:Hphi}], respectively.
In \cref{fig:energy-gaps} we plot the energy gaps for the second excited state of the $0$-$\pi$ Hamiltonian, $\Delta_{20}^{0\mathrm{-}\pi}$, and the first excited state of the $\varphi$-mode Hamiltonian, $\Delta_{10}^\varphi$.
Here we find that the energy gap for the $0$-$\pi$ Hamiltonian is less than the energy gap for the $\varphi$-mode Hamiltonian when $E_{C_\theta} < 2E_L$, and that the two energy gaps become equal for $E_{C_\theta} \ge 2E_L$.
The reason for this is that when $E_{C_\theta} < 2E_L$, the excitations in the $0$-$\pi$ Hamiltonian are plasmon-like with an energy gap given by $\Delta^{0\mathrm{-}\pi}_{02} \approx 4\sqrt{E_J E_{C_\theta}}$. 
In contrast, the excitations of \cref{eqn:Hphi} are fluxon-like with an energy gap given by $\Delta_{10}^\varphi \approx 4\sqrt{E_L E_{C_\varphi}}$. 
In \cref{fig:energy-gaps} we also plot the value of the Josephson energy in the $0$-$\pi$ Hamiltonian, which is chosen to be the value that optimises the degeneracy of the qubit at zero external flux (see \cref{app:optimal-EJ}). 
This shows that when $E_{C_\theta} < 2E_L$, then $2E_J \lesssim E_{C_\varphi}$, and therefore  $\Delta^{0\mathrm{-}\pi}_{02} \lesssim \Delta_{10}^\varphi$. 
On the other hand, when $E_{C_\theta} \ge 2E_L$ then the excitations in the $0$-$\pi$ Hamiltonian become fluxon-like and the gaps become the same as in the $\varphi$-mode Hamiltonian. 
Therefore, we conclude that the Born-Oppenheimer approximation is only a good approximation when $E_{C_\theta} < 2E_L$ and $2E_J < E_{C_\varphi}$.

We now estimate $E_0(\theta)$ under the above assumptions. 
With $2E_J < E_{C_\varphi}$, \cref{eqn:Hphi} corresponds to a harmonic oscillator with frequency $\omega_\varphi = 4\sqrt{ E_{C_\varphi} E_L}/\hbar$ that is perturbed by the $\theta$-dependent tunnelling term $-2E_J\cos\theta \cos\varphih$. 
Therefore, the ground-state energy of \cref{eqn:Hphi} may be approximated by the first and second-order corrections to the harmonic oscillator frequency,
\begin{equation}
    E_0(\theta) \approx \frac{\hbar\omega_\varphi}{2} + E_0^{(1)} + E_0^{(2)}. \label{eqn:E0theta}
\end{equation}
These corrections are obtained through perturbation theory and may be simplified as follows,
\begin{subequations}
    \begin{align}
        E_0^{(1)} &= -2E_J\bra{0}\cos\varphih\ket{0} \\
        &= -2E_J e^{-\pi Z_\varphi / 2R_Q}  \cos\theta, \label{eqn:E01-2}\\
        E_0^{(2)} &= -\frac{4E_J^2}{\hbar \omega_\varphi}\sum_{n=1}^\infty \frac{\abs{\bra{n} \cos\varphih \ket{0}}^2}{n}   \cos^2\theta\\
        &= -\frac{4E_J^2}{\hbar \omega_\varphi} e^{-\pi Z_\varphi/R_Q} \sum_{k=1}^\infty \frac{(\pi Z_\varphi / R_Q)^{2k}}{2k(2k)!}  \cos^2\theta \label{eqn:E02-2}\\
        &\sim -\frac{2E_J^2 R_Q}{\pi \hbar \omega_\varphi Z_\varphi}  \cos^2\theta \label{eqn:E02-3}\\
        &= -\frac{E_J^2}{2 E_{C_\varphi}}  \cos^2\theta, \label{eqn:E01-4}
    \end{align}
\end{subequations}
where $\ket{n}$ and $Z_\varphi/R_Q = \sqrt{ E_{C_\varphi}/\pi^2 E_L} = \sqrt{L/4 C_J}$ are the eigenstates and impedance for the unperturbed harmonic oscillator, respectively.
In \cref{eqn:E02-2} we used the matrix element expressions given in \cref{eqn:cos-phi-mat-elts}, and in \cref{eqn:E02-3} we used the asymptotic equivalence given in \cref{eqn:asymptotic-equivalence} with $\alpha = 1$.
The latter is justified because $Z_\varphi/R_Q \gg 1$. 

Promoting $\theta$ back to an operator, and substituting \cref{eqn:E0theta,eqn:E01-2,eqn:E01-4} into \cref{eqn:HBO}, the effective one-dimensional Hamiltonian is
\begin{equation}
    \H_\text{BO} \approx 4 E_{C_\theta} \n_\theta^2 - 2E_J e^{-\pi Z_\varphi / 2 R_Q} \cos\thetah - \frac{E_J^2}{4 E_{C_\varphi}} \cos 2\thetah,
\end{equation}
where we have used that $2\cos^2\thetah = \cos2\thetah + 1$ and neglected constant-energy terms. 
We see that this matches the heuristic model [\cref{eqn:H-heuristic}], where the effective degree of freedom $\alphah$ may be identified with the $\theta$-mode of the $0$-$\pi$ qubit, and the effective charging and Josephson energies are given by
\begin{equation}
    E_{C_\alpha} =  E_{C_\theta}, \quad E_{J_\alpha} = 2E_J e^{-\pi Z_\varphi / 2 R_Q}, \quad E_{J_{2\alpha}} = \frac{E_J^2}{4 E_{C_\varphi}}.
\end{equation}
Importantly, the $2e$-tunnelling term is exponentially suppressed in $Z_\varphi$ relative to the $4e$-tunnelling term, agreeing with our intuition for the $0$-$\pi$ qubit being an approximate $\pi$-periodic Josephson element.
	
\subsubsection{Tight-binding approximation}\label{app:tight-binding-regime}

As discussed above, when $E_{C_\theta} \ge 2E_L$, the Born-Oppenheimer approximation is no longer valid and $2E_J$ becomes larger than $E_{C_\varphi}$. 
Now $E_J$ is the dominant energy in the $0$-$\pi$ qubit and we use a tight-binding approximation instead of the Born-Oppenheimer approximation to reduce the Hamiltonian to a one-dimensional model. 
A two-dimensional tight-binding model was also used to derive an effective model for a multimode qubit in Ref.~\cite{Smith2022}. 
We follow a similar approach.

To begin with, we express the Hamiltonian in the basis of flux eigenstates localised at the minima of the potential $-2E_J\cos\theta\cos\varphi$. 
These are located at the positions $(\theta, \varphi) = (m_\theta\pi, m_\varphi\pi)$, where $m_\theta \in \{0,1\}$ and $m_\varphi \in \Z$ such that $m_\theta + m_\varphi \in 2\Z$. 
We then enumerate these minima by a single site index $m = 2\lfloor m_\varphi/2 \rfloor + m_\theta$. 
This facilitates the following one-dimensional tight-binding Hamiltonian
\begin{equation}\label{eqn:TB-1}
\begin{aligned}
    \H_\text{TB} = &{} \sum_{m \in \Z} \Bigl[ E_L (\pi m)^2 \ketbra{m}{m} \\
    &- t_1 \left(\ketbra{m}{m+1} + \text{h.c.}\right) - t_2 \left(\ketbra{m}{m+2} + \text{h.c}\right) \Bigr], 
\end{aligned}
\end{equation}
where $\ket{m}$ denotes the flux eigenstate at the $m$-th cosine well, and $t_1$ and $t_2$ represent nearest- and next-nearest-neighbour tunnelling rates, respectively.
The next-nearest-neighbour tunnelling path is purely in the $\varphi$ direction, and its value is well-approximated by the formula for one-dimensional tunnelling when $E_{C_\varphi} \gg E_{C_\theta}$~\cite{Koch2007} 
\begin{equation} \label{eqn:t2}
    t_2 = \frac{8 E_{C_\varphi}}{\sqrt{\pi}} \left(\frac{4E_J}{E_{C_\varphi}}\right)^{3/4} e^{-4\sqrt{ E_J/E_{C_\varphi}}}.
\end{equation} 
The nearest-neighbour tunnelling rate could be calculated from a two-dimensional WKB or instanton calculation along a numerically computed tunnelling path~\cite{Smith2020,Zwiebach2022}.
In \cref{app:tunnelling-rates} we numerically compute its tunnelling path and show that the tunnelling rate scales as 
\begin{equation} \label{eqn:t1}
    t_1 \sim e^{-2(\sqrt{2} - 1) \sqrt{2E_J/E_{C_\theta}} - \pi(\sqrt{2} - 1)\sqrt{E_J/2 E_{C_\varphi}}}.
\end{equation}
Note that whilst the distance for nearest-neighbour tunnelling is smaller than the next-nearest-neighbour tunnelling path, its rate is exponentially suppressed in $\sqrt{E_{C_\varphi}/E_{C_\theta}}$ relative to $t_2$ due to the component of its tunnelling path in the $\theta$ direction. 
To obtain the exact value of $t_1$, we numerically fit the spectrum for the one-dimensional model to the two-dimensional Hamiltonian with $t_2$ given by the formula in \cref{eqn:t2}.
We show in \cref{app:tunnelling-rates} that its numerically fitted value is consistent with the scaling in \cref{eqn:t1}.

Introducing the operators $\hat{m} = \sum_{m \in \Z} \ketbra{m}$ and $\cos(k\hat{p}_m) = \frac{1}{2}\sum_{m \in \Z} \left(\ketbra{m}{m+k} + \text{h.c.}\right)$, \cref{eqn:TB-1} may be written as
\begin{equation} \label{eqn:HTB}
    \H_\text{TB} = E_L\pi^2 \hat{m}^2 - 2 t_1 \cos(\p_m) - 2 t_2 \cos(2\p_m).
\end{equation}
By identifying $\hat{m}$ with $\n_\alpha$ and $\p_m$ with $\alphah$, this takes the form of \cref{eqn:H-heuristic} where the effective capacitive and Josephson energies are given by
\begin{equation} \label{eqn:TB-hamiltonian-params}
    E_{C_\alpha} = \pi^2 \frac{E_L}{4}, \quad E_{J_\alpha} = 2t_1, \quad E_{J_{2\alpha}} = 2t_2.
\end{equation}
Note that here flux and charge have swapped roles relative to the original interpretation of \cref{eqn:H-heuristic}. 
This is similar to fluxonium where a periodic fluxon basis is used~\cite{Koch2009}. 
We briefly comment that our model differs from that used in Refs.~\cite{Smith2020,Smith2022} in that the phase variable appearing in their $\pi$-periodic term may be identified with one of the phase variables in their original Hamiltonian rather than a quasi-charge variable in the dual conjugate basis.

In this tight-binding model, $\thetah$ may be replaced by $\pi\n_\alpha$ in the interaction terms \cref{eqn:zero-pi-interaction-exp-suppression,eqn:zero-pi-zeta-interaction} since $n_\alpha$ is even when $\theta = 0$ and $n_\alpha$ is odd when $\theta = \pi$. This yields \cref{eqn:zero-pi-interaction-exp-suppression-alpha,eqn:zero-pi-zeta-interaction-alpha}, which we used for the numerical results in \cref{sec:numerical-simulations-zero-pi}. Furthermore, since $\varphih$ becomes $\pi \n_\alpha$ in the effective model, the disorder term $\delta E_L \varphih \zetah$ becomes $\pi \delta E_L \n_\alpha \zetah$, which we used for the numerical results in \cref{sec:circuit-disorder}.

To compare the effective models, we compute their spectral errors for the first five excited states, defined as
\begin{equation} \label{eqn:spectral-error}
    \varepsilon_\mathrm{s} = \frac{1}{5} \sum_{i=1}^5 \frac{\abs{ E_i^\mathrm{1D} - E_i^\mathrm{2D} - E_0^\mathrm{1D} + E_0^\mathrm{2D}}}{E_i^{2D} - E_0^\mathrm{2D}},
\end{equation}
where $E_i^{\mathrm{1D}/\mathrm{2D}}$ denotes the $i$-th eigenstate of the 1D/2D Hamiltonian. \Cref{fig:spectral-errors} shows the spectral errors for both effective models as a function of $Z_\zeta/R_Q$. Consistent with the regime of validity shown in \cref{fig:energy-gaps} and discussed in \cref{app:born-oppenheimer-approx}, we find that the Born-Oppenheimer model is a better approximation to the two-dimensional Hamiltonian when $E_{C_\theta} < 2E_L$, and the tight-binding model is a better approximation when $E_{C_\theta} \ge 2E_L$. Moreover, this transition occurs at the point where $2E_J \approx E_{C_\varphi}$, consistent with the assumptions made in each model. In particular,  \cref{fig:spectral-errors} shows that the tight-binding model is the better effective model when $Z_\zeta/R_Q \ge 1$. Since this is the relevant regime for a protected gate, we use this as the effective model to simulate a gate with the $0$-$\pi$ qubit in \cref{sec:zero-pi}.

\subsubsection{Calculation of tunnelling action}
\label{app:tunnelling-rates}

\begin{figure}
    \centering
    \includegraphics[width=\linewidth]{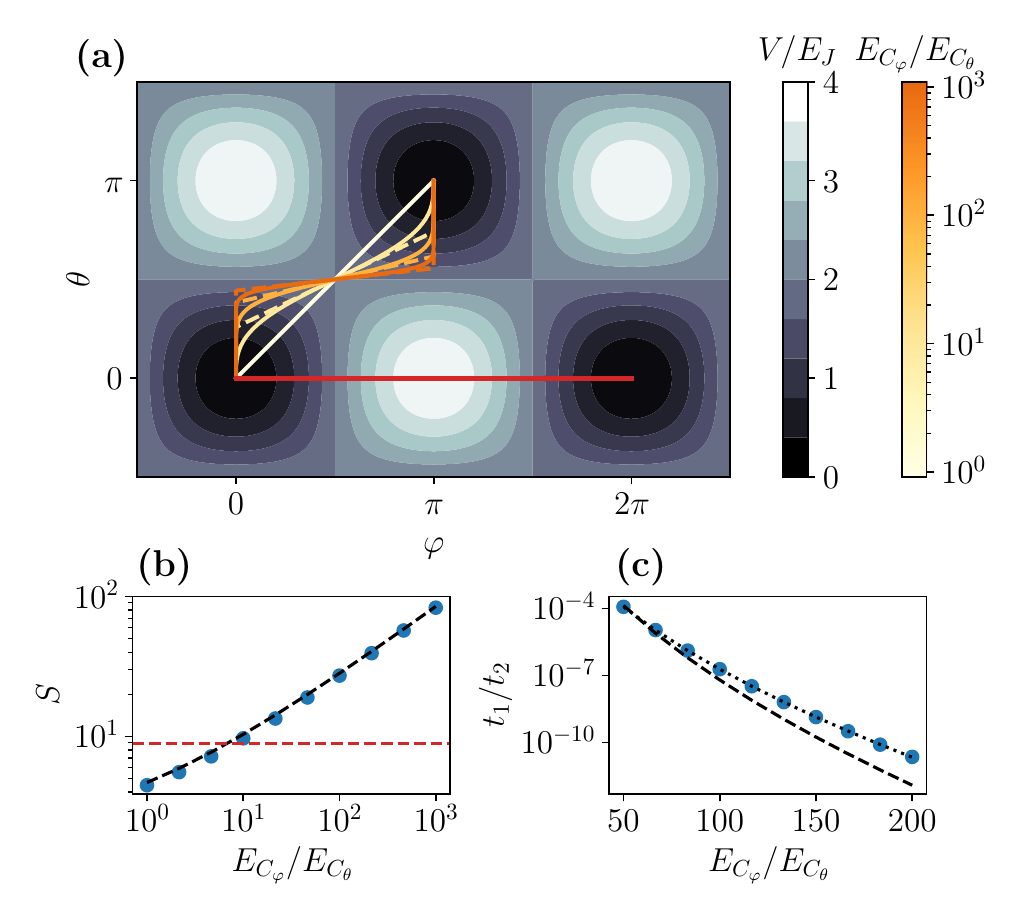}%
    \subfloat{\label{fig:tunnelling-paths}}%
    \subfloat{\label{fig:action}}%
    \subfloat{\label{fig:tunnelling-rates}}%
    \caption{Tight-binding model tunnelling paths and rates. 
    (a) Nearest- and next-nearest-neighbour tunnelling paths for a $0$-$\pi$ qubit with $E_J/E_{C_\varphi} = 5$ and different charging energy ratios $E_{C_\varphi}/E_{C_\theta}$, in the $Z_\zeta \to \infty$ ($E_L = 0$) limit. 
    The numerically computed nearest-neighbour tunnelling paths for different charging energy ratios are shown in the orange solid lines and a piecewise linear approximation for each path is shown by the dashed lines. 
    The next-nearest-neighbour tunnelling path is shown in the red line.
    (b) Classical action $S$ for the tunnelling paths shown in panel (a) as a function of the charging energy ratio $E_{C_\varphi}/E_{C_\theta}$. 
    Blue denotes the numerically computed action for the nearest-neighbour tunnelling paths shown in the solid orange lines in panel (a)
    The black dashed line denotes the theoretical expression for the nearest-neighbour action in \cref{eqn:nearest-neighbour-action}, calculated using the dashed lines in panel (a).
    The horizontal red dashed line denotes the action for the next-nearest-neighbour tunnelling in \cref{eqn:next-nearest-neighbour-action}, corresponding to the red line in panel (a).
    (c) Ratio of nearest- to next-nearest-neighbour tunnelling rates $t_1 / t_2$ as a function of the charging energy ratio $E_{C_\varphi}/E_{C_\theta}$. 
    The blue circles denote the numerically computed tunnelling rates from fitting the one-dimensional Hamiltonian in \cref{eqn:H-alpha} to the two-dimensional Hamiltonian in \cref{eqn:zero-pi-theta-phi} with $Z_\zeta/R_Q = 10$ and $E_J/E_{C_\varphi} = 5$ [the same data as in \cref{fig:tunnelling-rates-mass-ratio}].
    The dashed line denotes the scaling from the estimate for the tunnelling rates computed in \cref{eqn:next-nearest-neighbour-action,eqn:nearest-neighbour-action}, and the dotted line is a fit to the blue circles.
    }
\end{figure}

Here we calculate the exponential scaling of the nearest- and next-nearest-neighbour tunnelling rates $t_1$ and $t_2$ for the tight-binding model, referred to as $E_{J_\alpha}$ and $E_{J_{2\alpha}}$, respectively, in the main text.
Their asymptotic scalings are given by $e^{-S}$, where $S$ is the classical action~\cite{Zwiebach2022,Benderskii1995}:
\begin{equation} \label{eqn:action-integral}
    S = \int_\ell \vect n \cdot d \vect \theta.
\end{equation}
Here, $\vect \theta = (\varphi, \theta)$ and $\vect n = (n_\varphi, n_\theta)$ are two-dimensional position and momentum vectors, and the integral is over the tunnelling path $\ell$ that minimises the action.

For these calculations, we consider the classical kinetic and potential energy corresponding to the two-dimensional $0$-$\pi$ Hamiltonian, \cref{eqn:zero-pi-theta-phi}, in the $E_L \to 0$ limit:
\begin{equation} \label{eqn:classical-T-and-V}
\begin{aligned} 
    T &= 4 E_{C_\varphi} n_\varphi^2 + 4 E_{C_\theta} n_\theta^2, \\
    V &= 2E_J(1 - \cos\varphi\cos\theta).
\end{aligned}
\end{equation}
We have also added $2E_J$ to the potential energy to ensure it is positive.

First, we compute the action for the next-nearest-neighbour tunnelling path.
Its path is a straight line in the $\varphi$ direction as shown by the red line in \cref{fig:tunnelling-paths}.
Therefore, \cref{eqn:action-integral} reduces to
\begin{equation} \label{eqn:next-nearest-neighbour-action-integral}
    S_2 = \int_0^{2\pi} n_\varphi \, d\varphi.
\end{equation}
Here, $n_\varphi$ may be found by equating $T$ and $V$ in \cref{eqn:classical-T-and-V} with $\theta = n_\theta = 0$, yielding $n_\varphi = \sqrt{E_J(1 - \cos\varphi) / 2E_{C_\varphi}}$.
Substituting into \cref{eqn:next-nearest-neighbour-action-integral} and computing the integrals results in 
\begin{equation} \label{eqn:next-nearest-neighbour-action}
    S_2 = 4\sqrt{\frac{E_J }{ E_{C_\varphi} }},
\end{equation}
which leads to the exponential scaling in \cref{eqn:EJ2,eqn:t2}.

Next, we compute the action for the nearest-neighbour tunnelling path.
In the case of equal charging energies, the nearest-neighbour tunnelling path is given by the diagonal line connecting adjacent minima, as shown by the white line in \cref{fig:tunnelling-paths}.
Solving the integral in \cref{eqn:action-integral} along this diagonal path $\theta = \varphi$ with $E_{C_\theta} = E_{C_\varphi}$ in \cref{eqn:classical-T-and-V} leads to $S_1 = S_2 / 2$.
This means that nearest-neighbour tunnelling is preferred over next-nearest-neighbour tunnelling when the charging energies are equal.

In the case of unequal charging energies, the tunnelling path is no longer a straight line.
To find the tunnelling paths for unequal charging energies we numerically solve the classical equations of motion corresponding to \cref{eqn:classical-T-and-V} with an inverted potential, following the approach of Refs.~\cite{Benderskii1995,Smith2020}.
The solid lines in \cref{fig:tunnelling-paths} show the numerically computed paths over a range of charging energy ratios $E_{C_\varphi}/E_{C_\theta}$. 

To obtain an estimate for the action corresponding to these paths, we approximate each path by the piecewise linear paths shown in the dashed lines in \cref{fig:tunnelling-paths}.
We set the slope of the diagonal segment to $\sqrt{E_{C_\theta}/E_{C_\varphi}}$, which we find is very close to the gradient of the numerically computed tunnelling paths at the saddle point $(\varphi, \theta) = (\pi/2, \pi/2)$.
This slope determines the angle of the momentum vector, and its magnitude is found by equating $T$ and $V$ in \cref{eqn:classical-T-and-V}.
The action along the diagonal segment is then found by computing the integral in \cref{eqn:action-integral} along this line.
The action for each vertical segment is found by setting $\varphi = n_\varphi = 0$ in \cref{eqn:classical-T-and-V} and computing the integral in \cref{eqn:action-integral} from $\theta = 0$ to $\theta = (1 - \sqrt{E_{C_\theta}/E_{C_\varphi}})\pi/2$.
Summing the contributions from each linear segment and expanding to leading order in $E_{C_\theta}/E_{C_\varphi}$ leads to the following action
\begin{equation} \label{eqn:nearest-neighbour-action}
    S_1 \approx 2(\sqrt{2} - 1)\sqrt{\frac{2E_J}{E_{C_\theta}}} 
    + (\sqrt{2} - 1)\pi\sqrt{\frac{E_J}{2E_{C_\varphi}}},
\end{equation}
which is the scaling given in \cref{eqn:t1,eqn:EJ1}.

In \cref{fig:action} we plot the numerically computed action for the nearest-neighbour tunnelling as a function of the charging energy ratio $E_{C_\varphi}/E_{C_\theta}$ at a fixed value of $E_J / E_{C_\varphi} = 5$ in blue along with the equation for $S_1$ in \cref{eqn:nearest-neighbour-action} in the black dashed line.
We find good agreement over the entire range of charging energy ratios and that the agreement improves with increasing $E_{C_\varphi}/E_{C_\theta}$.
We also plot the action for next-nearest-neighbour tunnelling $S_2$ from \cref{eqn:next-nearest-neighbour-action} in the red dashed line.
This shows that the nearest-neighbour action exceeds the next-nearest-neighbour action for $E_{C_\varphi} / E_{C_\theta} \ge 10$.
Therefore, next-nearest-neighbour tunnelling is preferred over nearest-neighbour tunnelling (meaning that the qubit behaves as an approximately $\pi$-periodic Josephson element) when $E_{C_\varphi} / E_{C_\theta} \ge 10$ for sufficiently large $E_J$.

In \cref{fig:tunnelling-rates} we compare the analytically calculated scalings for the tunnelling rates to the numerically fitted values plotted in \cref{fig:tunnelling-rates-mass-ratio}.
Recall that the latter are obtained by fitting the one-dimensional Hamiltonian \cref{eqn:H-alpha} to the two-dimensional \cref{eqn:zero-pi-theta-phi} with $E_J / E_{C_\varphi} = 5$ and $Z_\zeta/R_Q = 10$, whereas the former are computed in the $Z_\zeta \to \infty$ limit.
The blue circles show the numerically fitted tunnelling rates and the black dashed line shows the scaling from \cref{eqn:next-nearest-neighbour-action,eqn:nearest-neighbour-action}.
A fit to the numerically computed tunnelling rates, shown by the dotted black line, shows that their scaling is 
\begin{equation}
    \frac{t_1}{t_2} \sim e^{-2.2\sqrt{E_{C_\varphi}/E_{C_\theta}}}.
\end{equation}
The prefactor in the exponent is close to the value obtained from \cref{eqn:nearest-neighbour-action,eqn:next-nearest-neighbour-action}, which is $2(\sqrt{2} - 1)\sqrt{2E_J/E_{C_\varphi}} \approx 2.7$ for $E_J/E_{C_\varphi} = 5$.
The smaller value from the numerical fit is likely due to the nonzero $E_L$, which will contribute more to the next-nearest-neighbour action than the nearest-neighbour action.

\bibliography{bibliography}

\end{document}